\documentclass[epj,nopacs]{svjour}
\usepackage{graphics}
\newcommand{\la}{\left<}
\newcommand{\ra}{\right>}

\newcommand{\nvecl}{\underline{n}_l}

\newcommand{\rvec}{\ensuremath{\underline{r}}}
\newcommand{\rvecl}{\ensuremath{\underline{r}_l}}

\newcommand{\ddiff}{\ensuremath{\mathrm{d}}}

\newcommand{\rx}{r_x}
\newcommand{\ry}{r_y}

\newcommand{\rl}{r_{l}}

\newcommand{\nlx}{n_{l,x}}
\newcommand{\nly}{n_{l,y}}

\newcommand{\rcut}{r_\mathrm{cut}}

\newcommand{\kB}{k_\mathrm{B}}
\newcommand{\Hhat}{\ensuremath{\hat{H}}}
\newcommand{\tauhat}{\ensuremath{\hat{\sigma}}}
\newcommand{\muA}{\ensuremath{\mu_\mathrm{A}}}
\newcommand{\muAhat}{\ensuremath{\hat{\mu}_\mathrm{A}}}

\newcommand{\muF}{\ensuremath{\mu_\mathrm{F}}}

\newcommand{\muFone}{\ensuremath{\mu_1}}
\newcommand{\muFtwo}{\ensuremath{\mu_0}}
\newcommand{\muSF}{\ensuremath{\mu_\mathrm{sf}}}

\newcommand{\tincr}{\delta t}
\newcommand{\tsamp}{\Delta t}
\newcommand{\tsampmax}{{\Delta t}_{\mathrm{max}}}
\newcommand{\tspacer}{{\Delta t}_{\mathrm{spac}}}
\newcommand{\ttemper}{{\Delta t}_{\mathrm{temp}}}

\newcommand{\xsamp}{\Delta x}

\newcommand{\xbf}{\ensuremath{\mathbf{x}}}

\newcommand{\Nc}{\ensuremath{N_\mathrm{c}}}

\newcommand{\Nk}{\ensuremath{N_\mathrm{k}}}

\newcommand{\Nm}{\ensuremath{N_\mathrm{m}}}

\newcommand{\Ocal}{\ensuremath{{\cal O}}}

\newcommand{\dvtot}{\ensuremath{\delta v^2_{\mathrm{tot}}}}
\newcommand{\svtot}{\ensuremath{\delta v_{\mathrm{tot}}}}
\newcommand{\dvint}{\ensuremath{\delta v^2_{\mathrm{int}}}}
\newcommand{\svint}{\ensuremath{\delta v_{\mathrm{int}}}}
\newcommand{\dvext}{\ensuremath{\delta v^2_{\mathrm{ext}}}}
\newcommand{\svext}{\ensuremath{\delta v_{\mathrm{ext}}}}
\newcommand{\dvgauss}{\ensuremath{\delta v^2_{\mathrm{G}}}}
\newcommand{\svgauss}{\ensuremath{\delta v_{\mathrm{G}}}}

\newcommand{\svpeak}{\ensuremath{\left.\delta v_{\mathrm{G}}/v\right|_{\mathrm{max}}}}

\newcommand{\cinf}{\ensuremath{c_{\infty}}}

\newcommand{\fcusp}{\ensuremath{f_\mathrm{cusp}}}
\newcommand{\fbeta}{\ensuremath{f_{\beta}}}
\newcommand{\csg}{\mbox{$c_s^{\mathrm{g}}$}}
\newcommand{\hsg}{\mbox{$h_s^{\mathrm{g}}$}}
\newcommand{\RA}{\ensuremath{R_\mathrm{A}}}
\newcommand{\Rinf}{\ensuremath{R_{\infty}}}
\newcommand{\Ta}{\ensuremath{T_2}}
\newcommand{\Tb}{\ensuremath{T_4}}
\newcommand{\Tc}{\ensuremath{T_3}}
\newcommand{\gDebye}{\ensuremath{g_{\mathrm{Debye}}}}

\newcommand{\pmax}{p_{\mathrm{max}}}

\newcommand{\fhat}{\hat{f}}

\newcommand{\Rstar}{R_\mathrm{M}}

\newcommand{\tauA}{\tau_\mathrm{A}}
\newcommand{\taustar}{\tau_\mathrm{M}}
\newcommand{\taualph}{\tau_{\alpha}}

\newcommand{\Snonerg}{\Delta_{\mathrm{ne}}}
\newcommand{\Tnonerg}{\tau_{\mathrm{ne}}}
\newcommand{\Tglass}{T_{\mathrm{g}}}
\newcommand{\Tdemix}{T_{\mathrm{frac}}}
\newcommand{\taualpha}{\tau_{\alpha}}

\newcommand{\taubasin}{\tau_{\mathrm{b}}}

\newcommand{\gamext}{\gamma_{\mathrm{ext}}}

\begin{document}

\title{Ensemble fluctuations matter for variances of macroscopic variables}

\author{G. George
\and L. Klochko
\and A.N. Semenov
\and J. Baschnagel
\and J.P.~Wittmer\thanks{joachim.wittmer@ics-cnrs.unistra.fr}
}
%\affiliation{Institut Charles Sadron, Universit\'e de Strasbourg \& CNRS, 23 rue du Loess, 67034 Strasbourg Cedex, France}
\institute{Institut Charles Sadron, Universit\'e de Strasbourg \& CNRS, 23 rue du Loess, 67034 Strasbourg Cedex, France}
\date{Received: date / Revised version: date}

\abstract{Extending recent work on stress fluctuations in complex fluids and amorphous solids 
we describe in general terms 
the ensemble average $v(\tsamp)$ and the standard deviation $\delta v(\tsamp)$ of the variance $v[\xbf]$ of time series $\xbf$ 
of a stochastic process $x(t)$ measured over a finite sampling time $\tsamp$.
Assuming a stationary, Gaussian and ergodic process, $\delta v$ is given by 
a functional $\svgauss[h]$ of the autocorrelation function $h(t)$.
$\delta v(\tsamp)$ is shown to become large and similar to $v(\tsamp)$ 
if $\tsamp$ corresponds to a fast relaxation process.
Albeit $\delta v = \svgauss[h]$ does not hold in general for non-ergodic systems,
the deviations for common systems with many microstates are merely finite-size corrections.
Various issues are illustrated for shear-stress fluctuations in simple coarse-grained model systems.}

%\date{\today}
\maketitle

\section{Introduction}
\label{sec_intro}
Let us consider a stochastic dynamical variable $x(t)$, a generalized coordinate
characterizing a large physical system, like certain density fields averaged over the system volume.
Ensembles of discrete time series 
$\xbf = \{x_i = x(t_i),i=1,\ldots,I\}$ 
are sampled with the data sequence taken at equally spaced times $t_i = i \tincr$ from $t_1=\tincr$ 
up to the ``sampling time" $\tsamp = I \tincr$.\footnote{We frequently switch between a discrete and 
a continuous representation $i \leftrightarrow t, I \leftrightarrow \tsamp, h_i \leftrightarrow h(t),
c_i \leftrightarrow c(t), \ldots$} % \cite{foot_disc_cont}.
We focus on the ensemble average $v$ and the standard deviation $\delta v$ 
of the (empirical) variance\footnote{The 
empirical variance is defined here without the usual ``Bessel correction" \cite{numrec}.
Equation~(\ref{eq_vxdef}) is the formal definition of $v[\xbf]$ which coincides
with the genuine variance of $x(t)$ only in the limit $\tsamp \propto I \to \infty$.} %foot_variance}
\begin{equation}
v[\xbf] \equiv \frac{1}{I} \sum_{i=1}^{I} x_i^2 - \frac{1}{I^2} \sum_{i,j=1}^{I} x_i x_j.
\label{eq_vxdef}
\end{equation}
Extending recent work on stress fluctuations 
\cite{SBM11,XWP12,WXP13,WXB15,WXBB15,WKB15,LXW16,WXB16,WKC16,ivan17a,ivan17c,ivan18,film18,lyuda19a}
we want to give a systematic and uncluttered overview of three general points 
of relevance for a large variety of problems in condensed matter 
\cite{FerryBook,GraessleyBook,DoiEdwardsBook,RubinsteinBook,HansenBook,GoetzeBook,ChaikinBook},
material modeling \cite{TadmorCMTBook,TadmorMMBook} 
and in computational physics \cite{AllenTildesleyBook,LandauBinderBook}.
One important motivation is that many physical quantities can be obtained by 
{\em equilibrium} molecular dynamics (MD) or Monte Carlo (MC) simulations \cite{AllenTildesleyBook,LandauBinderBook} 
using fluctuation relations \cite{Lebowitz67}. Studying how the respective variances $v$ and their 
standard deviations $\delta v$ evolve with the computational feasible length $\tsamp$ of the production runs of the simulations
is thus of particular interest.

%\paragraph*{First key point.}
We assume here
that $x(t)$ is a {\em stationary} stochastic process respecting the time-translational invariance 
\cite{vanKampenBook}. 
Our first point is that the expectation value $v$ for sampling times $\tsamp$ smaller 
then the terminal relaxation time $\tau$ is not necessarily 
a $\tsamp$-independent constant as often tacitly assumed.\footnote{The longest relaxation time $\tau$ 
of glass-forming liquids is generally called $\taualpha$ \cite{HansenBook,GoetzeBook,FerryBook}.} 
This is seen (Sec.~\ref{theo_stationary}) from the ``stationarity relation" \cite{WXB15,WKB15,WKC16,ivan17c,ivan18,film18,lyuda19a}
\begin{equation}
v = \frac{2}{I^2} \sum_{i=1}^{I-1} (I-i)  \ h_i 
\mbox{ with } h_{i-j} = \la (x_i-x_j)^2 \ra/2
\label{eq_key_1}
\end{equation}
being the autocorrelation function (ACF) characterizing the mean-square displacements of 
the data entries $x_i$.\footnote{The ensemble average $\la \ldots \ra$ may be computed by 
taking the arithmetic average over $\Nc$ independently prepared and sampled configurations $c$. 
For ergodic systems it is equivalent to sample over $\Nk$ sub-intervals of length $\tsamp$ 
of a very long trajectory of length $\tsampmax \gg \tau$.} % \cite{foot_ensaver}.
Hence, $v$ generally depends on $I$ or $\tsamp$ and this is especially relevant
if the ACF $h_i = h(t_i)$ increases strongly for $t \approx \tsamp$. 

%\paragraph*{Second key point.}
%
Our second and most central point concerns the standard deviation $\delta v$ of $v[\xbf]$.
It has been observed for shear-stress fluctuations \cite{WKC16,ivan17c,ivan18,film18,lyuda19a}
that $\delta v$ may become rather large and of the order of the mean value $v$
if $h(t)$ varies strongly for $t \approx \tsamp$, 
i.e. the mean behavior standard experimental or theoretical work focuses on 
\cite{FerryBook,DoiEdwardsBook,TadmorCMTBook,GoetzeBook}
gets masked by strong fluctuations.
Reworking Ref.~\cite{lyuda19a} this can be simply understood assuming a stationary {\em Gaussian} 
stochastic process which implies that
\begin{eqnarray}
\delta v  & = & \svgauss[h] \mbox{ with } \label{eq_key_2} \\
\dvgauss[h] & \equiv & \frac{1}{2I^4} \sum_{i,j,k,l=1}^{I} \ g_{ijkl}^2 \ \mbox{ and } \nonumber \\
g_{ijkl} & \equiv & (h_{i-j} + h_{k-l}) - (h_{i-l} + h_{j-k}).
\nonumber
\end{eqnarray}
as shown in Sec.~\ref{theo_wick}. By analyzing the functional $\svgauss[h]$ it will be seen 
(Secs.~\ref{theo_properties} and \ref{theo_models})
that while $\delta v(\tsamp)$ must remain small for $h(t \sim \tsamp) \approx$ constant, 
$\delta v(\tsamp)$ becomes generally large if $\tsamp$ is similar to the characteristic time of an efficient 
relaxation pathway corresponding to a strong change of $h(t)$ for $t \approx \tsamp$.
%

%\paragraph*{Third key point.}
%
Our third key point emphasizes one limitation of Eq.~(\ref{eq_key_2}) 
which hinges on the ergodicity of the stochastic process. 
If the system is (strictly or in practice) {\em non-ergodic}, i.e. if independently 
created trajectories $c$ are restricted to different meta-basins of the generalized phase space,
this implies as shown in Sec.~\ref{theo_nonerg} that
\begin{equation}
\delta v(\tsamp) \to \Snonerg = \mbox{constant} \mbox{ for } \tsamp \gg \Tnonerg \gg \taubasin
\label{eq_key_3}
\end{equation}
with $\taubasin$ being the typical relaxation time of the meta-basins,
$\Tnonerg$ a crossover time defined below and 
$\Snonerg$ the static standard deviation of the quenched variances $v_c$ of the configurations $c$. 
In this limit $\delta v(\tsamp)$ must thus differ from $\svgauss(\tsamp) \propto 1/\sqrt{\tsamp}$ for $\tsamp \gg \taubasin$.
However, as argued in Sec.~\ref{theo_V}, in the common case where the observables $x(t)$ average over many, 
more or less decoupled microstates, the quenched $v_c$ become similar with increasing system size and, hence, 
$\delta v \to \svgauss[h]$ in the macroscopic limit even for non-ergodic systems.
 
%\paragraph*{Outline.}
%
Various relations and issues discussed theoretically in Secs.~\ref{theo_station}-\ref{theo_V}
are illustrated for different coarse-grained model systems in Sec.~\ref{sec_shear} and Appendix~\ref{app_shear}.
The paper concludes in Sec.~\ref{sec_conc} with a summary and an outlook to future work.
Numerically more convenient reformulations of Eq.~(\ref{eq_key_2}) are given in Appendix~\ref{app_reformulations}.
The definitions of the instantaneous shear stress and the corresponding
Born-Lam\'e coefficient are reminded in Appendix~\ref{app_affine}. 
The three coarse-grained models simulated are presented in Appendix~\ref{app_algo} 
together with some technical details related to the data processing (Appendix~\ref{app_algo_data}).

\section{Stationary stochastic processes}
\label{theo_station}
%\section{Theoretical considerations}
%\label{sec_theo}

%\input{theo_intro}
\subsection{Introduction}
\label{theo_intro}

Having measured and stored the $I$ entries $x_i$ of a time series $\xbf = \{x_i,i=1,\ldots,I\}$ 
various functionals $\Ocal[\xbf]$ may be computed, e.g., the moments 
$m_{\alpha\beta}[\xbf] \equiv ( \sum_{i=1}^{I} x_i^{\alpha}/I)^{\beta}$.
As stated in the Introduction we focus in this work on the variance $v[\xbf] = m_{21}[\xbf] - m_{12}[\xbf]$.
Note that $v[\xbf]=0$ for $I=1$.
It is also useful to consider functionals with a discrete time lag $s$ 
(with $s=0,\ldots,I-1$) such as the ``gliding average" 
\cite{AllenTildesleyBook}
\begin{equation}
\csg[\xbf] \equiv \frac{1}{I-s} \sum_{i=1}^{I-s} c_{s,i} \mbox{ with } c_{s,i} = x_{i+s} x_i \label{eq_csxdef} 
\end{equation}
and correspondingly for $\hsg[\xbf]$ with
\begin{equation}
h_{s,i} = \frac{1}{2}\left(x_{i+s} - x_i\right)^2  = \frac{x_{i}^2+x_{i+s}^2}{2} - c_{s,i}.
\label{eq_hsxdef} 
\end{equation}
Obviously, $c_0^{\mathrm{g}}[\xbf]=m_{21}[\xbf]$ and $h_0^{\mathrm{g}}[\xbf]=0$.
Averages over a given time series are called {``\em $t$-averages"}.
Since the functionals $\Ocal[\xbf]$ are obtained in general from {\em correlated} data entries,
ensemble averages $\la \ldots \ra$ of fluctuation-type functionals may depend 
on the sampling time $\tsamp$. This is not the case for ``{\em simple averages}" \cite{AllenTildesleyBook,WXB16,WKC16} 
for which the ensemble average over independent trajectories and the $t$-average {\em commute}.
For instance, we have
\begin{equation}
m_{\alpha1} = 
\la \frac{1}{I} \sum_{i=1}^{I} x_i^{\alpha} \ra = \frac{1}{I} \sum_{i=1}^{I} \la x_i^{\alpha} \ra 
\propto \tsamp^0 
\label{eq_commute}
\end{equation}
since the ensemble average $\la x_i^{\alpha} \ra$ is $\tsamp$-independent.
Interestingly, the commutation of both averaging-operators is not possible for 
$m_{\alpha\beta}$ with $\beta \ne 1$.
An argument $\tsamp$ often marks below a property being {\em not} a simple average.
\subsection{Stationarity}
\label{theo_stationary}

%\paragraph*{Correlation functions.}
We suppose that the time series is taken from a {\em stationary} stochastic process whose 
joint probability distribution does not change when shifted in time
\cite{vanKampenBook}. Correlation functions such as $\la x_i x_j\ra$ thus only depend 
on the difference $s=|i-j|$ of the discrete indices $i$ and $j$. We thus define 
\begin{equation}
c_s = \la  \csg[\xbf] \ra \mbox{ and } h_s = c_0-c_s = \la \hsg[\xbf] \ra
\label{eq_cij_hij}
\end{equation}
with $0 \le s < I$ in terms of $\csg[\xbf]$ and $\hsg[\xbf]$ 
defined in Sec.~\ref{theo_intro}. Note that both $c_s$ and $h_s$ are simple averages,
i.e. they do not depend on $\tsamp$ \cite{WKB15,WKC16}.
Note also that $c_0 = m_{21} = \la m_{21}[\xbf] \ra$ and $h_0 = 0$.
See Sec.~\ref{theo_gauss_dh} for a subtle point related to the fluctuations $\delta c_s$ and $\delta h_s$.
 
%\paragraph*{Expectation value $v(\tsamp)$.}
%
Due to the assumed stationarity, the ensemble average $v = \la v[\xbf] \ra$ of Eq.~(\ref{eq_vxdef}) becomes
\cite{LandauBinderBook,AllenTildesleyBook,WXB15,WKB15,WKC16,ivan17a,ivan17c,ivan18,film18,lyuda19a} 
\begin{eqnarray}
v(\tsamp) 
& = & \frac{1}{I} \sum_{i=1}^{I} \la x_i^2 \ra \nonumber \\
     & - & 
\left( \frac{1}{I^2} \sum_{i=1}^{I} \la x_i^2 \ra 
+ \frac{2}{I^2} \sum_{k=1}^{I-1} (I-k) \la x_{k+1} x_1 \ra \right) \nonumber \\
& = &  c_0 \ (1- I^{-1}) - \frac{2}{I^2} \sum_{k=1}^{I-1} (I-k) c_k \nonumber \\
     & = & \frac{2}{I^2} \sum_{i=1}^{I-1} (I-i) \ h_i \label{eq_vb}
\end{eqnarray}
as already stated in the Introduction, Eq.~(\ref{eq_key_1}).
Note that in the last step it was used that $h_s = c_0 - c_s$ and 
\begin{equation}
\frac{2}{I^2} \sum_{k=1}^{I-1} (I-k) =  1-1/I.
\label{eq_help1}
\end{equation}
In statistical mechanics Eq.~(\ref{eq_vb})
is closely related to the equivalence of the Green-Kubo
and the Einstein relations for transport coefficients
\cite{HansenBook,LandauBinderBook,AllenTildesleyBook,ivan18,film18}.\footnote{This 
may be better seen from the continuum representation of Eq.~(\ref{eq_Rt2M}) rewritten as
$$\int_0^t \ddiff s \ R(s) = \frac{\ddiff}{\ddiff t} [(M(t)t^2/2].$$
If the left-hand side converges to a constant $\eta$ for $t \to \infty$ 
this implies $M(t)t^2 \to 2 \eta t$.
}
Albeit the mentioned $\tsamp$-dependence is well known \cite{AllenTildesleyBook,LandauBinderBook}
it is emphasized here for systematic reasons and since $\tsamp$-effects for such
fluctuations are rarely checked \cite{SBM11,WXP13}.
We also remind \cite{WKC16,ivan17a,ivan17c,ivan18,film18,lyuda19a}
that in the continuum limit for large $I = \tsamp/\delta t$, Eq.~(\ref{eq_vb}) reads
\begin{equation}
v(\tsamp) = \frac{2}{\tsamp^2} \int_0^{\tsamp} \ddiff t \  (\tsamp-t)  \ h(t) 
\label{eq_vc}
\end{equation}
with $h(t)$ being the continuum limit of $h_s$.
This result may be restated equivalently using the inverse relation 
$h(t) = [v(t) t^2/2]^{\prime\prime}$ with a prime denoting a derivative with respect to time
\cite{ivan18,lyuda19a}.
Using that $m_{21}$ is a simple average Eq.~(\ref{eq_vc}) implies that
\begin{equation}
m_{12}(\tsamp) = m_{21} - 
\frac{2}{\tsamp^2} \int_0^{\tsamp} \ddiff t  \ (\tsamp-t)  \ h(t). \label{eq_m12tsamp}
\end{equation}
The ensemble averages $v(\tsamp)$ and $m_{12}(\tsamp)$ thus depend in general 
on the sampling time $\tsamp$. 
However, the $\tsamp$-dependence disappears, it $h(t)$ becomes constant.
For instance, this is the case, if $h(t)$ plateaus in an intermediate,
sufficiently large, time window, i.e. $h(t) \approx h_p = c(0)-c_p$ with
$h_p$ and $c_p$ being constants. We then have
\begin{eqnarray}
v(\tsamp) & \approx & h_p = c(0)-c_p = \mbox{constant}, \nonumber \\
m_{12}(\tsamp) & \approx & m_{21} - h_p = c_p = \mbox{constant}.
\label{eq_plateau}
\end{eqnarray}
Equation~(\ref{eq_plateau}) also holds, if $c(t)$ tends to a constant for times much longer
than the terminal relaxation time $\tau$ of the system. % \cite{foot_taualpha}.
Then, $c_p$ in Eq.~(\ref{eq_plateau}) is replaced by the long-time limit
$\cinf = \lim_{t\to \infty} c(t) = \lim_{\tsamp \to \infty} m_{12}(\tsamp)$.
 
\subsection{Linear response and generalized modulus}
\label{theo_linear_response}

The functions $h(t)$ and $c(t)$ can be related to the linear response
to an external perturbation conjugate to $x(t)$. 
Let $R(t)$ denote the linear response function of the
system to a weak external perturbation that is instantaneously switched on at $t=0$
and held constant for $t > 0$. By virtue of the fluctuation-dissipation theorem
one can show that \cite{HansenBook,DoiEdwardsBook,WXB15,WKC16}
\begin{equation}
R(t) = R_A - h(t) = (R_A - c(0)) + c(t)
\label{eq_Rt}
\end{equation}
with $R_A = R(0)$ being a constant characterizing the initial response of the
system after the external perturbation is applied.\footnote{Equation~(\ref{eq_Rt}) 
holds if the perturbation is a ``deformation". In the case of an externally applied 
``force" it becomes $R(t)=h(t)$ \cite{DoiEdwardsBook}.} %\cite{foot_Rt}.
For elastic properties this constant is given by a Born-Lam\'e affine modulus
(Appendix~\ref{app_affine}) \cite{WKC16,ivan17a,ivan17c,ivan18,film18,lyuda19a}.
$R(t)$ is a simple average just as $h(t)$ and $c(t)$.
Note that
\begin{equation}
\Rinf \equiv \lim_{t \to \infty} R(t) = R_A - c(0) + \cinf
\label{eq_Rinf}
\end{equation}
may in general be finite.
We rewrite now Eq.~(\ref{eq_vc}) in terms of $R(t)$ as
\begin{equation}
M(\tsamp) \equiv R_A - v(\tsamp) = 
\frac{2}{\tsamp^2} \int_0^{\tsamp} \ddiff t \ (\tsamp-t) R(t)
\label{eq_Rt2M}
\end{equation}
with $M(\tsamp)$ being the ``generalized dynamical modulus" \cite{WKC16}.
Although this modulus does in general depend on $\tsamp$, it becomes constant
%\begin{equation}
$M(\tsamp) \to \Rinf$ % = R_A-c(0)+\cinf \mbox{ for } \tsamp/\tau \to \infty.
for $\tsamp/\tau \to \infty$.
%\label{eq_Minf}
%\end{equation}
%
Being a second integral over $R(t)=[M(t)t^2/2]^{\prime\prime}$, % \cite{foot_GK_EH}, 
$M(\tsamp)$ is a smoothing function statistically better behaved than $R(t)$
and containing in general information about both the reversibly stored energy and 
the dissipation processes.

\begin{figure}[t]
\centerline{\resizebox{0.9\columnwidth}{!}{\includegraphics*{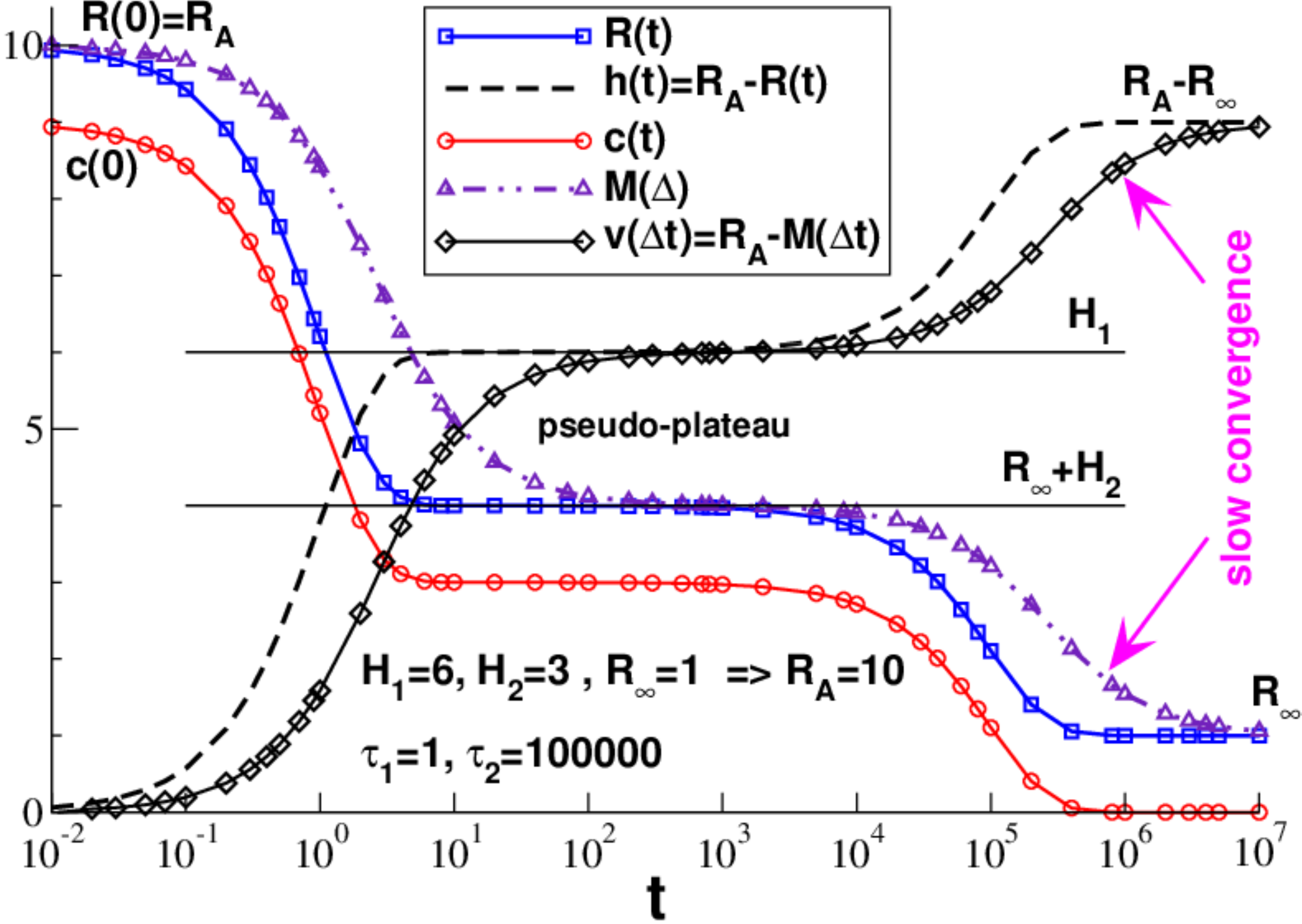}}}
\caption{Illustration of properties discussed in Sec.~\ref{theo_gen_Maxwell}
using a two-modes version of Eq.~(\ref{eq_Rt_genMaxwell})
with amplitudes $\Rinf=1$, $H_1=6$, $H_2=3$, i.e. $R(0)=\RA = \Rinf + H_1 + H_2 = 10$,
and relaxation times $\tau_1=1$ and $\tau_2=100000$.
$h(t)$ and $c(t)$ are given by Eq.~(\ref{eq_Rt}), where we have additionally set $\cinf=0$,
and $M(\tsamp)$ and $v(\tsamp)$ by means of Eqs.~(\ref{eq_Rt2M}) and (\ref{eq_M_MW}).
The two solid horizontal lines mark the intermediate pseudo-plateau for $\tau_1 \ll t \ll \tau_2$.
$v(\tsamp)$ and $M(\tsamp)$ are seen to converge much more slowly to the respective plateau values than 
the corresponding response functions $h(t)$ and $R(t)$.
}
\label{fig_theo_stationary}
\end{figure}
 
\subsection{Generalized Maxwell model}
\label{theo_gen_Maxwell}
%\paragraph*{Generalized Maxwell model.}

Response functions are often fitted using the generalized Maxwell model 
\begin{equation}
R(t) = \Rinf +  \sum_{p=1}^{\pmax} H_p \exp(-t/\tau_p)
\label{eq_Rt_genMaxwell}
\end{equation}
with $H_p$ and $\tau_p$ being, respectively, the amplitude and the relaxation
time of the mode $p$ \cite{FerryBook,RubinsteinBook}. 
Note that $R(0) = \Rinf + \sum_{p=1}^{\pmax} H_p$. % and $\lim_{t\to \infty} R(t) = \Rinf$.
(Commonly, one considers logarithmic time scales for $R(u)$ with $u \propto \log(t)$
and the modes are distributed logarithmically in time \cite{FerryBook,Provencher1982}.)
Using Eq.~(\ref{eq_Rt2M}) this implies \cite{WKC16}
\begin{equation}
M(\tsamp) = \Rinf + \sum_{p=1}^{\pmax} H_p \ \gDebye(\tsamp/\tau_p) 
\label{eq_M_MW}
\end{equation}
with $\gDebye(x) = 2\left[\exp(-x)-1+x \right]/x^2$
being the Debye function well known in polymer science \cite{DoiEdwardsBook,RubinsteinBook}.
Figure~\ref{fig_theo_stationary} presents both $R(t)$ and $M(\tsamp)$ 
for a generalized Maxwell model with two modes with $\tau_1 \ll \tau_2$.
The upper solid horizontal line indicates an intermediate pseudo-plateau, Eq.~(\ref{eq_plateau}).
Note also that $h(t) \approx v(\tsamp) \approx \RA - \Rinf$ for $t\approx\tsamp \gg \tau_2$.
Since $v(\tsamp)$ and $M(\tsamp)$ are second integrals over $h(t)$ and $R(t)$,
they converge less rapidly to the respective intermediate or terminal plateau values.
($h(t)$ being a monotonically increasing function implies $h(t) > v(t)$ and $R(t) < M(t)$.)
As shown by Fig.~\ref{fig_theo_stationary},
the determination of a plateau value by means of Eq.~(\ref{eq_vc}) or
Eq.~(\ref{eq_Rt2M}) may thus be tedious \cite{WXP13,WXB15,WKC16}.

\section{Ergodic Gaussian processes}
%\subsection{Ergodic Gaussian stochastic processes}
\label{theo_gauss}

\subsection{Gaussian variables}
\label{theo_gauss_variables}
Let us consider a Gaussian variable $y$ of variance $\sigma^2$.
Since $\la (y-\la y \ra)^4 \ra = 3 \sigma^4$ we have
\begin{equation}
\la z^2 \ra - \la z\ra^2 = 2 \sigma^4 \mbox{ for } z = (y-\la y \ra)^2,
\label{eq_y4gauss}
\end{equation}
i.e. the variance of the variance $z$ of $y$ is twice the squared variance of $y$.
We assume now that the time series $\xbf$ is a Gaussian process \cite{vanKampenBook}.
(The main physical reason why this assumption holds for many systems is discussed in Sec.~\ref{theo_V}.)
The mean $m_{11}[\xbf]$ is thus a Gaussian variable and Eq.~(\ref{eq_y4gauss}) holds for 
$y = m_{11}[\xbf]$.
Assuming that $\la y \ra = m_{11} = 0$ by symmetry or by shifting of the data and 
using that $m_{\alpha 1}[\xbf]^{\beta} = m_{\alpha\beta}[\xbf]$ this implies \cite{lyuda19a}
\begin{equation}
\delta m_{12}^2 = m_{14} - m_{12}^2 
= 2 (\delta m_{11}^2)^2 = 2 m_{12}^2.
\label{eq_mu1fluctu}
\end{equation}

\subsection{${\bf \delta c}$ and ${\bf \delta h}$ for Gaussian processes}
\label{theo_gauss_dh}

Let us next discuss the typical fluctuations of the ACFs $c_s$ and $h_s$
defined in Sec.~\ref{theo_stationary}. There are two meaningful ways to define the
variances. One characterizes the fluctuations of $\csg[\xbf]$ and $\hsg[\xbf]$ by means of
\begin{eqnarray}
\delta \csg^2(I) & = & \la \csg[\xbf]^2 \ra - \la \csg[\xbf] \ra^2 \label{eq_csg_fluctu}\\
\delta \hsg^2(I) & = & \la \hsg[\xbf]^2 \ra - \la \hsg[\xbf] \ra^2. \label{eq_hsg_fluctu}
\end{eqnarray}
This allows to get the variances and the error bars for the numerical most accurate way 
to compute $c_s$ and $h_s$. The trouble with this definition is that, 
since the gliding averages are performed first and since the data entries 
$x_i$ are correlated in time, Eq.~(\ref{eq_csg_fluctu}) and Eq.~(\ref{eq_hsg_fluctu})
depend on these correlations in an intricate way.\footnote{The variances increase with $s$ since the 
number of data used for the gliding average decreases linearly with $s$.} %\cite{foot_dcdh_gliding}. 
This may mask the fact that the data have a Gaussian distribution.
A second way to characterize the fluctuations is to measure in a first step $c_{s,i}$ and $h_{s,i}$ (cf. Sec.~\ref{theo_intro}),
to take then the ensemble averages
\begin{equation}
\delta c_{s,i}^2 = \la c_{s,i}^2 \ra - \la c_{s,i} \ra^2
\mbox{ and }
\delta h_{s,i}^2 = \la h_{s,i}^2 \ra - \la h_{s,i} \ra^2
\label{eq_cshs_fluctu}
\end{equation}
and only as the last step (last loop) 
to take the arithmetic average over all $I-s$ possible indices $i$, 
i.e.
\begin{equation}
\delta c_s^2 = \frac{1}{I-s} \sum_{i=1}^{I-s} \delta c_{s,i}^2, \
\delta h_s^2 =  \frac{1}{I-s} \sum_{i=1}^{I-s} \delta h_{s,i}^2.
\label{eq_cshs_fluctu_ga}
\end{equation}
Assuming $\xbf$ to be Gaussian,
$y = (x_{i+s}-x_i)/\sqrt{2}$ is a Gaussian variable of zero mean.
According to Eq.~(\ref{eq_y4gauss}) this implies the important relation 
\begin{equation}
\delta h_s^2 = \la y^4 \ra - \la y^2 \ra^2 = 2 \la y^2 \ra^2 = 2 h_s^2.
\label{eq_key_Gauss}
\end{equation}
In a similar way we find: $\delta c_s^2 = c_0^2 +c_s^2$.
For the fluctuations of $R_s = \RA-h_s$ with $\RA$ being
constant Eq.~(\ref{eq_key_Gauss}) yields in turn $\delta R_s^2 = 2 h_s^2$. 
The latter relation may even hold if $\RA$ is not strictly constant. 
This is relevant for the Born-Lam\'e coefficients considered 
in Sec.~\ref{sec_shear} and Appendix~\ref{app_shear}.
\subsection{${\bf \delta v = \svgauss[h]}$ for Gaussian processes}
\label{theo_wick}

We turn now to the derivation of Eq.~(\ref{eq_key_2})
for the variance $\delta v^2 \equiv \la v[\xbf]^2 \ra - \la v[\xbf] \ra^2$. 
Using Eq.~(\ref{eq_vxdef}) this may be written 
\begin{eqnarray}
\delta v^2 & = & \Ta + \Tb - \Tc \mbox{ with } \label{eq_dv2_terms} \\
\Ta & \equiv & \delta m_{21}^2 = \la m_{21}[\xbf]^2 \ra - \la m_{21}[\xbf] \ra^2 \nonumber \\
    & =      &  \frac{1}{I^2} \sum_{ij} \la x_i^2 x_j^2 \ra 
- \frac{1}{I^2} \sum_{ij} \la x_i^2 \ra \la x_j^2\ra \nonumber \\
\Tb & \equiv & \delta m_{12}^2 = \la m_{12}[\xbf]^2 \ra - \la m_{12}[\xbf] \ra^2 \nonumber \\
    & =      & \frac{1}{I^4} \sum_{ijkl} \la x_i x_j x_k x_l \ra
- \frac{1}{I^4} \sum_{ijkl} \la x_i x_j \ra \la x_k x_l\ra \nonumber \\
\Tc & \equiv & 2 \ \mbox{cov}(m_{21},m_{12}) \nonumber\\
    & \equiv & 2 \left(\la m_{21}[\xbf] m_{12}[\xbf] \ra - \la m_{21}[\xbf] \ra \la m_{12}[\xbf] \ra
 \right) \nonumber \\
    & =      & \frac{2}{I^3} \sum_{ikl} \la x_i^2 x_k x_l \ra
- \frac{2}{I^3} \sum_{kl} \la x_i^2 \ra \la x_k x_l \ra  \nonumber
\end{eqnarray}
where the sums run over all $I$ data entries.
As we have assumed that the stochastic process is stationary and Gaussian,
Wick's theorem must hold \cite{vanKampenBook,DoiEdwardsBook}
\begin{eqnarray}
\hspace*{-.8cm}\la x_i x_j x_k x_l \ra & = & \nonumber\\
& & \hspace*{-2.cm} 
\la x_i x_j \ra \la x_k x_l \ra 
+ \la x_i x_k \ra \la x_j x_l \ra
+ \la x_i x_l \ra \la x_j x_k \ra.
\label{eq_Wick}
\end{eqnarray}
Setting in addition $c_{i-j} = \la x_i x_j \ra$
it is thus readily seen that the three terms in Eq.~(\ref{eq_dv2_terms})
can be rewritten as 
\begin{eqnarray}
\Ta(\tsamp) & = & 
\frac{2}{I^2} \sum_{ij} c_{i-j}^2  
\label{eq_Ta}\\
\Tb(\tsamp) & = & 
\frac{2}{I^4} \left( \sum_{ij} c_{i-j} \right)^2
\label{eq_Tb}\\
\Tc(\tsamp) & = & 
\frac{4}{I^3} \sum_{s,i,j} c_{i-s} c_{j-s}.
\label{eq_Tc}
\end{eqnarray}
Note that $\Tb = \delta m_{12}^2 = 2 m_{12}^2$ in agreement with Eq.~(\ref{eq_mu1fluctu}).
Numerical more convenient reformulations of $\Ta$, $\Tb$ and $\Tc$
are given in Appendix~\ref{app_reformulations}.
% 
%\subsection{Useful reformulations}
%\label{theo_reformulations}
%
Importantly, Eqs.~(\ref{eq_dv2_terms},\ref{eq_Ta},\ref{eq_Tb},\ref{eq_Tc}) 
are equivalent to the more compact formula \cite{lyuda19a}
\begin{eqnarray}
\dvgauss[c] & = & \frac{1}{2I^4} \sum_{i,j,k,l} g_{ijkl}^2 
\mbox{ with} \label{eq_key_c_foursum} \\
g_{ijkl} & = & (c_{i-j} + c_{k-l}) - (c_{i-l} +c_{j-k}) \nonumber
\end{eqnarray}
which looks rather similar as Eq.~(\ref{eq_key_2}).
That this holds can be verified by straightforward expansion of Eq.~(\ref{eq_key_c_foursum}).
Note that the squared terms $c_{i-j}^2+\ldots$ with two different indices contribute to $\Ta$,
the terms $c_{i-j}c_{k-l}+\ldots$ with four different indices to $\Tb$ and
the terms $c_{i-j}c_{i-l}+\ldots$ with three different indices to $\Tc$.

With $a$ and $b$ being real constants it follows directly from Eq.~(\ref{eq_key_c_foursum}) that
\begin{equation}
\svgauss[a]=0 \mbox{ and } \svgauss[b (f-a)] = |b| \ \svgauss[f]
\label{eq_ci_shifted}
\end{equation}
for any function $f(t)$.
Specifically, $\svgauss[c]=\svgauss[h]$. 
This demonstrates finally that Eq.~(\ref{eq_key_2}) 
is equivalent to Eq.~(\ref{eq_key_c_foursum}) and, hence, to
Eqs.~(\ref{eq_dv2_terms},\ref{eq_Ta},\ref{eq_Tb},\ref{eq_Tc}).
It may also be useful to replace $c(t)$ by $c(t) -\cinf$
or --- for thermodynamic equilibrium systems --- 
by the linear response function $R(t)$, Eq.~(\ref{eq_Rt}).
We discuss now in Sec.~\ref{theo_properties} some general properties of $\svgauss[f]$
and in Sec.~\ref{theo_models} the behavior of $\svgauss[f]$ for various test functions $f(t)$
not necessarily being ACFs.

\subsection{Some general properties of ${\bf \svgauss[f]}$}
\label{theo_properties}

Assuming a constant function $f(t) = a$ one obtains 
from either Eqs.~(\ref{eq_Ta},\ref{eq_Tb},\ref{eq_Tc}) or using the
corresponding continuum relations that
\begin{equation}
2 \Ta = 2 \Tb = \Tc = 4 a^2,
\label{eq_const_ct}
\end{equation}
i.e. $\dvgauss = \Ta+\Tb-\Tc$ must vanish in agreement with Eq.~(\ref{eq_ci_shifted}).
This is of relevance for very short sampling times $\tsamp$ where $f(t) \approx f(0)=f_0$ or 
if $f(t)$ has an intermediate plateau extending over several orders of magnitude.
The summand $g_{ijkl}^2$ in Eq.~(\ref{eq_key_c_foursum}) 
must remain small, if $f(t)$ is not rigorously, but only nearly constant.
The typical summand $g^2$ can be estimated by the typical slope on logarithmic time scales 
\cite{lyuda19a}
\begin{equation}
g(\tsamp) \approx f(\tsamp)-f(\tsamp/2) \approx \left.df(t)/d\log(t))\right|_{t\approx \tsamp}. 
\label{eq_g_shorttimes}
\end{equation}
One thus expects 
\begin{equation}
\dvgauss[f] \approx \sum_{ijkl} g_{ijkl}^2/I^4 \approx g(\tsamp)^2.
\label{eq_f_shorttimes}
\end{equation}
For instance, $f(t)$ may decrease for $t\ll \tau$ as $f(t) \approx b \exp(-(t/\tau)^{\beta}) + f_{\infty}$ 
with constants $\beta > 0$. Equations~(\ref{eq_g_shorttimes}) and (\ref{eq_f_shorttimes}) lead then to
\begin{equation}
\svgauss[f] \approx |b| (\tsamp/\tau)^{\beta} \mbox{ for } \tsamp \ll \tau.
\label{eq_cbeta_small_tsamp} 
\end{equation}
In the opposite limit of very large $\tsamp \gg \tau$, the
leading scaling dependence is obtained by replacing in Eqs.~(\ref{eq_Ta_3}-\ref{eq_Tc_3})
the upper integration bounds by $\tau$ and $f(t)$ by $a \approx f(\tau) - f_{\infty}$ 
using Eq.~(\ref{eq_ci_shifted}).  This implies 
\begin{equation}
\Ta \approx a^2 \ \tau/\tsamp, \Tb \approx \Tc \approx a^2 \ (\tau/\tsamp)^2. 
\label{eq_Ti_large_time}
\end{equation}
In other words, $\dvgauss$ is dominated for $\tsamp/\tau \gg 1$ by $\Ta=\delta m_{21}^2$,
i.e. $\svgauss \propto 1/\sqrt{\tsamp}$ as expected for $\tsamp/\tau$ uncorrelated subintervals.
Adding heuristically the short and the long time behavior, Eq.~(\ref{eq_f_shorttimes})
and Eq.~(\ref{eq_Ti_large_time}), yields the phenomenological approximation \cite{lyuda19a}
\begin{equation}
\dvgauss[f] \approx g(\tsamp)^2 + (f(\tau)-f_{\infty})^2 \ (\tau/\tsamp)
\label{eq_f_pheno}
\end{equation}
which is useful for processes with one main dominant relaxation process.

\subsection{${\bf \svgauss[f]}$ for test functions ${\bf f(t)}$}
\label{theo_models}

\subsubsection{Introduction}

%\paragraph*{\textcolor{red}{Introduction.}}
To illustrate some properties of the non-linear functional $\svgauss[f]$ 
we discuss now several test functions $f(t)$.
Not all presented $f$ belong to the space of legitimate ACFs $c$ or $h$ of stationary stochastic processes.
We remind \cite{HansenBook,GoetzeBook} that a legitimate ACF may not change too strongly 
(especially not discontinuously)
and must not violate the Wiener-Khinchin theorem on the power spectrum of the signal
stating that the Fourier transform (FT) of $c(t)$ is given by the squared FT of $x(t)$ 
\cite{HansenBook,GoetzeBook,AllenTildesleyBook}.
A general (necessary and sufficient) criterion for a function $f(t)$ to be a legitimate ACF is thus 
\cite{HansenBook,GoetzeBook}\footnote{According to Bochner's theorem $\fhat(\omega) \ge 0$
if and only if $f(t)$ is a positive-definite function, i.e. all eigenvalues
of the matrix $g_{i,j}=f(t_i-t_j)$ are non-negative \cite{GoetzeBook}.}
\begin{equation}
\fhat(\omega) \equiv \int_0^{\infty} f(t) \cos(\omega t) \ddiff t \ \ge 0 \mbox{ for any real } \omega.
\label{eq_ft_legACF}
\end{equation}
This ensures that 
$f(0) \ge |f(t)| \ge 0$ and $\fhat(\omega=0) = \int_0^{\infty} \ddiff t f(t) \ge 0$.
Taking advantage of the affine transform Eq.~(\ref{eq_ci_shifted}) 
we often set without loss of generality $f(0)=1$ and $f(t) \to 0$ for $t \to \infty$.
If there is only one characteristic time it is also set to unity.

\begin{figure}[t]
\centerline{\resizebox{.9\columnwidth}{!}{\includegraphics*{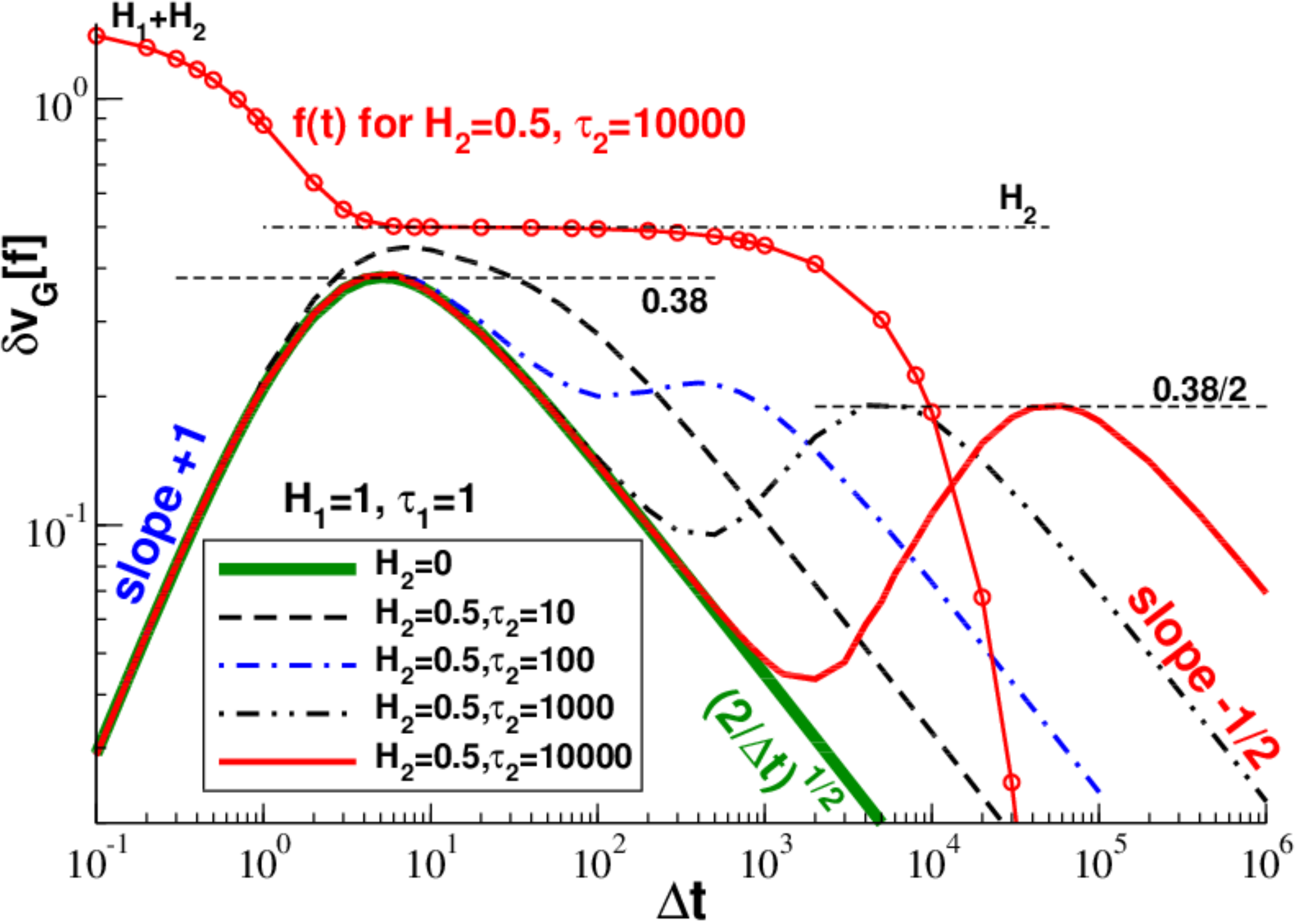}}}
\caption{$\svgauss[f]$ {\em vs.} $\tsamp$ for the one-mode Maxwell model with $H_1=\tau_1=1$ (bold solid line)
revealing a maximum at $\tsamp \approx 5$ and a final decay $\svgauss \approx \sqrt{2/\tsamp}$.
The other data refer to the two-step relaxation model Eq.~(\ref{eq_ft_twostep}) with $H_2 = 0.5$. 
%All data are for $H_1=\tau_1=1$.
Also given is $f(t)$ for $\tau_2=10000$ (solid line with circles).
$\svgauss[f]$ becomes bimodal with increasing $\tau_2/\tau_1$ with
a minimum slightly below $\tau_2$ and a second separate maximum at $\approx 5 \tau_2$.
}
\label{fig_ft_maxwell}
\end{figure}

\subsubsection{Maxwell model}
%\paragraph*{\textcolor{red}{Maxwell model.}}
One of the few cases where $\svgauss[f]$ can be calculated analytically is the Maxwell model (Debye decay) $f(t) = \exp(-t)$.
This model is especially of relevance for the self-assembled network systems considered below in Sec.~\ref{shear_TSANET}.
Since $\fhat(\omega) = 1/(1+\omega^2) >0$ for all $\omega$, $f(t)$ is a legitimate ACF as expected.
%and it is useful to also indicate the analytical solution for this important case. 
Note first that $v(\tsamp) = 1 - \gDebye(\tsamp)$ with $\gDebye(\tsamp)$ 
being the Debye function introduced in Sec.~\ref{theo_stationary}, Eq.~(\ref{eq_M_MW}).
The three contributions $\Ta$, $\Tb$ and $\Tc$ to $\dvgauss[f]=\Ta+\Tb-\Tc$ are
\begin{eqnarray}
\Ta & = & 2\gDebye(2\tsamp), \ \Tb = 2\gDebye(\tsamp)^2, \label{eq_ft_dvgauss_MW} \\
\Tc & = & \frac{4}{\tsamp^3} \times \nonumber \\
& & \hspace*{-.0cm} \left[- e^{-2\tsamp} + (2\tsamp+8) e^{-\tsamp}  + 4 \tsamp - 7 \right]. \nonumber
\end{eqnarray}
Since $\gDebye(x) \approx 2/x$ for large $x$ we have $\svgauss \approx \sqrt{2/\tsamp}$ for large $\tsamp$.
The analytical solution for the Maxwell model is indicated by a bold solid line in Fig.~\ref{fig_ft_maxwell}.
This exact result may be used for testing the numerical determination of $\svgauss[f]$ by means of 
Eqs.~(\ref{eq_Ta_3},\ref{eq_Tb_3},\ref{eq_Tc_3}). 

\subsubsection{Two-step relaxation}
%\paragraph*{\textcolor{red}{Two-step relaxation.}}
%
In view of the presented simulations it is useful to discuss 
an example for systems with {\em two} relaxation processes similar to Fig.~\ref{fig_theo_stationary}. 
Of interest is the limit where $f(t)$ develops an intermediate plateau $f(t) \approx f_p$
for $\tau_1 \ll t \ll \tau_2$ with $\tau_1$ corresponding to a fast, local process 
and $\tau_2$ to a slow, collective relaxation. 
One expects $\svgauss(\tsamp)$ to become bimodal with a first maximum around $\tau_1$ followed 
by a $1/\sqrt{\tsamp}$-decay and a second maximum around $\tau_2$ followed by a second $1/\sqrt{\tsamp}$-decay.
The minimum between both maxima should systematically become deeper with increasing plateau width. 
Figure~\ref{fig_ft_maxwell} presents numerically obtained $\svgauss[f]$-data for
\begin{equation}
f(t) = H_1 \exp(-t/\tau_1) + H_2 \exp(-t/\tau_2)
\label{eq_ft_twostep}
\end{equation}
with $H_1 = \tau_1 = 1$ and $H_2=0.5$ for the amplitude of the second mode.
As for all generalized Maxwell models 
\begin{equation}
\fhat(\omega) = \sum_{p=1}^{\pmax} \frac{H_p \tau_p}{1+(\omega \tau_p)^2} > 0,
\label{eq_ft_MWfreq}
\end{equation}
i.e. Eq.~(\ref{eq_ft_twostep}) is a legitimate ACF.
We scan $\tau_2$ over several orders of magnitude as indicated in the figure. 
We indicate $f(t)$ for the longest second relaxation time, 
$\tau_2=10000$, at the top of the figure (solid line with circles).
For large $\tau_2/\tau_1$ one observes for $\svgauss(\tsamp)$ two well separated maxima of same 
shape but different amplitudes $\propto H_p$. 
Note that the ratio of the two dashed horizontal lines is $H_1/H_2=2$.
The decay from both maxima is given by $\svgauss \approx H_p \sqrt{2 \tau_p/\tsamp}$.

\begin{figure}[t]
\centerline{\resizebox{.9\columnwidth}{!}{\includegraphics*{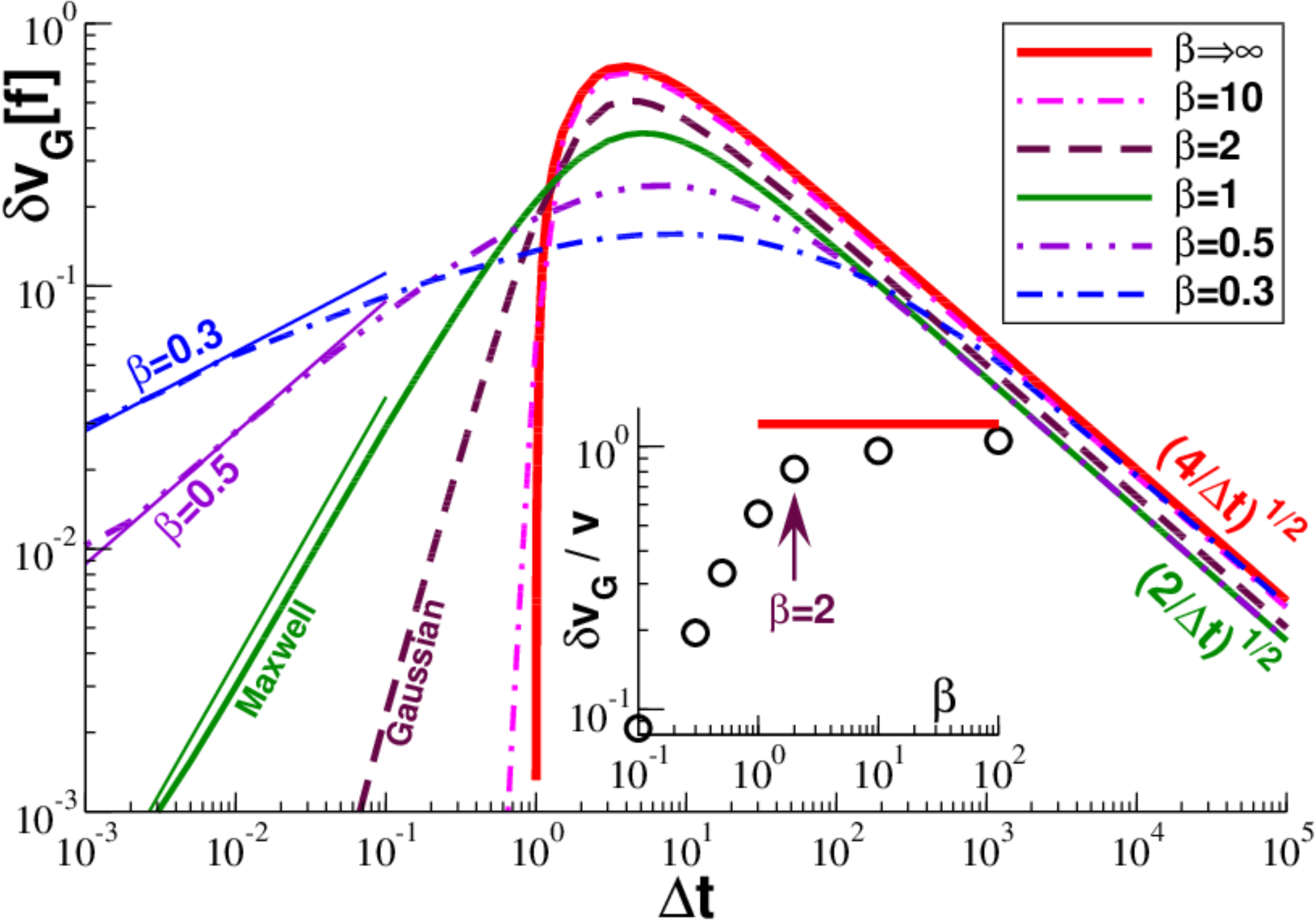}}}
\caption{$\svgauss[f]$ {\em vs.} $\tsamp$ for $f(t)=\fbeta(t)\equiv\exp(-t^{\beta})$ with
$\beta=1$ corresponding to the one-mode Maxwell model, $\beta=2$ to a Gaussian and
$\beta \to \infty$ to the cusp model $\fcusp(t) \equiv H(t)-H(t-1)$.
Only exponents $\beta \le 2$ correspond to legitimate ACFs.
Note that $\svgauss \propto \tsamp^{\beta}$ for $\tsamp \ll 1$ 
(thin solid lines for $\beta=0.3$, $0.5$ and $1$)
and $\svgauss \propto 1/\sqrt{\tsamp}$ for $\tsamp \gg 1$.
Inset: $\svpeak$ vs. $\beta$. 
The vertical arrow marks the ratio $\approx 0.82$ for $\beta=2$,
the horizontal line the ratio $\approx 1.21$ for $\beta \to \infty$.
}
\label{fig_ft_beta}
\end{figure}

\subsubsection{Stretched and compressed exponentials}
%\paragraph*{\textcolor{red}{Stretched and compressed exponentials.}}
%
Another natural generalization of the one-mode Maxwell model ($\beta=1$) is seen in Fig.~\ref{fig_ft_beta}
where we present $\svgauss[f]$ for $f(t) = \fbeta(t) \equiv \exp(-t^{\beta})$.
$\fbeta(t)$ is a ``stretched" exponential for $\beta < 1$ and a ``compressed" exponential for $\beta > 1$.
It can be readily checked numerically that Eq.~(\ref{eq_ft_legACF}) only holds for $\beta \le 2$ but not for larger 
exponents $\beta$ which do not correspond to ACF of stationary stochastic processes.
To see this let us just mention two cases.
Since $\fhat(\omega) \propto \exp(-\omega^2/4)$ for $\beta=2$, Eq.~(\ref{eq_ft_legACF}) holds for the Gaussian model 
and it thus also does for even more gently decreasing (less compressed) functions with $\beta < 2$.
On the other hand $\fbeta(t)$ becomes for $\beta \to \infty$ equivalent to the cusp singularity $\fcusp(t) \equiv H(t)-H(t-1)$.
($\svgauss[\fcusp]$ can be readily calculated analytically and this exact formula is used in Fig.~\ref{fig_ft_beta}.)
The cusp singularity is not a legitimate ACF since $\fhat = \sin(\omega)/\omega$ may be negative, 
i.e. Eq.~(\ref{eq_ft_legACF}) does not hold.
As may be seen from the main panel, all $\svgauss[\fbeta]$ have a maximum between
$\tsamp \approx 4$ (large $\beta$) and $\tsamp \approx 10$ (small $\beta$).
As expected from $\fbeta(t) \to \fcusp(t)$ for $\beta \to \infty$, it is seen that $\svgauss(\beta)$ becomes 
increasingly similar to the standard deviation of the cusp model (bold solid line),
i.e. the peaks become systematically higher, sharper and more lopsided with increasing $\beta$.
The power-law slopes $\beta$ (thin solid lines) observed for $\tsamp \ll 1$ 
are expected from Eq.~(\ref{eq_cbeta_small_tsamp}).
All models decrease as $\svgauss \approx \sqrt{a/\tsamp}$ for large $\tsamp$
in agreement with Eq.~(\ref{eq_Ti_large_time}).
The amplitude $a$ of this ultimate decay is the {\em largest} for the cusp model ($a=4)$ 
and the {\em smallest} for the Maxwell model ($a=2$).
    
The inset of Fig.~\ref{fig_ft_beta} shows the ratio $\svpeak$ taken at the maximum of 
$\svgauss(\tsamp)$ for a broad range of the exponent $\beta$.
This shows a monotonic increase with $\beta$ approaching from below the ratio $\approx 1.21$ of the cusp model (bold horizontal line).
The ratio is $\approx 0.55$ for the Maxwell model and $\approx 0.82$ for the Gaussian (vertical arrow).
Importantly, the standard deviations thus become of the same order as the average behavior
for the most rapidly decaying legitimate ACFs with $\beta \le 2$.

\begin{figure}[t]
\centerline{\resizebox{.9\columnwidth}{!}{\includegraphics*{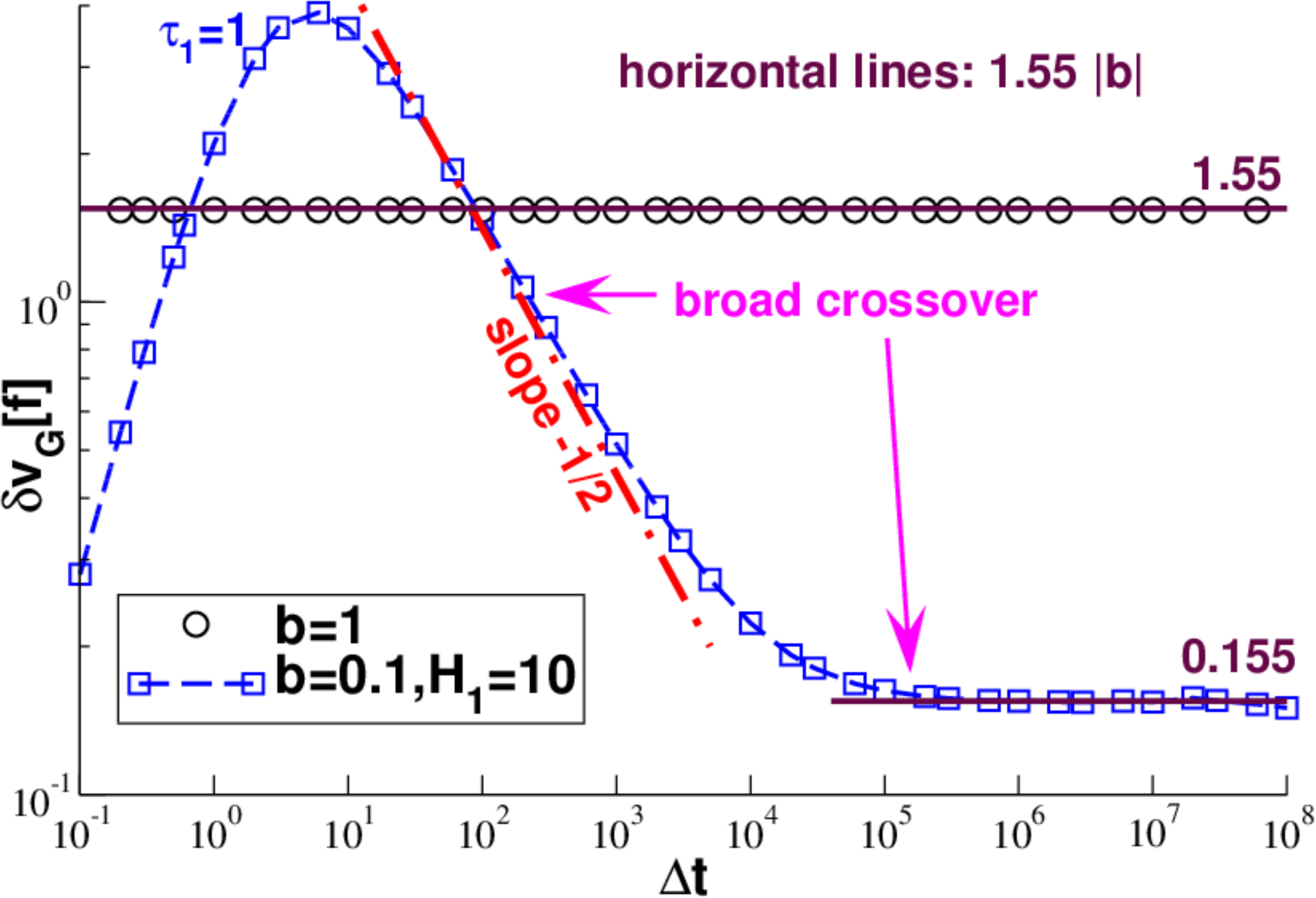}}}
\caption{$\svgauss[f]$ {\em vs.} $\tsamp$ for logarithmic creep.
The circles indicate Eq.~(\ref{eq_ft_creepA}) for $b=1$, 
the squares Eq.~(\ref{eq_ft_creepB}) for $H_1=10$, $\tau_1=\tau_2=1$, $b=0.1$ and $\tau_3=10^{10}$
and the dash-dotted line the $1/\sqrt{\tsamp}$-decay of the short-time Maxwell model.
The arrows mark the broad crossover between the Maxwell model and the plateau expected for $b=0.1$.
}
\label{fig_ft_creep}
\end{figure}

\subsubsection{Logarithmic creep}
%\paragraph*{\textcolor{red}{Logarithmic creep.}}
%
Logarithmically slow ACFs are expected for hopping processes in systems with a broad
distribution of barriers and are generally observed in glass-forming fluids 
\cite{FerryBook,HansenBook,GoetzeBook}.
The general scaling relation Eq.~(\ref{eq_f_shorttimes}) suggests 
\begin{equation}
\svgauss[f] \approx 1.55 |b| \mbox{ if } f(t) \approx a  - b \ln(t)
\label{eq_ft_creepA}
\end{equation}
holds over a sufficiently broad intermediate time window.
The indicated prefactor $1.55$ is needed for the discussion of $\svgauss(\tsamp)$ 
for thin polymer films in Appendix~\ref{app_shear_film}.
Obviously, this value is not given by the scaling relation but by numerically computing $\svgauss[\ln(t)]$
as shown by circles in Fig.~\ref{fig_ft_creep}.
Due to the affinity relation Eq.~(\ref{eq_ci_shifted}),
this result corresponds to an amplitude $|b|=1$ and does not depend on the shift constant $a$.
Obviously, a legitimate ACF cannot diverge for $t \to 0$ and $t \to \infty$ and $f(t)=a-b \ln(t)$
cannot hold in these limits for both mathematical and physical grounds. 
To demonstrate that Eq.~(\ref{eq_ft_creepA}) may hold for an intermediate time window of a legitimate ACF
we are thus free to use, e.g., a generalized Maxwell model, Eq.~(\ref{eq_Rt_genMaxwell}), 
fitted (by inverse Laplace transformation \cite{Provencher1982})
to an intermediate creep. (Due to Eq.~(\ref{eq_ft_MWfreq}) this yields directly a legitimate ACF.)
More simply we may improve $f(t)=a-b \ln(t)$ by adding suitable continuous cutoffs.
As shown by the squares in Fig.~\ref{fig_ft_creep} we use
\begin{equation}
f(t) = H_1 e^{-t/\tau_1} + a - b \ln(t) \ (1-e^{-t/\tau_2}) \ e^{-t/\tau_3} 
\label{eq_ft_creepB}
\end{equation}
with $H_1=10$ and $\tau_1=1$ for the Maxwell model added to mimic the typical microscopic relaxation
and $\tau_2=1$ and $\tau_3=10^{10}$ setting, respectively, the lower and the upper cutoff of the logarithmic creep.
($\tau_3$ is irrelevant for the presented $\tsamp$-range and the constant $a$ is arbitrary.)
The strong Maxwell mode dominates $\svgauss(\tsamp)$ below $\tsamp \approx 10^3$. Interestingly, as marked by the 
left arrow deviations from the $1/\sqrt{\tsamp}$-decay (dash-dotted line) expected for the Mawell mode are already
observed at $\tsamp \approx 10^2$. 
Only after a broad crossover regime (about three decades) the plateau (solid horizontal line)
expected from Eq.~(\ref{eq_ft_creepA}) is reached.
The latter model demonstrates how a rather small additional logarithmic creep may lead to strong
deviations from an expected $1/\sqrt{\tsamp}$-decay.

\section{Non-ergodic stochastic processes}
%\subsection{Non-ergodic stochastic processes}
\label{theo_nonerg}

Our key relation Eq.~(\ref{eq_key_2}) and its various reformulations may obviously fail if one
of the stated or implicit assumptions does not apply for the particular ensemble of time series.
For instance, strong non-Gaussian contributions may be present in
a specific time or frequency range leading to the failure of Wick's theorem, Eq.~(\ref{eq_Wick}).
We want in this subsection to address an important assumption not yet explicitly stated.
In fact it was assumed that the stochastic process under consideration is ergodic, 
i.e. all independently created trajectories, called here ``configurations", are able to explore 
given enough time the complete (generalized) phase space. 
The averages which appear in Wick's theorem, can thus be either obtained
by averaging over independent configurations $c$ or by averaging ofter subsets of one
extremely long trajectory.
To see that this condition matters let us consider a strictly non-ergodic system where the configurations $c$
are trapped in subspaces of the total phase space 
(since the terminal relaxation time $\tau$ of the system diverges). 
If $t$ and $\tsamp$ exceed the typical relaxation time $\taubasin$ of these basins, 
$h(t)$ and $v(\tsamp)$ must become constant.
As shown in Sec.~\ref{theo_properties} and Sec.~\ref{theo_models},
$\svgauss \propto 1/\sqrt{\tsamp}$ for $\tsamp \gg \taubasin$. 
At variance to this $\delta v \to \Snonerg$ becomes constant with 
\begin{equation}
\Snonerg^2 \equiv \mbox{var}(v_c) = \frac{1}{\Nc} \sum_{c=1}^{\Nc} v_c^2 - 
\left(\frac{1}{\Nc} \sum_{c=1}^{\Nc} v_c\right)^2
\label{eq_Snonerg_def}
\end{equation}
being the variance of the $\Nc$ quenched variances $v_c = \lim_{\tsamp \to \infty} v[\xbf_c]$
of the independent configurations.
Obviously, $\Snonerg$ vanishes for identical $v_c$.
This holds indeed for ergodic systems for $\tsamp \gg \tau$ (with the finite $\tau$ replacing $\taubasin$),
but in general not for non-ergodic systems. 

On the other hand, for small $\tsamp$ the non-ergodicity constraint should not matter much
and one expects $\delta v^2 \approx \dvgauss$.
Interpolating between both $\tsamp$-limits a useful approximation for non-ergodic systems may be written as
\begin{equation}
\delta v^2(\tsamp) \approx \dvgauss(\tsamp) + \Snonerg^2 \mbox{ for } 
\taubasin \ll \tsamp \ll \tau
\label{eq_dv_gen}
\end{equation}
motivated by the idea that $\delta v^2$ is the sum of two variances describing the
independent fluctuations within each configuration and between the different configurations.
Moreover, Eq.~(\ref{eq_dv_gen}) suggests the operational definition
\begin{equation}
\svgauss(\tsamp\stackrel{!}{=}\Tnonerg) = \Snonerg
\label{eq_Tnonerg_def}
\end{equation}
identifying $\Tnonerg$ as the crossover time between both limits.
Quite generally, $\Tnonerg \gg \taubasin$.\footnote{For volume-averaged density fields
$\taubasin$ must be to leading order system-size independent
while $\Tnonerg$ diverges in the macroscopic limit \cite{fluctuGaussB}.}

A rigorous justification of the above interpolation formula Eq.~(\ref{eq_dv_gen}) 
will be given elsewhere \cite{fluctuGaussB}. We only outline here the general idea. 
To understand the discrepancy between $\delta v$ and $\svgauss$ for (strictly) non-ergodic systems
it is necessary to introduce an extended ensemble of time series $\xbf_{ck}$ 
where for each of the $\Nc$ independent configurations $c$ one samples $\Nk$ time series $k$
\cite{Yoshino12}.\footnote{The time series $k$ may be obtained by first tempering
the configuration $c$ over a time interval $\ttemper \gg \taubasin$
and by sampling then $\Nk$ time intervals $\tsamp$ separated
by constant spacer time intervals $\tspacer \gg \taubasin$.
$\Nk$ is assumed to be arbitrarily large and the $k$-averaged
properties $\svint$ and $\svext$ do thus neither depend on
$\Nk$ nor the total sampling time $\tsampmax= \Nk (\tsamp+\tspacer)$.
The latter point may become a delicate issue if the non-ergodicity
constraint ($\tau \to \infty$) is not strictly obeyed.}
Obviously, the time series $k$ of the same configuration $c$ are correlated (being all confined in the same basin) 
and $k$-averaged expectation values and variances may then depend on the configuration $c$. 
It thus becomes relevant in which order $c$-averages over configurations $c$
and $k$-averages over time series $k$ of a given configuration $c$ are performed.
Three variances of $v[\xbf_{ck}]$ must be distinguished:
the total variance $\dvtot = \dvint + \dvext$
and its contributions $\dvint$, the typical internal variance within the meta-basins,
and $\dvext$, characterizing the dispersion between the basins.
The present paper focuses on the total standard deviation $\svtot$. (The index $tot$ is dropped elsewhere.)
Importantly, {\em if} the trajectory of each configuration $c$ remains essentially Gaussian,
Wick's theorem can be applied to $\svint$ as before. This implies $\svint \approx \svgauss$.
Moreover, since $\svext(\tsamp) \approx \Snonerg$ for $\taubasin \ll \tsamp \ll \tau$, this leads to Eq.~(\ref{eq_dv_gen}).

Variances due to independent physical causes are naturally additive.
We remind that the variance of the blackbody radiation is 
the sum of a variance describing the Rayleigh-Jeans part of the spectrum
(wave aspect) and of a variance describing the Wien part (discrete particle aspect) \cite{Einstein09}.
Interestingly, as in the blackbody radiation analogy,
the two contributions $\svint$ (internal basin fluctuations)
and $\svext$ (fluctuations between basins) to $\svtot$ have also different statistics.
This is manifested by their different system size dependences as will be shown now.

\section{System-size effects}
%\subsection{Microscopic variables and system-size effects}
\label{theo_V}
Due to the central limit theorem \cite{vanKampenBook} the stochastic process of many systems 
is to a good approximation Gaussian since the data entries $x_i$ are averages over $\Nm \gg 1$ 
microscopic (often unknown or inaccessible) contributions $x_{im}$.
Specifically, we shall consider below the instantaneous shear stress 
$\tauhat_i = \int \ddiff \rvec \ \tauhat_{i,\rvec} /V$ 
being the volume average over the local shear stress $\tauhat_{i,\rvec}$. 
For such intensive field averages $\Nm$ corresponds to the number of local volume elements $dV$ computed, 
i.e. $\Nm \approx V/dV$.
Albeit these microscopic contributions $x_{im}$ may be correlated, i.e. they may not all fluctuate independently, 
the fluctuations of the $x_i$ commonly decrease with increasing $\Nm$.
Since $v \propto 1/\Nm$ for uncorrelated variables $x_{im}$, it is often useful to incorporate 
this reference in the definition of the data entries by rescaling $x_i \Rightarrow \sqrt{\Nm} x_i$. 
(This is done in Sec.~\ref{shear_intro} by rescaling the stress by $\sqrt{V}$.) 
For perfectly uncorrelated microscopic variables $x_{im}$ subject to a finite quenched random field this leads to
\begin{eqnarray}
v \propto  h \propto \svgauss & \propto & \Nm^0 \mbox{ and } \label{eq_xim_independent} \\
\Snonerg \equiv \lim_{\tsamp\to \infty} \delta v(\tsamp) & \propto & \Nm^{-\gamext} \mbox{ with } \gamext =1/2.
\nonumber %\label{eq_xim_dv}
\end{eqnarray}
Due to Eq.~(\ref{eq_key_2}) the $\Nm$-independence of $\svgauss$ is implied by the $\Nm$-independence of $h$.
To see that $\gamext=1/2$ it is sufficient to compute ($\tsamp$-independent) ensemble averages $\la \ldots \ra_c$
for the different quenched meta-basins $c$ compatible with the non-ergodicity constraint ($\tau \to \infty$).
(Note that $\la \ldots \ra_c$ is obtained for $\Nk \to \infty$.) 
%From $v_c = \la x^2 \ra_c - \la x \ra_c^2$ we then get  the variance $\Snonerg^2 = \mbox{var}[v_c]$.
Substituting $x = \sum_m x_{m}/\Nm$, setting $v_{cm} = \la x_m^2 \ra_c - \la x_m \ra_c^2$
and using that the microstates $m$ are decorrelated yields
\begin{equation}
v_c \equiv \la x^2 \ra_c - \la x \ra_c^2 = \frac{1}{\Nm} \times \left(\frac{1}{\Nm} \sum_m v_{cm} \right). 
\label{eq_theo_V_v_c}
\end{equation}
$v$ is then the ensemble average over all configurations $c$.
Using that also the variances $v_{cm}$ of each microstate are decorrelated we obtain in turn
\begin{equation}
\Snonerg^2 \equiv \mbox{var}[v_c]  = \frac{1}{\Nm^3} \times \left( \frac{1}{\Nm} \sum_m \mbox{var}[v_{cm}] \right).
\label{eq_theo_V_Snonerg}
\end{equation}
Since the $m-$averages (brackets) in Eq.~(\ref{eq_theo_V_v_c}) and Eq.~(\ref{eq_theo_V_Snonerg}) become $\Nm$-independent,
this implies $\Snonerg^2 \approx v^2/\Nm$ which in turn confirms $\gamext=1/2$. %, Eq.~(\ref{eq_xim_independent}).
Equation~(\ref{eq_xim_independent}) also holds for fluctuating density fields with a finite 
$V$-independent correlation length $\xi$ for sufficiently large systems ($V \gg \xi^d$). 
In this case $\Nm$ in Eq.~(\ref{eq_xim_independent}) is simply replaced by the number
of independent subvolumes $V/\xi^d$.
A smaller exponent $\gamext < 1/2$ is expected for long-range and scale-free spatial correlations. 
In agreement with Eq.~(\ref{eq_Tnonerg_def}) and assuming $\svgauss \propto 1/\tsamp^{\beta}$
with $\beta \approx 1/2$ we have 
\begin{equation}
\Tnonerg \propto \Nm^{\gamext/\beta},
\label{eq_Tnonerg}
\end{equation}
i.e. the crossover time increases with $\Nm$ and $\Snonerg^{1/\beta} \propto 1/\Tnonerg$.
Details will be given elsewhere \cite{fluctuGaussB}.
The generally important point is here that $\Snonerg$ decreases and $\Tnonerg$ increases with the system size 
{\em if} $\gamext > 0$ and thus $\delta v \to \svgauss$ for $\Nm \to \infty$ according to Eq.~(\ref{eq_dv_gen}).

\section{Shear-stress fluctuations}
\label{sec_shear}

\subsection{Introduction}
\label{shear_intro}

\begin{figure}[t]
\centerline{\resizebox{0.9\columnwidth}{!}{\includegraphics*{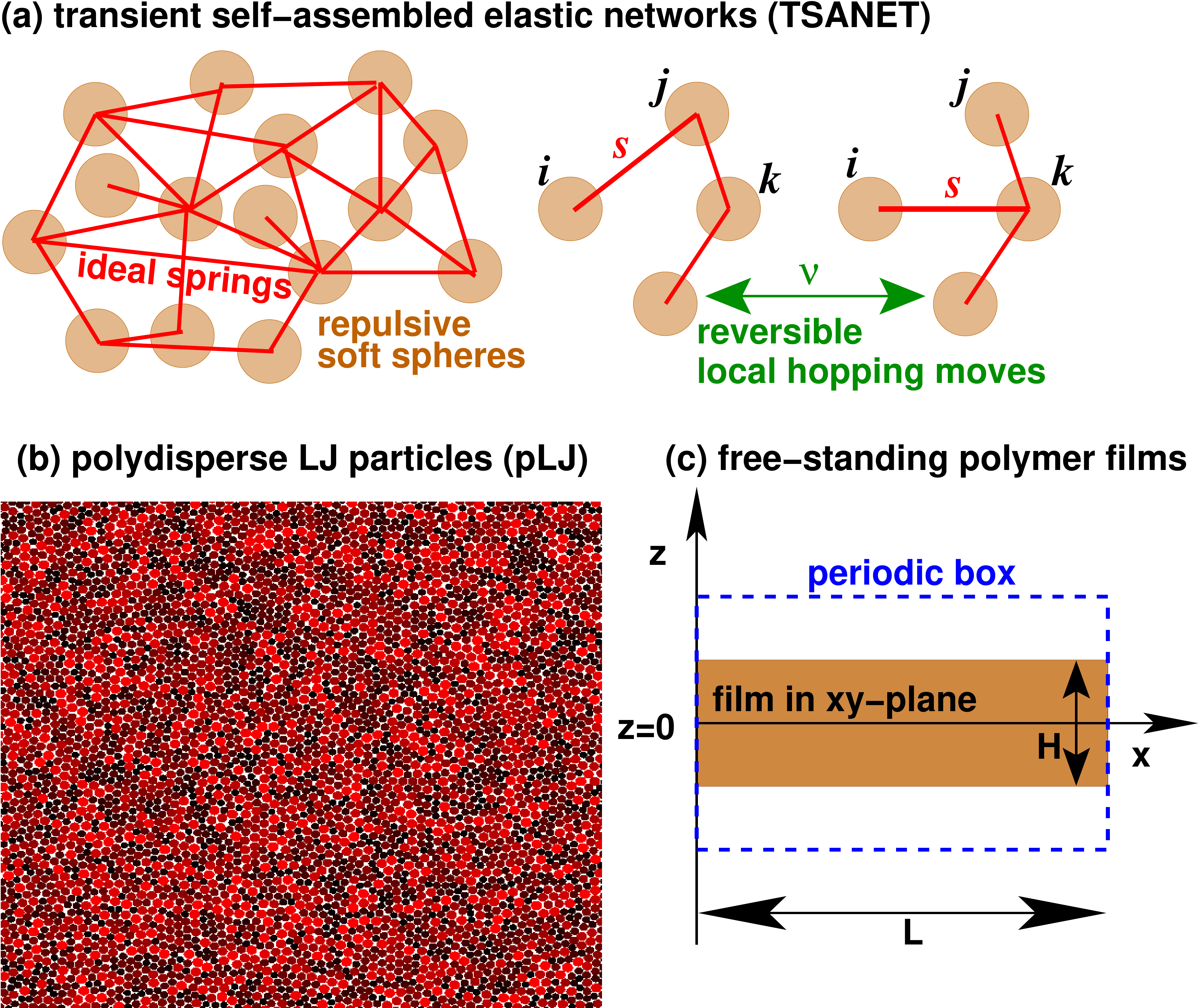}}}
\caption{Model systems considered computationally:
{\bf (a)}
Transient self-assembled elastic networks (TSANET) created by reversibly breaking and
recombining springs with an attempt frequency $\nu$ per spring.
The spring $s$ thus connects the beads $i$ and $j$ on the left and the beads $i$ and $k$ on the right.
Results presented in Sec.~\ref{shear_TSANET}.
{\bf (b)}
Monte Carlo (MC) simulations of polydisperse Lennard-Jones (pLJ) particles
with larger beads being red, smaller beads darker.
Results presented in Appendix~\ref{app_shear_pdLJ}.
{\bf (c)}
Molecular dynamics (MD) simulations of thin free-standing films of glass-forming polymers.
Results presented in Appendix~\ref{app_shear_film}.
}
\label{fig_algomod}
\end{figure}

The theoretical results presented above %in Sec.~\ref{sec_theo} 
should be useful for the analysis of general time series $\xbf$ of 
stochastic processes which are essentially stationary and Gaussian. 
We illustrate this for the shear-stress fluctuations measured numerically
for the coarse-grained model systems sketched in Fig.~\ref{fig_algomod}. 
See Appendix~\ref{app_algo} for further details of the model systems
and Appendix~\ref{app_affine} for the definition of the instantaneous shear stress $\tauhat$
and the corresponding instantaneous affine shear modulus $\muAhat$.
The stochastic process $x(t)$ is obtained by rescaling 
\begin{equation}
\tauhat(t) \Rightarrow x(t) \equiv \sqrt{\beta V} \tauhat(t)
\label{eq_xi_rescale}
\end{equation}
with $\beta=1/T$ being the inverse temperature (setting Boltzmann's constant $\kB$ to unity)
and $V$ the (two- or three-dimensional) volume of the system.
With this rescaling $v[\xbf]$, Eq.~(\ref{eq_vxdef}), characterizes the empirical 
shear-stress fluctuations of the time series and the expectation value $v(\tsamp)$ 
is equivalent to the ``shear-stress fluctuation" $\muF(\tsamp)$ considered in 
previous publications on the stress-fluctuation formalism for elastic moduli
\cite{WXP13,WKC16,ivan17c,ivan18,film18,lyuda19a,fluctuGaussA}.
For consistency with the theoretical considerations %Sec.~\ref{sec_theo} 
we keep the general notations defined above. 
Since these notations differ from the ones widely used for sheared elastic bodies
\cite{FerryBook,RubinsteinBook,DoiEdwardsBook,HansenBook,WXP13,WKC16,ivan17a,ivan17c,ivan18,lyuda19a}
the following correspondence list may be useful to the reader:
\begin{eqnarray}
v(\tsamp)      & \leftrightarrow & \muF(\tsamp) \equiv  \muFtwo - \muFone(\tsamp) \label{eq_muF} \\
m_{21}         & \leftrightarrow & \muFtwo \equiv  \beta V \la \overline{\tauhat^2} \ra \label{eq_muFtwo} \\
m_{12}(\tsamp) & \leftrightarrow & \muFone(\tsamp) \equiv  \beta V \la \overline{\tauhat}^2 \ra  \label{eq_muFone}\\
\RA            & \leftrightarrow & \muA = \la \muAhat \ra  \label{eq_muA} \\
R(t)           & \leftrightarrow & G(t)=\muA-h(t) \label{eq_Gtdef}\\
M(\tsamp)      & \leftrightarrow & \muSF(\tsamp) = \muA - \muF(\tsamp)  \label{eq_mudef}\\
\delta v       & \leftrightarrow & \delta \muF \label{eq_smuFdef}
\end{eqnarray}
The overbars on the right-hand sides denote the average over a given time series.
Note that Eq.~(\ref{eq_Gtdef}) is the fluctuation dissipation relation \cite{DoiEdwardsBook}
for the shear-stress relaxation after an infinitesimal change of the shear strain,
i.e., $R(t)$ is the ``shear relaxation function",
Eq.~(\ref{eq_mudef}) is the corresponding relation for the ``generalized shear modulus" \cite{WXP13}.
We also remind \cite{WXP13,WXB15,WXBB15} that the {\em additional} assumption 
$\RA = m_{21}$ together with the identities $c(0)=m_{21}$ and $h(t)=c(0)-c(t)$ imply that
\begin{equation}
R(t) = c(t) \mbox{ and } M(\tsamp) = m_{12}(\tsamp).
\label{eq_liquid_condition}
\end{equation}
While $\RA=m_{21}$ holds indeed under liquid equilibrium conditions,
this may become incorrect in general \cite{WXP13,WXB15,WXBB15}.
In order to test Eq.~(\ref{eq_key_2}) we compare the standard deviation $\delta v(\tsamp)$, 
lumping all $\Nc \times \Nk$ time series together (cf. Appendix~\ref{app_algo_data}), 
with $\svgauss[R]=\svgauss[h]=\svgauss[c]$ obtained by means of Eqs.~(\ref{eq_Ta_2},\ref{eq_Tb_2},\ref{eq_Tc_2})
using the measured ACFs.

\subsection{Self-assembled networks}
\label{shear_TSANET}
%\input{shear_TSANET}

%\paragraph*{Maxwell fluid.}
The TSANET model described in Appendix~\ref{app_algo_TSANET}
is from the rheological point of view very similar to patchy colloids
\cite{Leibler13,Kob13} or ``vitrimers" \cite{Leibler11}.
Rheologically similar self-assembled transient networks may also be formed by hyperbranched
polymer chains with sticky end-groups \cite{Friedrich10} or microemulsions bridged by
telechelic polymers \cite{Porte03,Safran06,Ligoure08}.
As shown in Figs.~6 and 7 of Ref.~\cite{WKC16}, TSANET is a simple Maxwell fluid,
i.e. the shear-stress relaxation modulus, computed by means of Eq.~(\ref{eq_Gtdef}) or 
Eq.~(\ref{eq_liquid_condition}), decays exponentially
\begin{equation}
R(t) \approx \Rstar \exp(-t/\taustar(\nu)) \mbox{ for } t \gg 1
\label{eq_Maxwell}
\end{equation}
with $\Rstar \approx 18$ being the plateau modulus set by the equilibrium shear modulus 
for permanent springs ($\nu=0)$ and $\taustar(\nu) \approx 16/\nu$ the Maxwell time.
One may use as scaling variable the reduced sampling time $\xsamp = \tsamp/\taustar$ for 
$1 \ll \tsamp \le \tsampmax$ to collapse data obtained for different $\tsamp$ and $\nu$ \cite{WKC16}.
The short-time decay of $R(t)$ from the initially imposed affine stain, $R(t=0)=\RA$,
to the plateau modulus $\Rstar$ is reasonably described by a compressed exponential ($\beta > 1$). 
A useful formula for $R(t)$ for all $t$ and $\nu$ is given by the two-mode approximation
\begin{equation}
R(t) = b \exp(-(t/\tauA)^{\beta}) + \Rstar \exp(-t/\taustar)
\label{eq_TSANET_Gtmodel}
\end{equation}
with amplitude $b \approx \RA \approx 33$, relaxation time $\tauA \approx 0.3$ and exponent $\beta \approx 1.5$.

\begin{figure}[t]
\centerline{\resizebox{.9\columnwidth}{!}{\includegraphics*{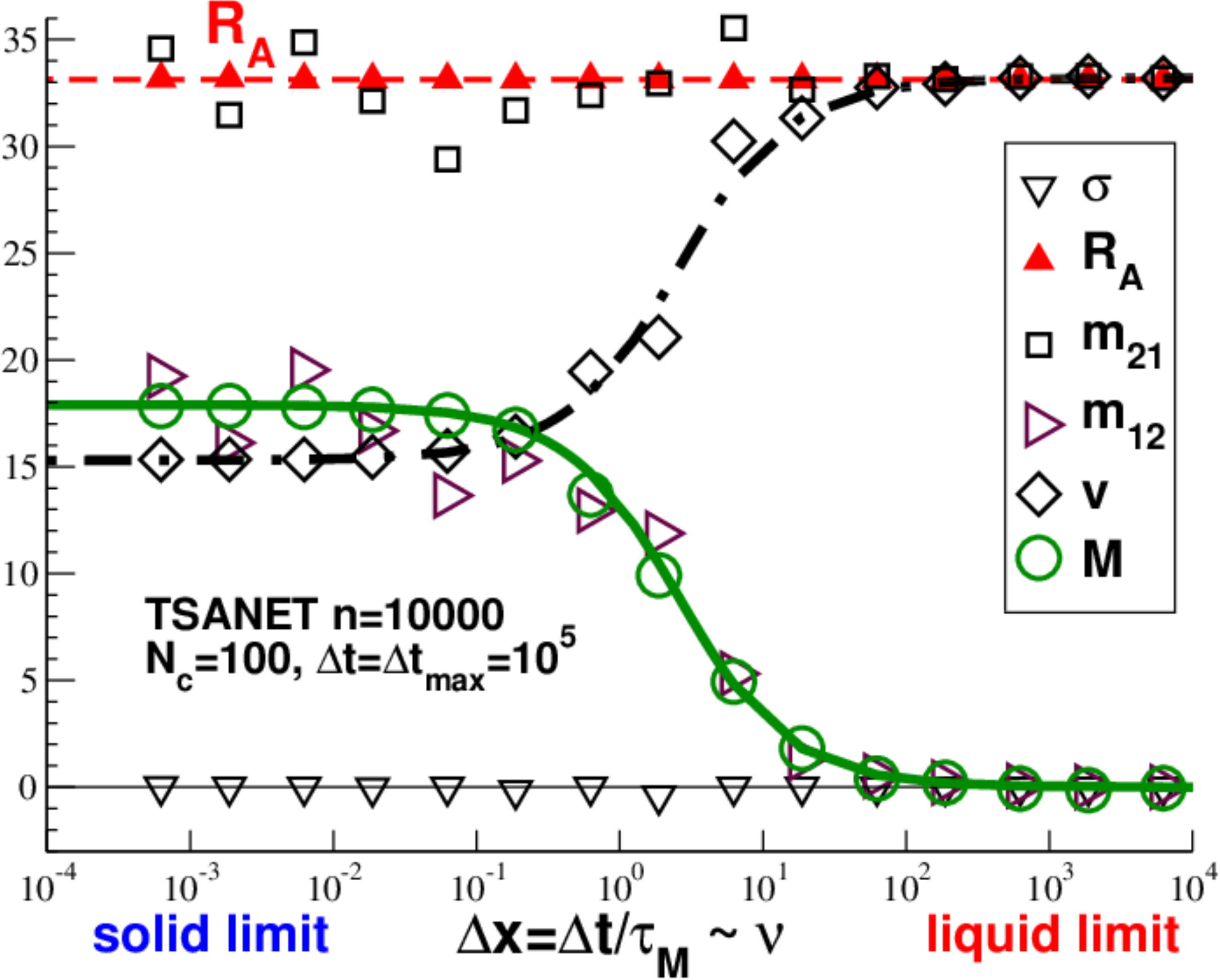}}}
\caption{Various properties obtained for the TSANET model {\em vs.} the reduced sampling time 
$\xsamp = \tsamp/\taustar$ for $\tsamp=10^5$: 
mean shear stress $\sigma$, affine shear modulus $\RA$, contributions $m_{21}$
and $m_{12}$ to the shear-stress fluctuation $v=m_{21}-m_{12}$
and stress-fluctuation formula $M = \RA-v$ for the shear modulus.
The prediction for a Maxwell model is indicated by the
bold solid line for $M$ and by the dashed-dotted line for $v$.
Note that $\RA \approx m_{21}$ and $M \approx m_{12}$ for all $\xsamp$.
}
\label{fig_TSANET_mean}
\end{figure}
Since the hopping moves changing the network connectivity obey detailed balance, 
changing $\nu$ leaves all truly static properties unchanged
as may be seen from Fig.~4 of Ref.~\cite{WKC16}.
For this reason all simple averages, Eq.~(\ref{eq_commute}), such as
the average shear stress $\sigma$, the average affine shear modulus $\RA$ 
or the moment $m_{21}$, should not depend on $\tsamp$ or $\xsamp$. 
That this is indeed the case can be seen from Fig.~\ref{fig_TSANET_mean}.
Note that $\sigma=0$ by symmetry.
$\RA$ and $m_{21}$ are roughly equal, $\RA \approx m_{21} \approx 33$,
albeit $m_{21}$ fluctuates more strongly for small $\nu$.\footnote{That two properties are equal 
on average does, of course, not imply that they are identical since other moments (fluctuations) 
may be different. This is the case for $\RA$ and $m_{21}$ which have different standard deviations 
$\delta \RA \ll \delta m_{21}$ as seen from Fig.~8 of Ref.~\cite{WKC16}.} % \cite{foot_TSANET_equal}. 
Also presented in Fig.~\ref{fig_TSANET_mean} are $m_{12}$, $v$ and $M$ which are all seen to 
depend on the reduced sampling time.
Note that $M \approx m_{12}$ in agreement with Eq.~(\ref{eq_liquid_condition}).
As expected for a Maxwell model according to Eq.~(\ref{eq_M_MW}), 
$M(\xsamp) = \Rstar \ \gDebye(\xsamp)$ holds (bold line). 
The corresponding relation 
$v(\xsamp)=\RA-\Rstar \ \gDebye(\xsamp)$ is indicated by the dash-dotted line.
Note that $\gDebye(x) \to 1$ for $x \to 0$ and $\gDebye(x) \to 2/x$ for $x \gg 1$.
This implies that in the liquid limit $M(\xsamp) \approx 2\Rstar/\xsamp$
and $v(\xsamp) \approx \RA - 2\Rstar/\xsamp$.
Importantly, the $\xsamp$-dependence of these properties is not due to aging
or equilibration problems but is caused by the finite time needed for the 
equilibrium fluctuations to explore the phase space 
reflecting the stress relaxation process $R(t)$.

\begin{figure}[t]
\centerline{\resizebox{.9\columnwidth}{!}{\includegraphics*{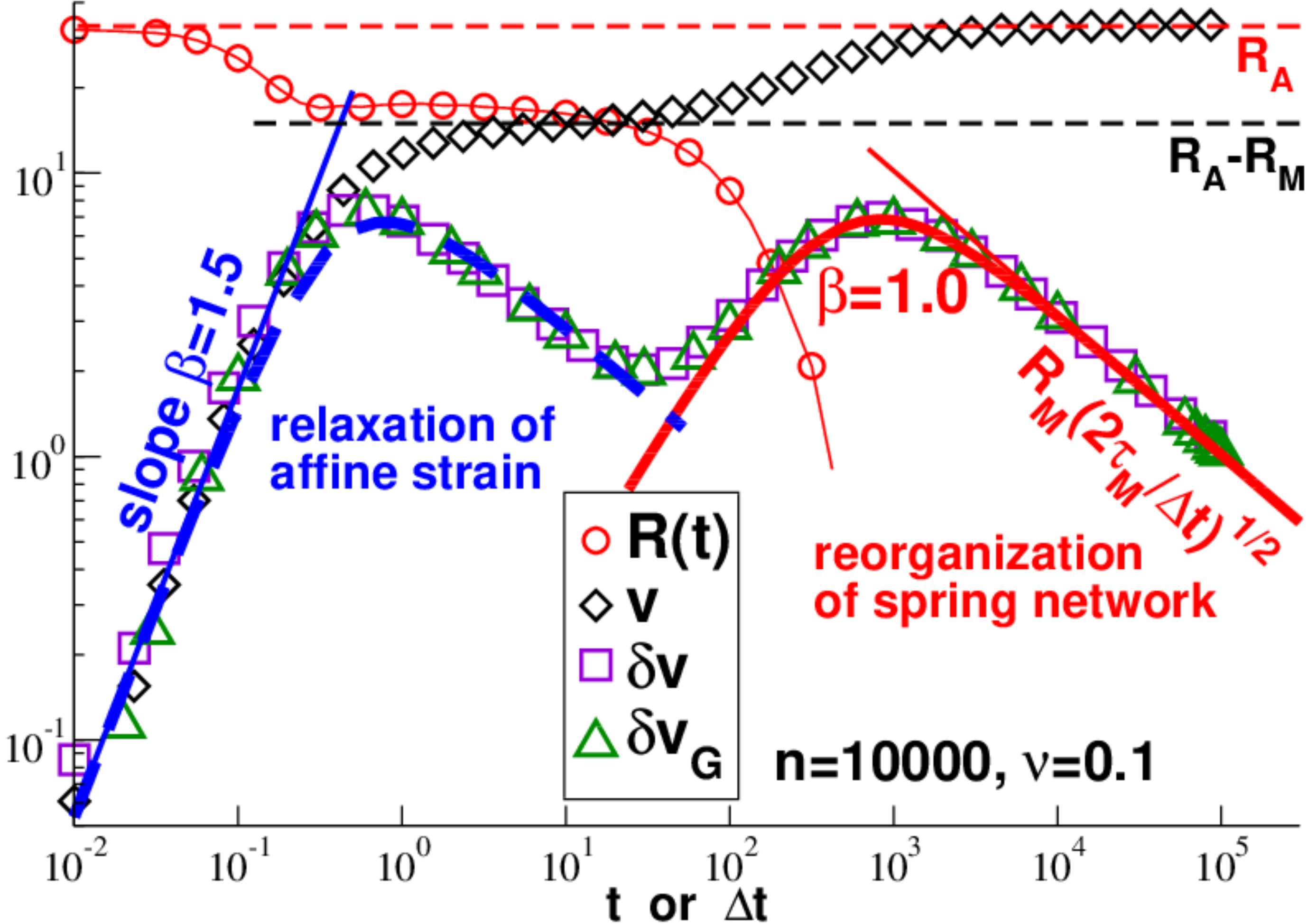}}}
\caption{Comparison of $\delta v(\tsamp)$ and $\svgauss(\tsamp)$ for a high hopping frequency $\nu=0.1$
confirming that $\delta v \approx \svgauss$ for all $\tsamp$.
Also included are the shear-stress relaxation function $R(t)$ and the shear-stress fluctuation $v(\tsamp)$.
Note that $v(\tsamp)$ has a shoulder below $\taustar \approx 160$ (lower dashed line)
and a plateau with $v \approx \RA$ for $\tsamp \gg \taustar$ (upper dashed line).
The thin solid line on the left indicates the power law $\svgauss \propto \tsamp^{\beta}$ with $\beta=1.5$
expected from Eq.~(\ref{eq_cbeta_small_tsamp}) and Eq.~(\ref{eq_TSANET_Gtmodel}), 
the thin solid line on the right the final $1/\sqrt{\tsamp}$-decay.
}
\label{fig_TSANET_dv_fhig}
\end{figure}

We turn now to the characterization of the standard deviation $\delta v$.
As shown in Fig.~\ref{fig_TSANET_dv_fhig} $\delta v \approx \svgauss$ is found to hold to 
high precision for all high hopping frequencies with $1 \ll \tsampmax/\taustar(\nu) \propto \nu$.
For $\nu=0.1$ both the short-time relaxation time $\tauA$, characterizing the relaxation of the affine strain, 
and the Maxwell time $\taustar(\nu) = 16/\nu$, characterizing the reorganization of the network, are relevant.
We have also included the shear-stress relaxation function $R(t)$ which is seen to vanish above $\taustar$.
Note that $v$ has a shoulder with $v \approx \RA-\Rstar$ for $\tsamp \ll \taustar$
and that $v \approx \RA$ for larger $\tsamp$, as expected in the liquid limit.
In agreement with Sec.~\ref{theo_models} we observe for this two-step relaxation 
process two well separated $\delta v$-maxima (cf. Fig.~\ref{fig_ft_maxwell}). 
The first relaxation process at $\tauA$ is described by a compressed exponential with $\beta=1.5$ (bold dashed line),
the second relaxation process due to the reorganization of the spring network
for larger $\tsamp$ by a Maxwell model ($\beta=1$) with relaxation time $\taustar$
(bold solid line).
In fact, for all not too small $\nu$ $\delta v$ is given 
by $\svgauss[R]$ assuming Eq.~(\ref{eq_TSANET_Gtmodel}) to hold.

\begin{figure}[t]
\centerline{\resizebox{.9\columnwidth}{!}{\includegraphics*{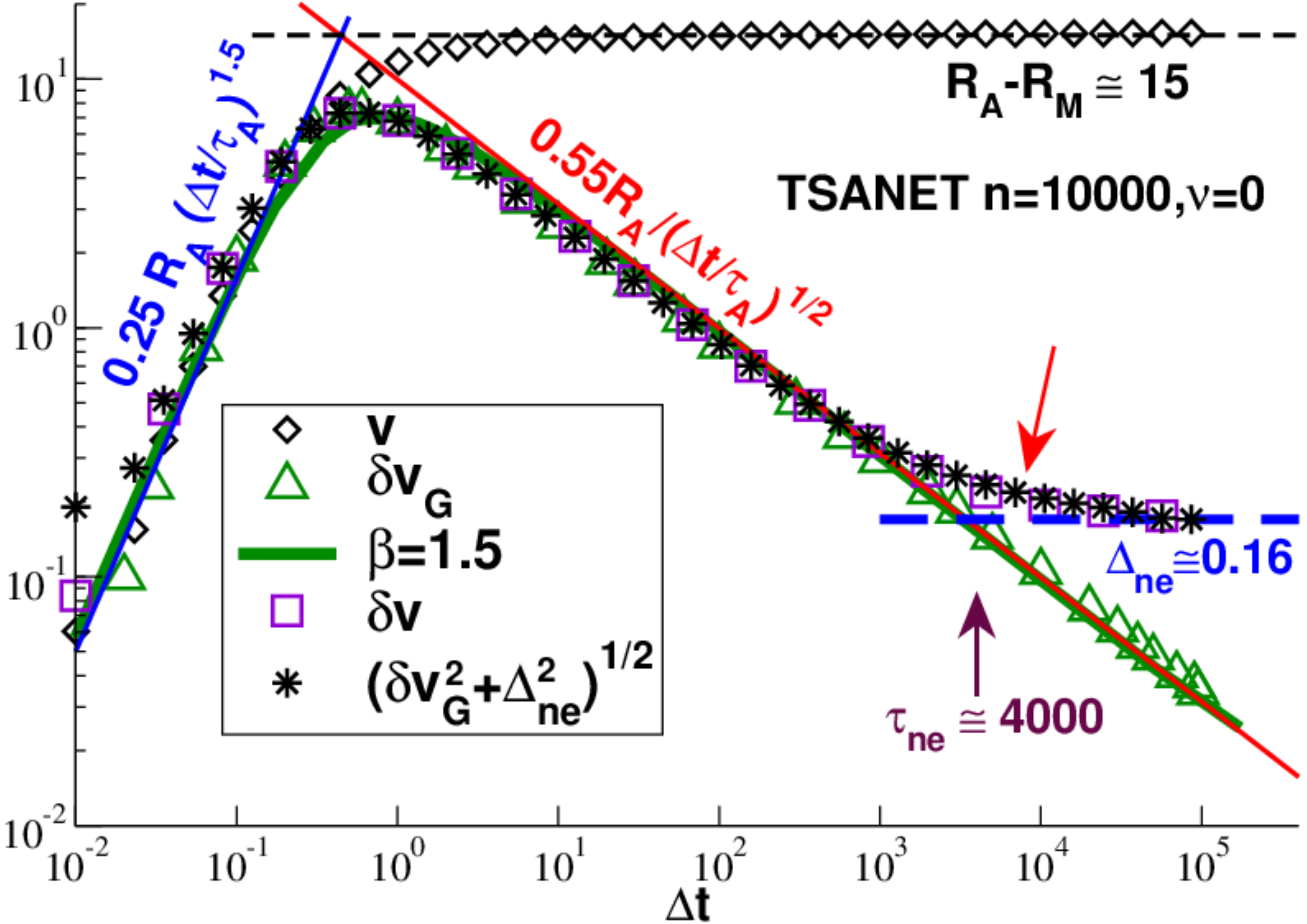}}}
\caption{Shear-stress fluctuation $v$ and the standard deviations $\delta v$ and $\svgauss$ {\em vs.} $\tsamp$ for $\nu=0$.
The bold solid line corresponds to $\svgauss$ obtained assuming a compressed exponential with $\beta=1.5$.
While $\delta v \approx \svgauss$ holds for $\tsamp \ll \Tnonerg \approx 4000$, 
$\delta v \to \Snonerg$ for larger $\tsamp$. 
The stars indicate Eq.~(\ref{eq_dv_gen}) using $\Snonerg=0.16$.
}
\label{fig_TSANET_dv_flow}
\end{figure}

As shown in Fig.~\ref{fig_TSANET_dv_flow} this becomes different for small $\nu$ 
due to quenched shear-stress fluctuations. We present here data obtained for a quenched network
with switched off hopping moves ($\nu=0$).
Also indicated is the shear-stress fluctuation $v$ (diamonds) which is seen to rapidly increase
for small $\tsamp$, corresponding to the relaxation of the imposed affine strain, and to level off
for $\tsamp \gg \tauA$ as indicated by the horizontal dashed line. Since in agreement with
Eq.~(\ref{eq_plateau}) $h(t)$ or $R(t)$ become also constant in this time regime (not shown)
this implies that $\svgauss$ must decay as $1/\sqrt{\tsamp}$ for $\tsamp \gg \tauA$.
This is confirmed by the $\svgauss[h]$-data (triangles) computed from the measured $h(t)$,
revealing after a first regime with $\svgauss \propto \tsamp^{\beta}$ and $\beta \approx 1.5$
the expected $1/\sqrt{\tsamp}$-decay. As shown by the bold solid line, a reasonably fit of $\svgauss$ 
for all $\tsamp$ is obtained using Eq.~(\ref{eq_TSANET_Gtmodel}).
(For $\nu =0$ the second term in Eq.~(\ref{eq_TSANET_Gtmodel}) is an irrelevant constant.)
While $\delta v$ is identical (within numerical precision) to $\svgauss$ for short $\tsamp$ it deviates for large $\tsamp$ 
where it levels off, $\delta v \to \Snonerg \approx 0.16$, as indicated by the bold dashed horizontal line.
As discussed in Sec.~\ref{theo_nonerg} the leveling-off is expected for a finite dispersion of the $v_c$.
The interpolation formula Eq.~(\ref{eq_dv_gen}) motivated in Sec.~\ref{theo_nonerg}
gives a reasonable approximation of $\delta v$ (stars) matching both limits for $\tsamp \gg \tauA$. 
To leading order, $\delta v$ is thus given by $\svgauss$ and, hence, by $h(t)$ or $R(t)$ {\em plus} an additional constant.
As indicated by the tilted arrow in Fig.~\ref{fig_TSANET_dv_flow}, Eq.~(\ref{eq_dv_gen}) slightly 
{\em overpredicts} $\delta v$ for intermediate $\tsamp$. This suggests that the constant $\Snonerg$ should be
replaced by the more general standard deviation $\svext(\tsamp)$, describing the $\tsamp$-depending dispersion 
between configurations,
approaching monotonically the large-$\tsamp$ limit $\Snonerg$ with increasing $\tsamp$ from below.  
This minor difference will be discussed elsewhere \cite{fluctuGaussB}.
Similar and complementary results for glass-forming polydisperse particles and
thin free-standing polymer films are reported in Appendix~\ref{app_shear}.

\subsection{System-size effects for ${\bf \delta v}$ and ${\bf \Snonerg}$}
\label{shear_V}

\begin{figure}[t]
\centerline{\resizebox{1.0\columnwidth}{!}{\includegraphics*{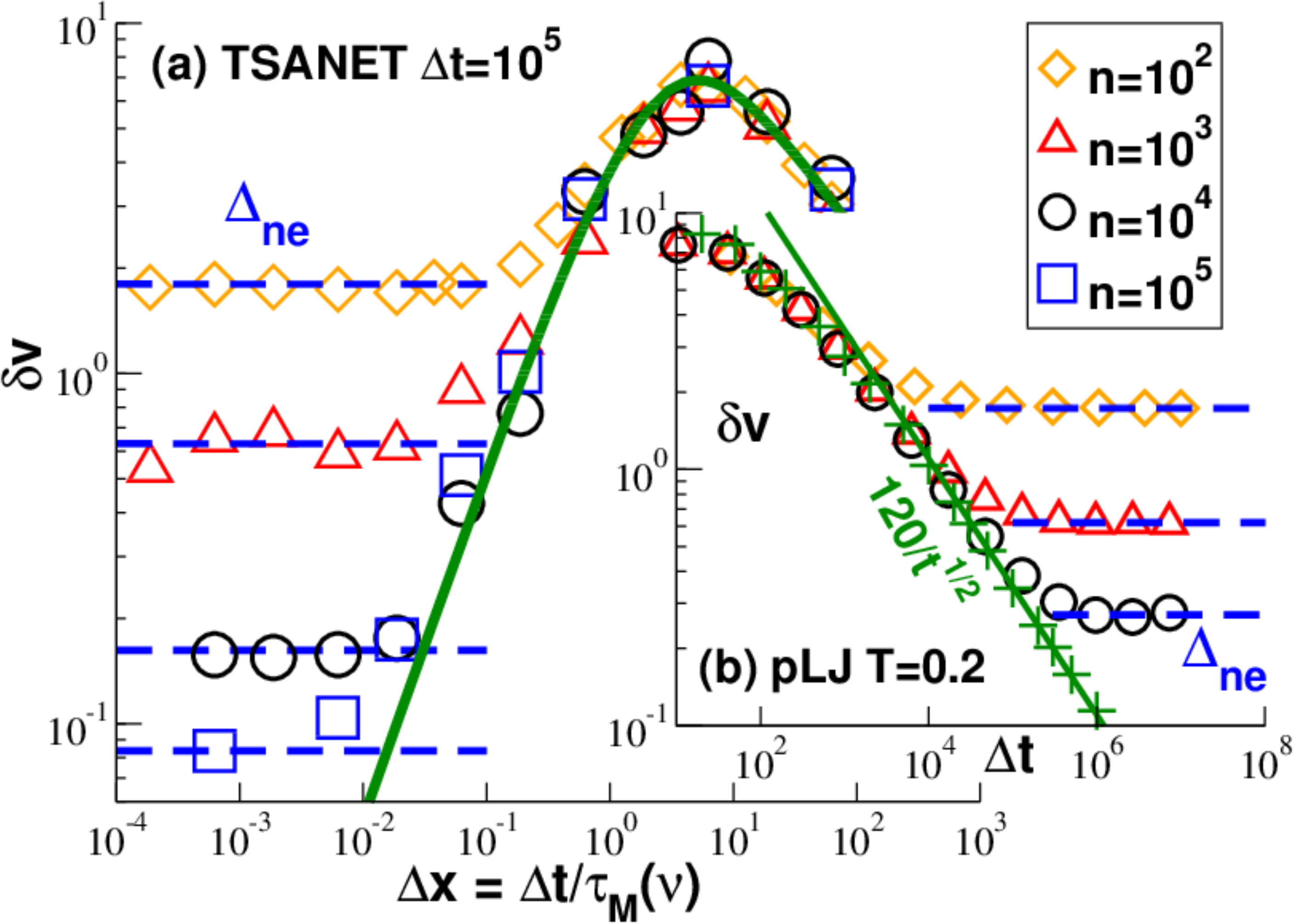}}}
\caption{$\delta v$ for different $n$ as indicated:
{\bf (a)} $\delta v(\xsamp)$ for TSANET networks with constant $\tsamp=10^5$. 
The bold solid line indicates $\svgauss$ for the Maxwell model, Eq.~(\ref{eq_ft_dvgauss_MW}), 
the dashed horizontal lines $\Snonerg$ obtained for quenched networks ($\nu=0$). 
{\bf (b)} $\delta v(\tsamp)$ for pLJ particles at $T=0.2$. 
$\svgauss$ obtained from $R(t)$ for $n=10^4$ is indicated by crosses, 
the asymptotic power-law slope $120/\sqrt{\tsamp}$ by the thin solid line.
The deviations from Eq.~(\ref{eq_key_2}) vanish with increasing $n$.
}
\label{fig_V_dv}
\end{figure}

We have focused up to now on the variation of the sampling time $\tsamp$,
the hopping frequency $\nu$ or the temperature $T$ (cf. Appendix~\ref{app_shear})
while keeping fixed other parameters such as the total number of beads $n$.
While most properties discussed above as $\RA$, $v$ or $h$ are defined as intensive properties, 
i.e. as we have checked their mean values do not or extremely weakly depend on $n$,  
this is less obvious for their respective standard deviations \cite{WKC16,Procaccia16,lyuda19a}.
We address here briefly the $n$-dependence of the standard deviation $\delta v$.
All presented systems have roughly the same number density $\rho$ of order unity, i.e. $n \approx V$.
$\delta v$ is presented in Fig.~\ref{fig_V_dv} for a broad range of $n$ for the TSANET model and the pLJ particles
(cf. Appendicies~\ref{app_algo_pdLJ} and \ref{app_shear_pdLJ}).
The TSANET data in panel (a) are plotted as a function of the reduced hopping 
frequency $\xsamp \propto \nu$ for our largest sampling time $\tsamp=\tsampmax=10^5$.
(The short-time behavior around $\tauA$ is thus irrelevant.)
The dashed horizontal lines indicate $\Snonerg$ obtained as in Fig.~\ref{fig_TSANET_dv_flow}
from the large $\tsamp$-limit of $\delta v$ for quenched networks.
The bold solid line represents $\svgauss$ for the one-mode Maxwell model.
Not shown for clarity are the $\svgauss[h]$ obtained 
for the different $n$ which are found to be essentially $n$-independent and very similar to the Maxwell model.
At variance to this $\delta v$ is only intensive for sufficiently large $\xsamp$ 
where $\delta v \approx \svgauss$ holds,
but not in the low-$\xsamp$ limit where $\delta v \to \Snonerg$.  
Panel (b) presents $\delta v$ as a function of $\tsamp$ for the pLJ particles.
Also given are $\svgauss[h]$-data for $n=10000$ (crosses).
The dashed horizontal lines indicate the plateau value $\Snonerg$ for each $n$.
$\Snonerg$ systematically decreases with $n$. 
$\delta v$ and $\svgauss$ thus become increasingly similar according to Eq.~(\ref{eq_dv_gen}).
We also note that a scaling collapse of $\delta v$ is achieved for both models by plotting 
$\delta v/\Snonerg(n)$ 
as a function of $\tsamp/\Tnonerg(n)$ with $\Tnonerg(n)$ determined according to Eq.~(\ref{eq_Tnonerg_def}).

\begin{figure}[t]
\centerline{\resizebox{0.9\columnwidth}{!}{\includegraphics*{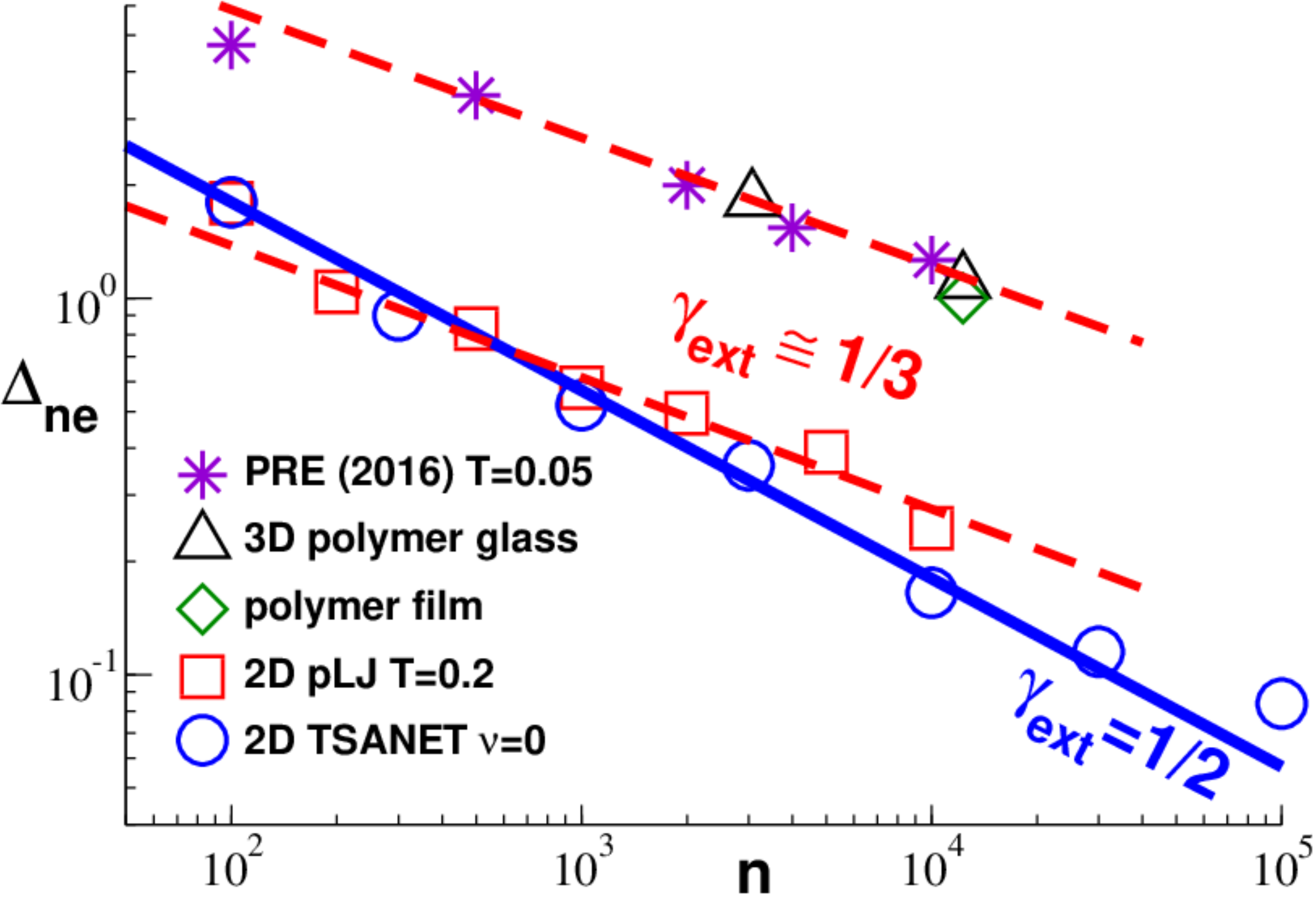}}}
\caption{$\Snonerg$ vs. $n$
for a binary LJ mixture at $T=0.05$ (stars) \cite{Procaccia16},
for a 3D polymer glass at $T \approx 0.1$ (triangles) \cite{lyuda19a},
for polymer films at $T=0.05$ (diamonds),
for the pLJ model at $T=0.2$ (squares) and
for the TSANET model at $\nu=0$ (circles).
$\Snonerg$ decreases with a power-law exponent $\gamext\approx 1/3$ for the amorphous glasses (dashed lines) 
and $\gamext=1/2$ for the TSANET model (bold solid line).
}
\label{fig_V_Snonerg}
\end{figure}

Figure~\ref{fig_V_Snonerg} summarizes the system-size dependence of $\Snonerg$ for several models.
We compare published results \cite{Procaccia16,lyuda19a} with new data obtained for the TSANET model ($\nu=0$), 
the pLJ particles ($T=0.2$) and the free-standing polymer films ($T=0.05$). % presented in Appendix~\ref{app_shear_film}.
The TSANET data (circles) are fitted by $\Snonerg \propto 1/n^{\gamext}$
with $\gamext=1/2$ (bold solid line). The observed ``strong self-averaging" \cite{LandauBinderBook} 
suggests that independent localized shear-stress fluctuations with a finite correlation length $\xi$ are 
responsible for $\Snonerg$ in agreement with Eq.~(\ref{eq_xim_independent}).
This finding is at variance to the somewhat smaller exponent $\gamext \approx 1/3$ 
suggested by recent simulation studies of 2D binary LJ mixtures (stars) \cite{Procaccia16}\footnote{The 
data of the binary LJ mixture scanned from the first panel of Fig.~2 of Ref.~\cite{Procaccia16} corresponds strictly 
speaking to the standard deviation $\delta M$ for the shear modulus $M=\RA-v$. Since the fluctuations 
of $\RA$ are negligible, however, as shown elsewhere \cite{WKC16,lyuda19a}, $\delta M \approx \delta v$.}
and of dense 3D polymer glasses (triangles) \cite{lyuda19a}.
This led us to conclude \cite{lyuda19a} that local elastic (structural) properties in amorphous systems may 
cause long-range spatial correlations and a diverging correlation length $\xi$.
As shown by the lower dashed line, the exponent $\gamext \approx 1/3$ is also compatible with the new data 
obtained for our extremely well equilibrated pLJ particles.
Unfortunately, only two system sizes have been probed for the 3D polymer system \cite{lyuda19a}.
This makes it difficult to assess whether $\gamext$ depends on the spatial dimension or not.
Simulations with a broader range of $n$ are currently sampled to corroborate this point
and to verify $\gamext$ in two and three dimensions.

\section{Conclusion}
\label{sec_conc}
We have discussed systematically the ensemble average $v(\tsamp)$ and 
the standard deviation $\delta v(\tsamp)$ of the variance $v[\xbf]$, 
Eq.~(\ref{eq_vxdef}), of a time series $\xbf$ measured over a sampling time $\tsamp$.
Our aim was to give an uncluttered overview of some relations which may be useful in different fields
where the stochastic processes are essentially, albeit perhaps not rigorously, both stationary and Gaussian.
We have emphasized first in Sec.~\ref{theo_station} that for stationary processes 
$v$ is given by a weighted sum (integral), 
Eq.~(\ref{eq_key_1}), over the ACF $h(t) = c(0)-c(t)$ (Sec.~\ref{theo_stationary}). 
Assuming an ergodic Gaussian process (Sec.~\ref{theo_gauss}) 
$\delta v$ was shown in Sec.~\ref{theo_wick} 
to be given by the functional $\svgauss[h]$, Eq.~(\ref{eq_key_2}).
As discussed in Sec.~\ref{theo_models}
the reduced standard deviation $\svgauss/v$ taken at the maximum of $\svgauss$
may become of order unity if $h(t)$ changes rapidly (Fig.~\ref{fig_ft_beta}), 
i.e. the average behavior $v$ gets masked by strong fluctuations.
As emphasized in Sec.~\ref{theo_nonerg} Eq.~(\ref{eq_key_2}) cannot hold
for non-ergodic systems with a finite dispersion of the frozen variances $v_c$ of the
different independent configurations $c$ since $\delta v$ must become constant,
$\Snonerg$, for $\tsamp \gg \Tnonerg$ while $\svgauss$ vanishes.
However, if the observable $x$ is the sum of many more or less decoupled 
microscopic variables the quenched $v_c$ become similar with increasing system size  
and, hence, $\Snonerg \to 0$, $\Tnonerg \to \infty$ and $\delta v \to \svgauss$ 
even for non-ergodic systems (Sec.~\ref{theo_V}).
%\footnote{In agreement with Eq.~(\ref{eq_Tnonerg_def}) 
%and assuming $\svgauss \propto 1/\tsamp^{\beta}$ with $\beta \approx 1/2$ we have 
%$\Tnonerg \propto \Nm^{\gamext/\beta}$, i.e. the crossover time increases with $\Nm$ and 
%$\Snonerg^{1/\beta} \propto 1/\Tnonerg$.}
%
 
In the computational part of this work (Sec.~\ref{sec_shear}) we have illustrated some of the relations 
%of Sec.~\ref{sec_theo} 
by applying them to the shear-stress fluctuations 
(Sec.~\ref{shear_intro}) in transient self-assembled networks (Sec.~\ref{shear_TSANET}).
Similar results are reported in Appendix~\ref{app_shear} 
for glass-forming polydisperse particles and free-standing polymer films.
Albeit it is non-trivial \cite{HansenBook,Barrat14b,lyuda18,lyuda19a} 
whether the shear-stress trajectories of these systems are sufficiently stationary and Gaussian,
all examples reveal qualitatively the same behavior:
\begin{itemize}
\item
all systems are (at least effectively) stationary as shown in panel (c) of Fig.~\ref{fig_film_SFF};
\item 
$\delta h(t)^2 \approx 2 h(t)^2$ holds to high precision (Fig.~\ref{fig_pdLJ_dh})
as expected for Gaussian processes;
\item
$\delta v \approx \svgauss$ holds within numerical precision in the ergodic limit for large 
hopping frequencies $\nu$ (Fig.~\ref{fig_TSANET_dv_fhig}) or temperatures $T$ 
(Figs.~\ref{fig_pdLJ_dv_T} and \ref{fig_film_dv}b);
\item
while $\svgauss \propto 1/\sqrt{\tsamp}$ vanishes in the non-ergodic limit,
$\delta v$ becomes constant, $\delta v \to \Snonerg > 0$, for large $\tsamp$ 
(Figs.~\ref{fig_TSANET_dv_flow}, \ref{fig_pdLJ_dv_tsamp} and \ref{fig_film_dv}a);
\item
$\Snonerg$ decreases with system size  (Figs.~\ref{fig_V_dv} and \ref{fig_V_Snonerg}) 
suggesting that Eq.~(\ref{eq_key_2}) becomes valid for macroscopic albeit non-ergodic elastic bodies.
\end{itemize}

The presented work was limited to the characterization of fluctuations and relaxation processes of 
stationary (equilibrium) stochastic processes, i.e. no external perturbation was applied to directly measure
the average response function $R(t)$ or the average modulus $M$ and their, respective, standard deviations $\delta R(t)$
and $\delta M$. 
Our claim that $\delta R / R$ or $\delta M/M$ must generally become large (of order unity)
for times where $R(t)$ strongly decays and that these ratios are, moreover, system-size independent 
may in fact be misleading for the {\em out-of-equilibrium} responses of real macroscopic materials. 
From the theoretical point of view it is an interesting question how to generalize
the fluctuation-dissipation relations, connecting the {\em average} linear out-of-equilibrium response 
to the {\em average} equilibrium relaxation \cite{DoiEdwardsBook,HansenBook,WKC16}, 
to describe the sample-to-sample fluctuations.
 
We have briefly discussed in Sec.~\ref{shear_V} the system-size effects for various properties 
focusing on $\delta v$ and the associated finite plateau value $\Snonerg$ for non-ergodic systems. 
As shown in Fig.~\ref{fig_V_Snonerg} two different exponents $\gamext$ characterize the decay of $\Snonerg$ with $n$
for the perfectly equilibrated TSANET model ($\gamext=1/2$) and the quenched amorphous glasses ($\gamext \approx 1/3$).
The clarification of this difference is beyond the scope of the present paper.
We shall also give elsewhere \cite{fluctuGaussB} a systematic description of the {\em two} different types 
of standard deviations briefly mentioned in Sec.~\ref{theo_nonerg} which must be distinguished for the 
complete characterization of fluctuations of ensembles of non-ergodic systems.
This will allow to further discuss the surprising ``breakdown of nonlinear elasticity in amorphous solids" 
\cite{Procaccia16}
---
based on the numerically observed divergence with system size for standard deviations associated 
with higher-order nonlinear analogs of the elastic shear modulus
---
claimed at variance to the every day experience that sufficiently large amorphous (plastic) bodies are
well behaved according to standard continuum mechanics \cite{FerryBook,TadmorCMTBook}.

\section*{Author contribution statement}
JB and JPW designed the research project.
The presented theory was gathered from different sources and further developed by ANS and JPW.
GG (polymer films), LK (pLJ particles) and JPW (TSANET) performed the simulations and the analysis of the data.
JPW wrote the manuscript, benefiting from contributions of all authors.

%\begin{acknowledgments}
\section*{Acknowledgments}
We are indebted to O.~Benzerara (ICS, Strasbourg) for helpful discussions and acknowledge important computational 
resources from the HPC cluster of the University of Strasbourg.

\appendix
%\clearpage
%\newpage
%\input{app_reformulations}
\section{Reformulations of ${\bf \svgauss[f]}$}
\label{app_reformulations}

Since for large $I$ the sums over two, three or even four indices stated in Sec.~\ref{theo_wick} 
rapidly become numerically unfeasible, it is of importance that the three terms 
$\Ta$, $\Tb$ and $\Tc$ of Eq.~(\ref{eq_dv2_terms}) can be simplified to single loops
\cite{lyuda19a}. The first two terms simply become
\begin{eqnarray}
\Ta(\tsamp) & = & \frac{2}{I^2} \left( I f_0^2 + 2 \sum_{k=1}^{I-1} (I-k) f_k^2 \right)
\label{eq_Ta_2}\\
\Tb(\tsamp) & = & \frac{2}{I^4} \left( I f_0 + 2 \sum_{k=1}^{I-1} (I-k) f_k \right)^2
\label{eq_Tb_2}
\end{eqnarray}
with $f_k = f(t=t_k)$ for an arbitrary function $f(t)$.
Let us define the sum $S(s,I) \equiv \sum_{i=1}^{I} f_{i-s}$. 
Note that $S(s,I)$ may be computed starting from $S(0,I)$ using the recursion relation
$S(s+1,I) = S(s,I) + f_s - f_{I-s}$. Using this the calculation of 
\begin{equation}
\Tc(\tsamp) = \frac{4}{I^3} \sum_{s=1}^{I} S(s,I)^2 
= \frac{8}{I^3} \sum_{s=1}^{I/2} S(s,I)^2
\label{eq_Tc_2} 
\end{equation}
becomes also of order ${\cal O}(I)$. Using the symmetry $S(s,I)=S(I-s+1,I)$ we have
assumed in the last step that $I$ is even.
In the continuum limit for large $I=\tsamp/\delta t$ the three terms further simplify to
\begin{eqnarray}
\Ta(\tsamp) & = & 
\frac{4}{\tsamp^2} \int_0^{\tsamp} \ddiff t \ (\tsamp-t) f(t)^2 \label{eq_Ta_3} \\
\Tb(\tsamp) & = & 
\frac{8}{\tsamp^4} \left( \int_0^{\tsamp} \ddiff t \ (\tsamp-t) f(t) \right)^2 \label{eq_Tb_3} \\
\Tc(\tsamp) & = & 
\frac{8}{\tsamp^3} \int_0^{\tsamp/2} \ddiff t \left( \eta(t)+\eta(\tsamp-t) \right)^2 \label{eq_Tc_3}
\end{eqnarray}
using $\eta(t) \equiv \int_0^t \ddiff u \ f(u) $ for the last contribution.

\section{Shear stress and Born-Lam\'e coefficient}
\label{app_affine}
Let us consider a small simple shear strain \cite{TadmorCMTBook} $\gamma$ in the $xy$-plane as it would be 
used to measure the shear-stress relaxation function $R(t)$ \cite{AllenTildesleyBook,WXB15,WXBB15,WKB15}.
Assuming that all particle positions $\rvec$ follow an imposed ``macroscopic" shear
in an {\em affine} manner according to $\rx \to \rx + \gamma \ \ry$
the Hamiltonian $\Hhat$ of a given configuration changes to leading order as \cite{WXBB15,WKC16,ivan17a,ivan18}
\begin{equation}
(\Hhat(\gamma)-\Hhat(\gamma=0))/V \approx \tauhat \gamma + \frac{1}{2} \muAhat \gamma^2
\mbox{ for } |\gamma| \ll 1.
\label{eq_Hhatexpand}
\end{equation}
The instantaneous shear stress $\tauhat$ and the instantaneous
Born-Lam\'e coefficient $\muAhat$ are thus defined as
\begin{eqnarray}
\tauhat & \equiv & \Hhat^{\prime}(\gamma)/V|_{\gamma=0} \label{eq_tauhatdef} \mbox{ and } \\
\muAhat & \equiv & 
\Hhat^{\prime\prime}(\gamma)/V|_{\gamma=0} =
\tauhat^{\prime}(\gamma)|_{\gamma=0} 
\label{eq_muhatdef}
\end{eqnarray}
where a prime denotes a functional derivative with respect to the affine small strain transform.
All properties considered here refer to the excess contributions due to the potential
part of the Hamiltonian, i.e. the ideal contributions are assumed to be integrated out.\footnote{This 
is called Monte Carlo gauge in previous work \cite{WXP13,WXB15}.
Note that an ideal gas has a vanishing shear modulus.
Thus, the ideal contributions to the stress-fluctuation formula $M = \RA - v$ rigorously cancel.}
%\cite{foot_idealcontri}.
%
Assuming a pairwise central conservative potential $\sum_l u(\rl)$
with $\rl$ being the distance between a pair of monomers $l$,
one obtains the excess contributions \cite{WXBB15,film18}
\begin{eqnarray}
\tauhat & = & \frac{1}{V} \sum_l \rl u^{\prime}(\rl) \ \nlx \nly   
\label{eq_tauexhat} \ \mbox{ 
and } \\
\muAhat & = & \frac{1}{V} \sum_l  \left( \rl^2 u^{\prime\prime}(\rl)
- \rl u^{\prime}(\rl) \right) \nlx^2 \nly^2 \nonumber \\
& + & \frac{1}{V} \sum_l \rl u^{\prime}(\rl) \ (\nlx^2+\nly^2)/2  
\label{eq_muexhat}
\end{eqnarray}
with $\nvecl = \rvecl/\rl$ being the normalized distance vector.
Note that Eq.~(\ref{eq_tauexhat}) is strictly identical to the
corresponding off-diagonal term of the Kirkwood stress tensor  \cite{AllenTildesleyBook}.
We have used a symmetric representation for the last term of Eq.~(\ref{eq_muexhat})
exchanging $x$ and $y$ for the affine transform and averaging over the equivalent
$x$ and $y$ directions. Note that this last term
automatically takes into account the finite normal pressure of the system.
Similar relations are obtained for the $xz$- and the $yz$-plane.
See Refs.~\cite{XWP12,lyuda19a} for the corresponding expression of the 
ensemble average of $\muAhat$
in terms of the pair correlation functions of the bonded and the non-bonded
interactions of the particles.
Please note that $\muAhat$ depends on the second derivative $u^{\prime\prime}(r)$ 
of the pair potential. As emphasized elsewhere \cite{XWP12},
impulsive corrections need to be taken into account due to this term if the first derivative 
$u^{\prime}(r)$ of the potential is not continuous. Unfortunately, this is the case at the 
cutoff $\rcut$ of the LJ potentials used for the pLJ beads (Appendix~\ref{app_shear_pdLJ}) and 
for the polymer films (Appendix~\ref{app_shear_film}).
The ``bare" $\muAhat$ is thus roughly about $0.2$ too high for both models and must be corrected \cite{XWP12}.
The ensemble average $\la \muAhat \ra$ is called $\muA$ in previous publications 
\cite{SBM11,XWP12,WXP13,WXB15,WXBB15,WKB15,LXW16,WXB16,WKC16,ivan17a,ivan17c,ivan18,film18,lyuda19a}
and $\RA$ in the present work.

\section{Numerical models and technical details}
\label{app_algo}

\subsection{Introduction}
\label{app_algo_intro}
Various issues discussed theoretically in %Sec.~\ref{sec_theo} 
Secs.~\ref{theo_station}-\ref{theo_V}
are illustrated in Sec.~\ref{sec_shear} 
and in Appendix~\ref{app_shear} for the fluctuations of shear stresses in simple coarse-grained model systems.
A sketch of the three\footnote{A forth example may be found in Ref.~\cite{lyuda19a} 
where three-dimensional glass-forming polymer melts are presented.} models studied 
is given in Fig.~\ref{fig_algomod}.
Standard MD and MC methods \cite{AllenTildesleyBook,LandauBinderBook}
are used and in some cases combined. The number of particles $n$ and the temperature $T$ are imposed
while the system volume $V$ is allowed to fluctuate in some cases.
Boltzmann's constant $\kB$, the typical size of the particles (beads) and
the particle mass are all set to unity and Lennard-Jones units \cite{AllenTildesleyBook}
are used throughout this work. For all systems we use periodic boundary conditions 
\cite{AllenTildesleyBook,LandauBinderBook} and the number density $\rho = n/V$ is close to unity.
The salient features of each model and some algorithmic details are given below.

\subsection{Transient self-assembled networks}
\label{app_algo_TSANET}

As explained in detail in Ref.~\cite{WKC16}, we use a simple model
for transient self-assembled networks (TSANET) in $d=2$ dimensions 
where repulsive ``harmonic spheres" \cite{Berthier10,Berthier11a} are reversibly 
bridged by ideal springs.
As shown in panel (a) of Fig.~\ref{fig_algomod}, it is assumed that the springs break 
and recombine locally with an MC hopping frequency $\nu$.
The particles are monodisperse and the temperature $T$ is set to unity.
The body of our numerical results has been obtained using periodic simulation boxes of 
linear size $L=100$ containing $40000$ springs and $n=10000$ beads, i.e. 
$\rho = n/V = 1$.\footnote{Due to the strong repulsion of the beads and the high number
density, the bead distribution is always macroscopically
homogeneous and the overall density fluctuations are weak.
This has been checked using snapshots and the standard
radial pair correlation function $g(r)$ and its Fourier
transform $S(q)$ as discussed elsewhere \cite{WKC16}.} % \cite{foot_TSANET_homogeneous}.
We also report in Sec.~\ref{shear_V} on data for the same number densities of particles 
and springs obtained for $n=V=L^2 = 100$, $300$, $1000$, $3000$, $30000$ and $100000$.
In addition to the MC moves, changing the connectivity matrix of the network,
standard MD simulation with a strong Langevin thermostat \cite{AllenTildesleyBook}
is used to move the particles effectively by overdamped motion.\footnote{This allows us to 
suppress long-range hydrodynamic modes otherwise relevant for two-dimensional systems.}
We systematically vary $\nu$ using a constant total sampling time $\tsampmax = 10^5$.
For each trajectory we store every $\tincr = 0.01$ instantaneous properties.
Our configurations are first equilibrated in the liquid regime ($\nu = 1$).
We then gradually decrease $\nu$ over several orders of magnitude. 
The hopping moves are switched off ($\nu = 0$) for quenched networks.
All properties are averaged over $\Nc=100$ configurations.

\subsection{Polydisperse Lennard-Jones particles}
\label{app_algo_pdLJ}
We present in Appendix~\ref{app_shear_pdLJ} data obtained for polydisperse Lennard-Jones (pLJ) particles 
in $d=2$ dimensions \cite{AllenTildesleyBook,WTBL02,TWLB02,XWP12,WXP13} where 
the interaction range $\sigma = (D_i + D_j)/2$ is set by the Lorentz rule 
\cite{HansenBook} with $D_i$ and $D_j$ being the diameters of the interacting particles $i$ and $j$. 
Following Refs.~\cite{WTBL02,TWLB02,XWP12,WXP13,lyuda20a} the particle diameters are uniformly 
distributed between $0.8$ and $1.2$. The standard particle number used is $n=10000$.
Additional results obtained for $n=100$, $200$, $500$, $1000$, $2000$ and $5000$ are given in Sec.~\ref{shear_V}.
See panel (b) of Fig.~\ref{fig_algomod} for a snapshot of a configuration at $T=0.2$.
By means of an MC barostat \cite{WXP13} we impose an overall pressure $P=2$ to 
equilibrate the systems at a given temperature $T$.
\begin{figure}[t]
\centerline{\resizebox{0.9\columnwidth}{!}{\includegraphics*{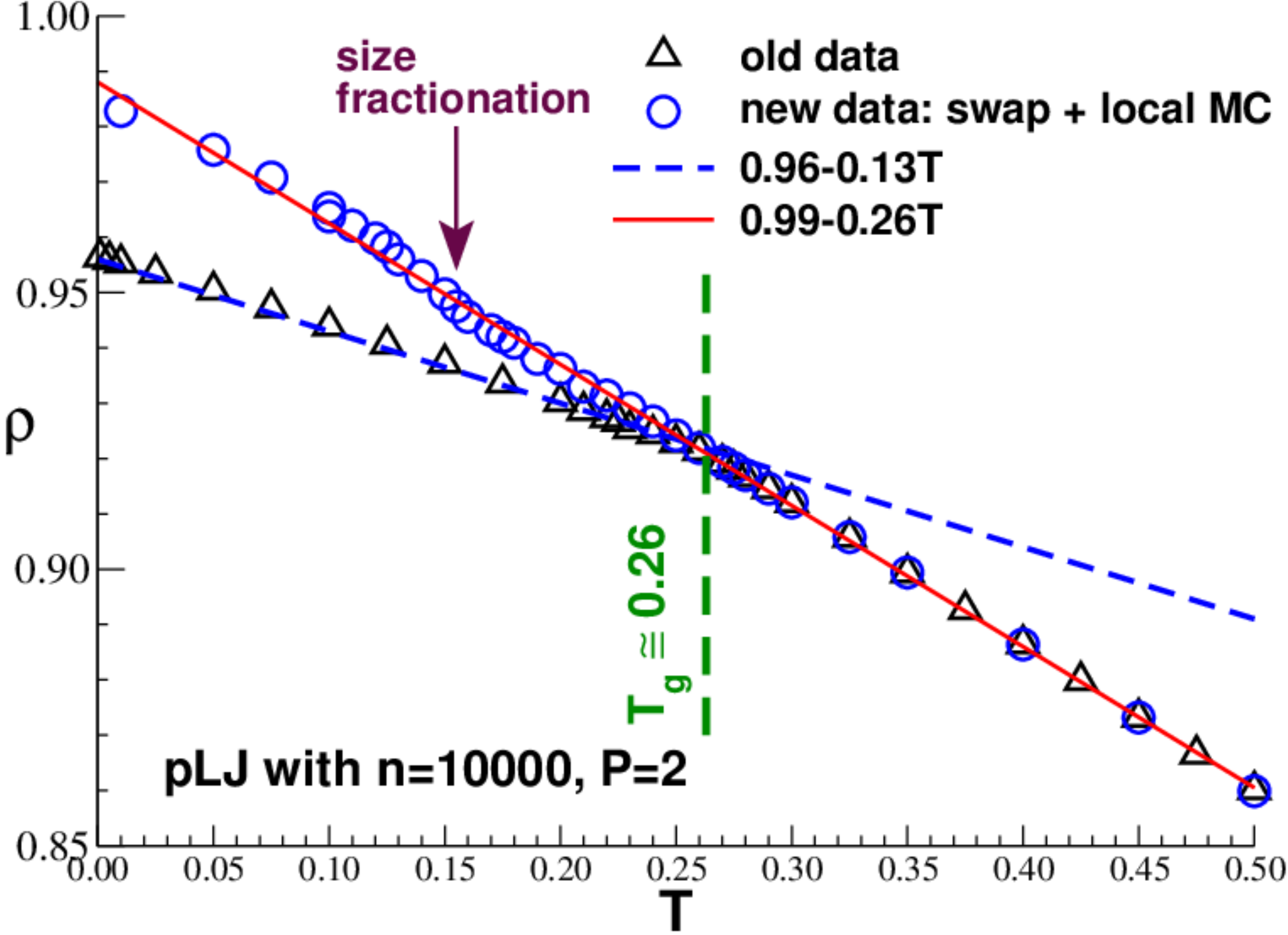}}}
\caption{Number density $\rho$ of the pLJ model at normal pressure $P=2$ 
as a function of temperature $T$. 
The old data (triangles) from Ref.~\cite{WXP13}, obtained from a continuous
cooling process (with rate $10^{-7}$) using local MC moves only, reveal two distinct 
linear slopes which were used to determine a glass transition temperature $\Tglass \approx 0.26$.
Using in addition swap moves much higher densities have been achieved (circles).
}
\label{fig_pdLJ_rho_T}
\end{figure}
As shown in Fig.~\ref{fig_pdLJ_rho_T}, the number density $\rho(T)$ of the older data
from Ref.~\cite{WXP13} reveal two distinct linear slopes which have been used
to define a glass transition temperature $\Tglass \approx 0.26$. For details see Ref.~\cite{WXP13}.
We present here new data \cite{lyuda20a} where we have used in addition MC swap moves \cite{Berthier17} 
exchanging pairs of particles of different sizes. 
Time is measured in terms of the local MC steps per particle as in Ref.~\cite{WXP13}.
We temper now each of the $\Nc=100$ independent configurations over $10^7$ MC steps 
at each temperature with switched on barostat and local and swap MC moves.
We then fix the volume and temper over again $10^7$ MC steps with local and swap moves and 
over additional $10^7$ MC steps only with local MC moves. Production runs only using local 
MC moves are then performed over $\tsampmax=10^7$. For temperatures between $T=0.21$ and
$T=0.25$ additional production runs have been done for $\Nc=20$ over $\tsampmax=10^8$ MC steps.
The data are normally sampled in intervals of $\tincr=10$ MC steps.
The number densities obtained from these new simulations are indicated in Fig.~\ref{fig_pdLJ_rho_T} (circles).
As can be seen, much higher densities are achieved at low temperatures
and $\rho(T)$ follows now the high-temperature slope (solid line) well below $\Tglass$.
Below $\Tdemix \approx 0.15$ the fractionation (demixing) of particles
of different size is observed, i.e. similar sized particles get closer. 
As shown in Appendix~\ref{app_shear_pdLJ},
the glass transition temperature $\Tglass \approx 0.26$, determined originally from the two 
density slopes, remains useful since it indicates roughly the temperature below which only using 
local MC moves the terminal relaxation time $\taualph(T)$ %\cite{foot_taualpha}
obtained from the relaxation of the shear-stress fluctuations exceeds $\tsampmax$.

\begin{figure}[t]
\centerline{\resizebox{.9\columnwidth}{!}{\includegraphics*{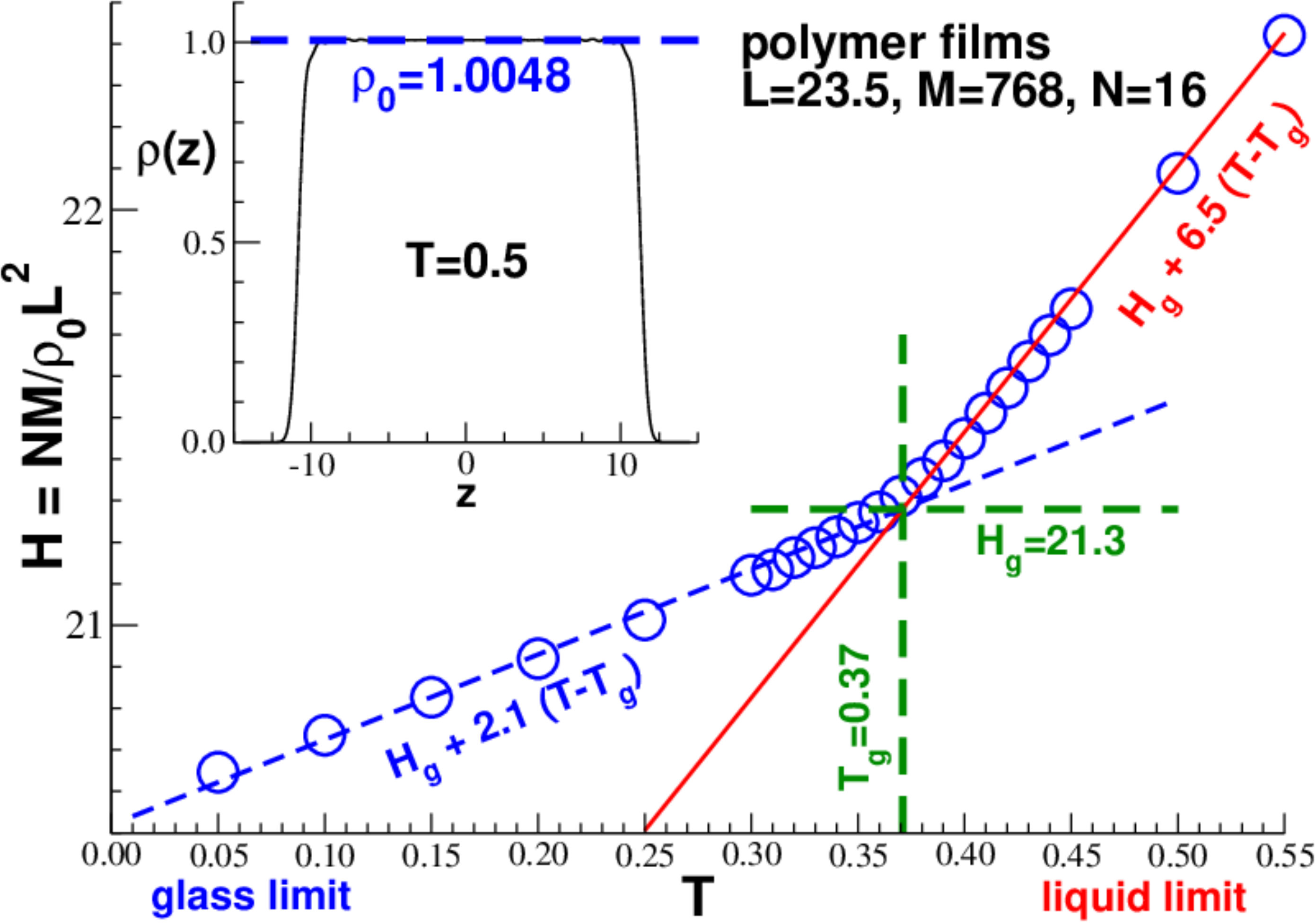}}}
\caption{Film thickness $H$ and glass transition temperature $\Tglass$ of free-standing polymer films.
Inset: 
Number density profile $\rho(z)$ for $T=0.5$ with $z=0$ corresponding to the center of mass of each film. 
Using the measured midplane density $\rho_0(T)$ (horizontal line) yields 
$H \equiv N M /\rho_0 L^2$ \cite{film18}.
Main panel: 
$H$ as a function of $T$. $\Tglass \approx 0.37$ is obtained from the intercept of the linear 
extrapolations of the glass (dashed line) and liquid (solid line) limits \cite{film18}.
}
\label{fig_film_H_T}
\end{figure}

\subsection{Free-standing polymer films}
\label{app_algo_film}

As sketched in panel (c) of Fig.~\ref{fig_algomod}, we study by means of MD simulation 
\cite{AllenTildesleyBook} of a bead-spring model \cite{LAMMPS} 
free-standing polymer films suspended parallel to the $(x,y)$-plane \cite{film18}. 
All unconnected monomers interact with a truncated and shifted LJ potential
while connected monomers are bonded by harmonic springs
\cite{LAMMPS,SBM11,Frey15,ivan17a,ivan17c,ivan18,lyuda19a,film18}.
The films contain $M=768$ monodisperse chains of length $N=16$, i.e. in total $n=12288$ monomers, 
in a periodic box of lateral box size $L=23.5$.
%
%\footnote{The vertical box size $L_z$ is chosen 
%sufficiently large ($L_z \gg H$) to avoid any interaction in this direction. The instantaneous stress 
%tensor vanishes outside the films. While this implies for all $z$-planes within the films that 
%the average vertical normal stress must vanish \cite{film18}, the tangential
%normal stresses must be finite at the film surface since otherwise the surface tension $\Gamma$
%\cite{AllenTildesleyBook} would vanish and the film be unstable.
%Note that $\Gamma \approx 1.7$ at $T \approx \Tglass$ decreasing weakly with $T$.
%Importantly, the surface tension gives rise to a non-negligible contribution
%to $\RA$ due to the last term of Eq.~(\ref{eq_muexhat}).
%The shear modulus $M$ does not vanish at high temperatures (as it must)
%if this term is not taken into account as in some studies \cite{Riggleman13}.} 
% \cite{foot_otherfilms,foot_boundary}.
%
Entanglement effects are irrelevant for the short chains
\cite{FerryBook,DoiEdwardsBook,RubinsteinBook,GraessleyBook}. 
As discussed in detail in Ref.~\cite{film18}, ensembles of $\Nc=100$ independent configurations 
are generated for a broad range of temperatures. % \cite{foot_boundary}.
Production runs are performed over $\tsampmax=10^5$ storing data each $\tincr=0.05$.
A central parameter for the description of our films is the film thickness $H$.
As shown in Fig.~\ref{fig_film_H_T}, using a Gibbs dividing surface construction and 
measuring the midplane density $\rho_0 = \rho(z \approx 0)$ the film thickness is defined 
as $H \equiv N M /\rho_0 L^2$ and the film volume as $V = H L^2$. 
As may be seen from the main panel, $H$ decreases monotonically upon cooling with 
two linear branches fitting the glass (dashed line) and the liquid (solid line) limits. 
$\Tglass \approx 0.37$ is estimated from the intercept of both asymptotes. 

\subsection{Data handling}
\label{app_algo_data}
As indicated above we equilibrate for each state point of the considered 
model $\Nc = 100$ independent configurations $c$.
This allows to probe all properties accurately.
For each configuration $c$ we compute and store {\em one} long trajectory 
with $\tsampmax/\tincr \approx 10^7$ data entries. 
Since we want to investigate the dependence of various properties on the sampling time $\tsamp$ 
we probe for each $\tsampmax$-trajectory $\Nk$ equally spaced subintervals $k$ of length 
$\tsamp \le \tsampmax$ with $I=\tsamp/\tincr$ entries. 
It is inessential for all properties discussed in the present work whether these subintervals do partially overlap or do not.
Since overlapping subintervals probe similar information it is, however, numerically not efficient to pack them too densely. 
We use below $\Nk = \tsampmax/\tsamp$, i.e. $\Nk$ and $\tsamp$ are thus coupled and the accuracy is better for small $\tsamp$. 

\section{Shear-stress fluctuations continued}
\label{app_shear}

\subsection{Polydisperse Lennard-Jones particles}
\label{app_shear_pdLJ}

\begin{figure}[t]
\centerline{\resizebox{.9\columnwidth}{!}{\includegraphics*{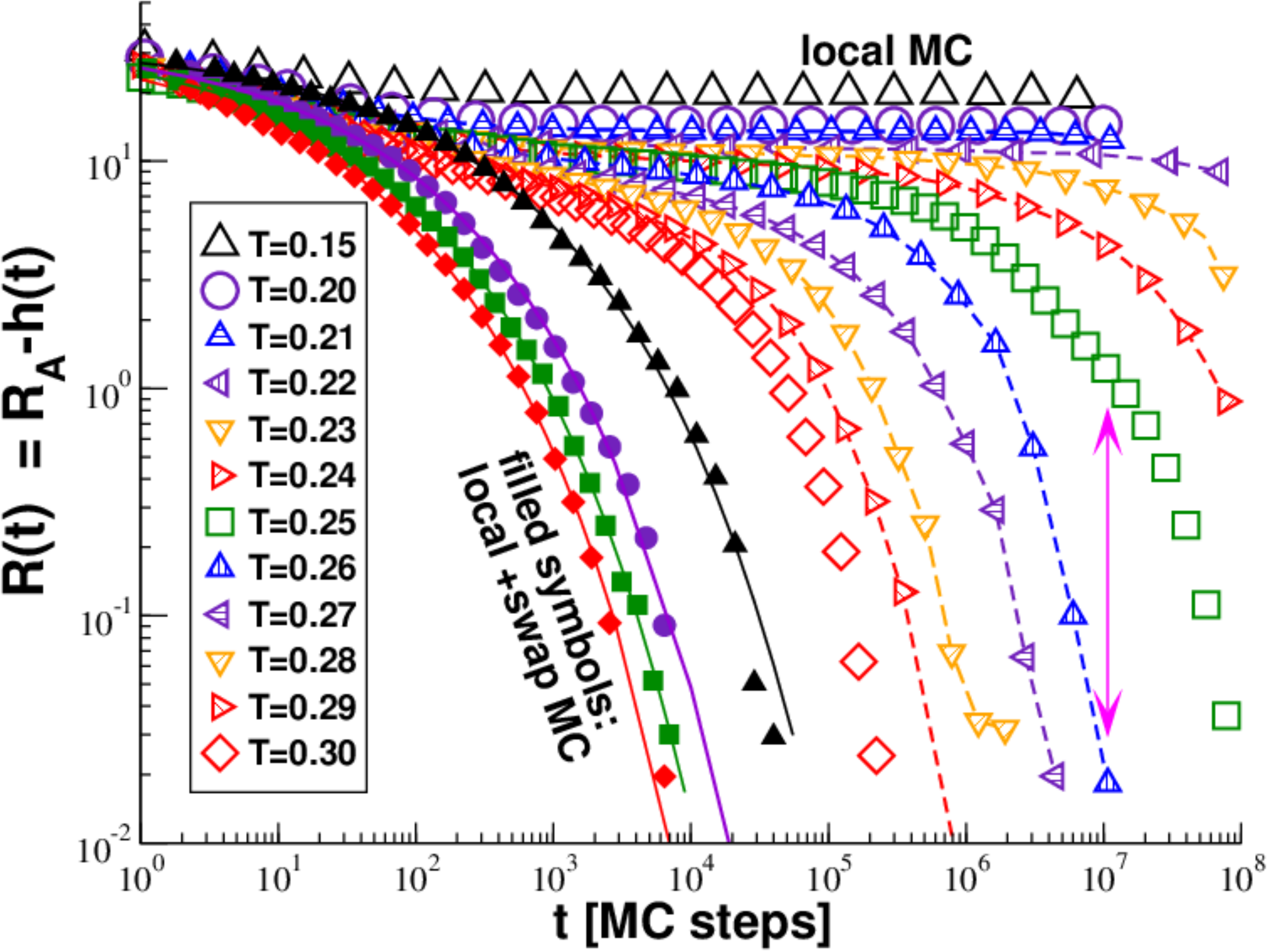}}}
%%\vspace*{-1.5cm}
\caption{Shear-stress relaxation function $R(t)$ for pLJ particles for several temperatures $T$
comparing the case with additional swap moves (filled symbols) with the standard method.
The thin solid lines indicate the phenomenological stretched exponential
$R(t) \approx 35 \exp(-(t/\taualpha)^{1/3})$ with $\taualpha \propto 1/T^{5/2}$
fitting the swapped systems for $T > \Tdemix$.
As shown by the double-headed arrow, 
$\taualpha(T)$, obtained with local moves only, becomes similar to $\tsampmax=10^7$ slightly below $\Tglass \approx 0.26$.}
\label{fig_pdLJ_Rt}
\end{figure}

We turn now to the shear-stress fluctuations of the pLJ model. 
As mentioned in Appendix~\ref{app_algo_pdLJ}, our configurations have been first tempered
and annealed at a constant pressure $P=2$ using in addition to the standard local MC moves 
\cite{LandauBinderBook,WXP13}
swap moves exchanging the particle diameters \cite{Berthier17}. That this changes
dramatically the stress relaxation and thus the equilibration of the configurations
can be seen in Fig.~\ref{fig_pdLJ_Rt} comparing the shear-stress relaxation function $R(t)$ 
for both methods at different temperatures. 
It is seen that the case with swap moves (filled symbols) decays orders of magnitude faster than the
standard method only using local moves.
(While all the data presented elsewhere for the pLJ model refers to productions runs
with $\tsampmax=10^7$ MC steps, we indicate here for the latter case also some temperatures 
sampled with $\tsampmax=10^8$.)
As a quick way to characterize the terminal relaxation time $\taualpha(T)$ one may set, say,
$R(t \approx \taualpha) \approx 0.1$. This implies for the swap moves that
$\taualpha \ll 10^5$ for all temperatures above $\Tdemix$. For instance, $\taualpha(T=0.2) \approx 10^4$
is three orders of magnitudes smaller than the production time $\tsampmax=10^7$.
This suggests that our pLJ samples are well equilibrated and one expects $\RA \approx m_{21}$ above $\Tdemix$. 
As may be seen from the open symbols, this equilibration does not change much the temperature
$\Tglass$ where the relaxation time $\taualpha$ exceeds the production time $\tsampmax=10^7$
if the swap moves are switched off. As emphasized by the double-headed arrow, a similar value
$\Tglass \approx 0.26$ is found as in our previous study \cite{WXP13}. 

\begin{figure}[t]
\centerline{\resizebox{.9\columnwidth}{!}{\includegraphics*{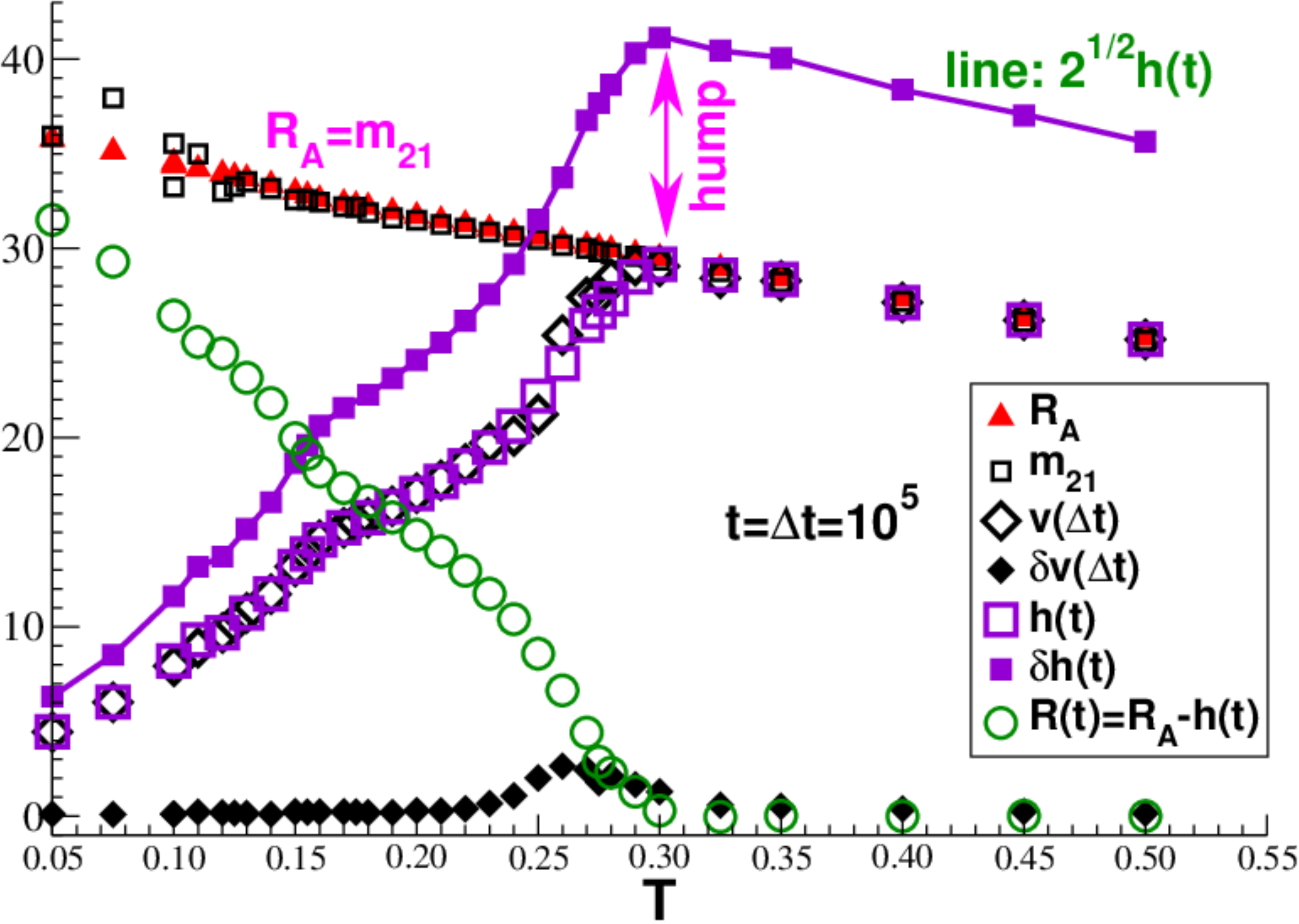}}}
\caption{Various properties obtained for the pLJ model taken at $t=\tsamp=10^5$ vs. $T$: 
affine shear modulus $\RA$, moment $m_{21}$, 
shear-stress fluctuation $v$ and its standard deviation $\delta v$,
mean-square displacement $h$ and its standard deviation $\delta h$
and shear-stress relaxation function $R=\RA-h$.
Note that $\RA \approx m_{21}$ and $h \approx v$ for all $T$ and
that $\delta h \approx 2^{1/2} h$ (bold solid line) holds in agreement with Eq.~(\ref{eq_key_Gauss}). 
}
\label{fig_pdLJ_dh}
\end{figure}

%\paragraph*{$T$-dependence of various expectation values.}
%
Figure~\ref{fig_pdLJ_dh} presents various properties of the pLJ model for one constant 
$t=\tsamp=10^5$ as functions of the temperature $T$. 
As expected for a well-equilibrated liquid,
% (but not necessarily for other mechanical or thermodynamically stable systems), 
$\RA \approx m_{21}$ holds at least down to $\Tdemix$.
(This was not the case for the older data in Ref.~\cite{WXP13}.)
Moreover, they decrease, more or less, linearly with increasing $T$.\footnote{Consistently with 
Ref.~\cite{lyuda19a}, $\RA(T) \approx m_{21}(T) \propto \rho(T)$.}
% \cite{foot_RA_rho}.
%
$\RA$ is the upper bound of the stress-fluctuation $v(\tsamp)$ for any $\tsamp$
since $M(\tsamp) \equiv \RA-v(\tsamp) \ge 0$.
Note that $v(T)$ increases first monotonically with increasing temperatures until it merges 
continuously at $T \approx 0.3$ with $\RA$ decreasing then together with $\RA$. 
This implies that after a monotonic decay at low $T$, the shear modulus $M(T)$ must also vanish 
continuously at $T \approx 0.3$ for $\tsamp=10^5$ (not shown). 
The relaxation function $R$, 
being related to $M$ through the general relation Eq.~(\ref{eq_Rt2M}),
thus vanishes continuously at about the same temperature. 
 
The mean-square displacement $h$ of the instantaneous shear-stresses 
(small, filled diamonds), Eq.~(\ref{eq_cij_hij}), is more or less identical to $v$ for all $T$.
(Closer inspection reveals that $v$ and $h$ slightly differ between $T \approx 0.25$ and $T \approx 0.3$.) 
The standard deviation $\delta h$ of $h$ was calculated using Eq.~(\ref{eq_cshs_fluctu_ga}).
% with the gliding average only added as the last step.
As discussed in Sec.~\ref{theo_gauss_dh}, 
Eq.~(\ref{eq_key_Gauss}) must hold for Gaussian stochastic processes.
As shown by the bold solid line, this is nicely confirmed to high precision.
A log-linear plot of the non-Gaussianity parameter $\alpha_2(t) = \delta h^2/2h^2-1$ 
demonstrates that $|\alpha_2(t)| \ll 0.02$ for all times and temperatures (not shown).
Moreover, since $h \approx v$ for all $T$ and $h = v = \RA$ above $T \approx 0.3$,
this implies the observed non-monotonic behavior of $\delta h(T)$. 
We have also checked that $\delta R \approx \delta h$ (not shown)
as one expects \cite{WKC16,lyuda19a} since $\RA$ barely fluctuates.
%
%Being one of the central results of this work, 
These findings confirm that Gaussian processes are dominant for all $T$,
even below the demixing transition at $\Tdemix$.

\begin{figure}[t]
\centerline{\resizebox{.9\columnwidth}{!}{\includegraphics*{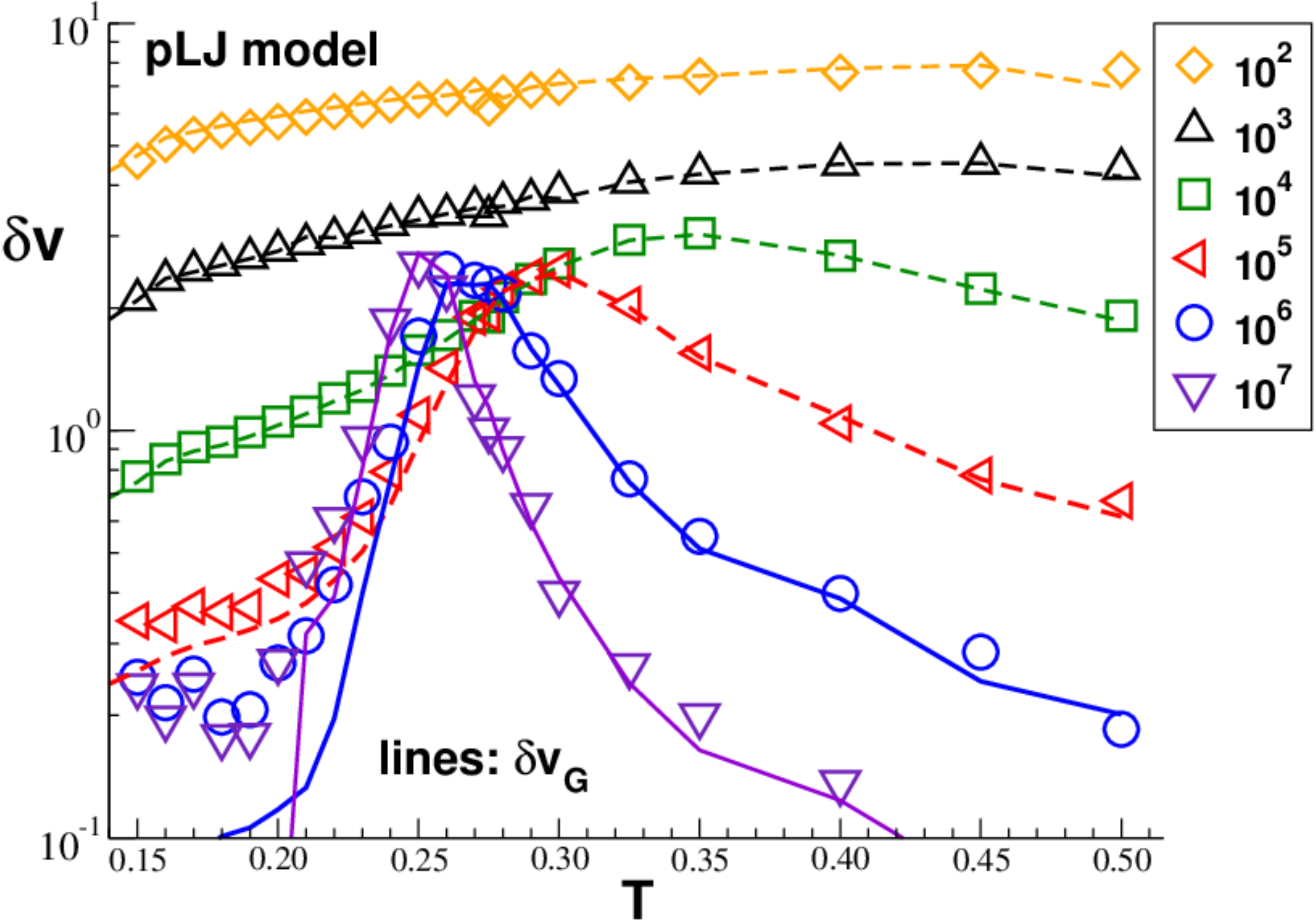}}}
%%\vspace*{-1.5cm}
\caption{$\delta v$ (open symbols) and $\svgauss$ (lines) as functions of $T$ for a broad range of $\tsamp$. 
While $\delta v \approx \svgauss$ holds for large $T$ and small $\tsamp$,
$\delta v$ is seen to become constant for low $T$ and high $\tsamp$.
An increasingly sharp maximum appears around $\Tglass$ shifting with increasing $\tsamp$ to lower $T$. 
}
\label{fig_pdLJ_dv_T}
\end{figure}

%\paragraph*{Standard deviation $\delta v$ vs. temperature $T$.}
%
Also indicated at the bottom of Fig.~\ref{fig_pdLJ_dh} is the standard deviation $\delta v$ of $v$.
Figure~\ref{fig_pdLJ_dv_T} compares $\delta v$ (open symbols) and $\svgauss$ (lines)
as functions of $T$ for a broad range of $\tsamp$.
Naturally, our data get less accurate for $\tsamp \to \tsampmax$.
The most important point is here that within numerical precision $\delta v \approx \svgauss$ for all $\tsamp$ 
for large $T$ even slightly below $\Tglass$.
With increasing $\tsamp$ the maximum of $\delta v(T) \approx \svgauss(T)$ becomes sharper and shifts to lower $T$.
Interestingly, $\delta v \approx \svgauss$ even holds for $T \ll \Tglass$, 
but only for $\tsamp \ll 10^5$. However, while $\svgauss$ decreases strongly upon cooling for larger $\tsamp$
this is not observed for $\delta v$ being only weakly $T$-dependent.
Thus, Eq.~(\ref{eq_key_2}) does not hold in this limit.
 
\begin{figure}
\centerline{\resizebox{.9\columnwidth}{!}{\includegraphics*{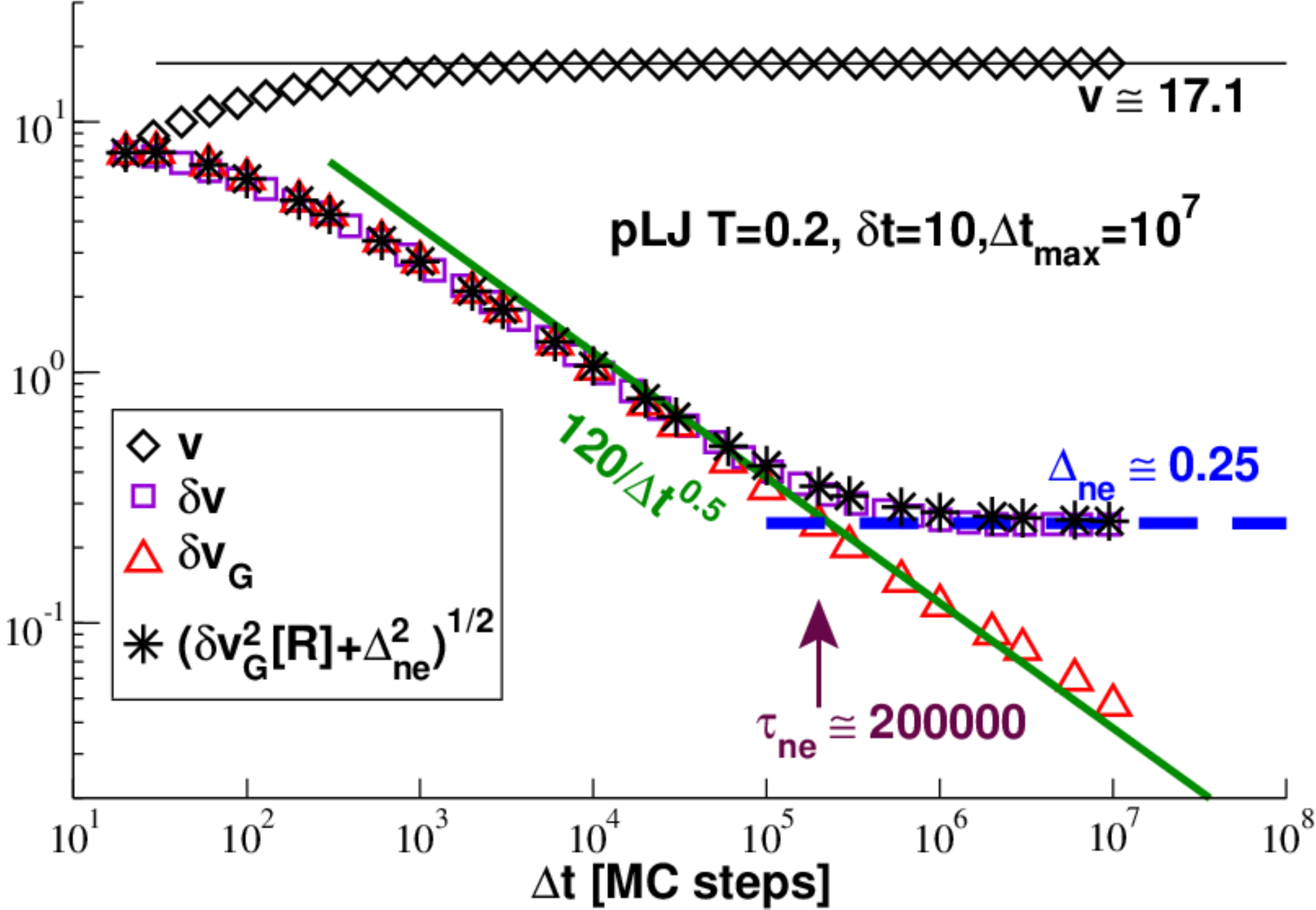}}}
%%\vspace*{-1.5cm}
\caption{Shear-stress fluctuation $\delta v$ and the corresponding standard deviations
$\delta v$ and $\svgauss$ {\em vs.} $\tsamp$ at $T=0.2$ which is well below $\Tglass$. 
$v(\tsamp)$ is seen to become constant above $\tsamp \approx 10^3$
where $v \approx h(t) \approx 17.1$. As indicated by the bold solid line,
$\svgauss \propto 1/\sqrt{\tsamp}$ as expected in this regime.
$\delta v$ levels off for $\tsamp \gg \Tnonerg \approx 200000$. 
}
\label{fig_pdLJ_dv_tsamp}
\end{figure}
 
Focusing on one low temperature ($T=0.2$) Fig.~\ref{fig_pdLJ_dv_tsamp} compares
$v$ with $\delta v$ and $\svgauss$ as functions of $\tsamp$. 
In agreement with Eq.~(\ref{eq_vc}), $v(\tsamp)$ increases monotonically becoming rapidly constant.
Since consistently with Eq.~(\ref{eq_plateau}) also $h(t)$ and $R(t)$ become constant (Fig.~\ref{fig_pdLJ_Rt}),
this implies $\svgauss(\tsamp) \propto 1/\sqrt{\tsamp}$ (bold solid line).
At variance to the decay of $\svgauss$,
$\delta v \to \Snonerg \approx 0.25$ for $\tsamp \gg \Tnonerg$ (dashed horizontal line) 
similar to the behavior for quenched TSANET systems (Fig.~\ref{fig_TSANET_dv_flow}).
As in Sec.~\ref{shear_TSANET} Eq.~(\ref{eq_dv_gen}) yields a reasonable 
approximation of $\delta v$ albeit the shifted data (stars) are slightly above $\delta v$ 
at $\tsamp \approx \Tnonerg$.
%This is consistent with $\Delta(\tsamp)$ approaching $\Snonerg$ from below
%as expected from the TSANET data (Fig.~\ref{fig_TSANET_s}). 
%
%The system-size effects of $\delta v$ and $\Snonerg$ for low-temperature pLJ particles 
%will be discussed in Sec.~\ref{shear_V}.

%\section{Shear-stress fluctuations in free-standing polymer films}
\subsection{Free-standing polymer films}
\label{app_shear_film}

\begin{figure}
\centerline{\resizebox{0.9\columnwidth}{!}{\includegraphics*{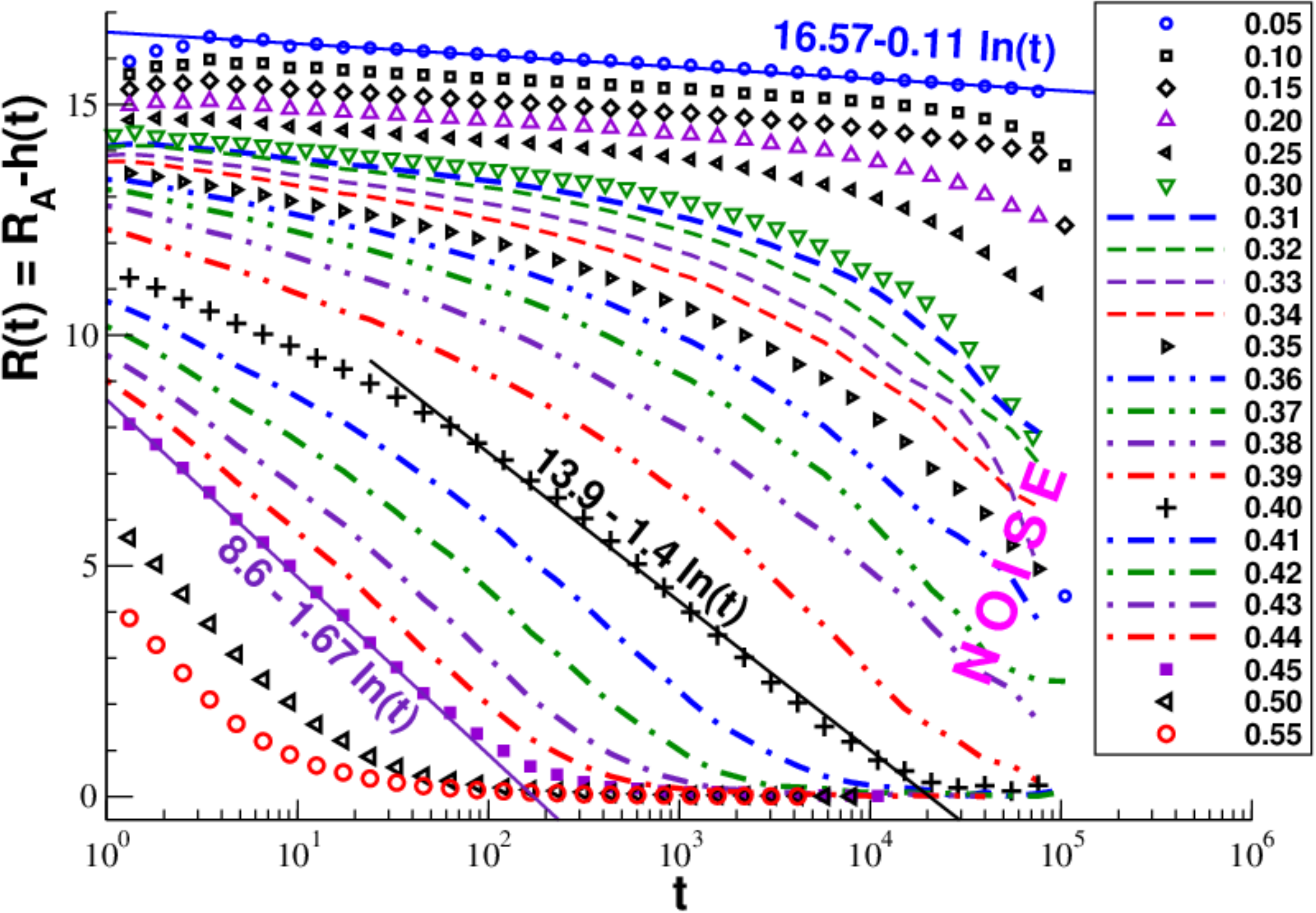}}}
%%\vspace*{-1.5cm}
\caption{$R(t) = \RA-h(t)$ for free-standing films
using half-logarithmic coordinates for a broad range of temperatures $T$. 
$R(t)$ increases continuously with decreasing $T$.
Logarithmic creep behavior (thin solid lines) with $R(t) \approx a - b \ln(t)$ is found
for $T \ll \Tglass$ and above the glass transition ($T \approx 0.45$).
}
\label{fig_film_Rt}
\end{figure}

%\paragraph*{Shear-stress relaxation modulus $R(t)$.}
%\label{film_intro}
The rheological properties of free-standing polymer films may be characterized in a computer 
experiment by means of the shear-stress relaxation function $R(t)$ as shown in 
Fig.~\ref{fig_film_Rt}.\footnote{Experimental studies on polymer films \cite{McKenna05,McKenna08}
rather investigate the (shear-strain) creep compliance $J(t)$ \cite{FerryBook}.
Since the Laplace transforms of $J(t)$ and $R(t)$ are reciprocally related
\cite{FerryBook}, if one function is precisely known, the other can be calculated in principle.}
%\cite{foot_Jt}. 
%
Albeit we average over $\Nc=100$ independent configurations 
it was necessary for the clarity of the presentation to use gliding averages,
Eq.~(\ref{eq_cij_hij}), i.e. the statistics becomes worse for $t \to \tsampmax=10^5$. 
%Without this strong averaging the data would appear too noisy around $\Tglass$.
%See Ref.~\cite{film18} for a discussion of the standard deviation $\delta G$ of $R(t)$.
Interestingly, it is clearly seen that $R(t)$ increases {\em continuously} with decreasing $T$
\cite{film18}.
%
%\footnote{Our results are consistent with the continuous decay of the storage modulus
%$G^{\prime}(\omega=const,T)$ shown in Fig.~6 of Ref.~\cite{Pablo05b}.
%Similar continuous behavior has also been reported for the Young modulus \cite{Riggleman13}.
%}
% foot_debate_film}
% with decreasing $T$ without any indication of the suggested 
%jump-singularity \cite{GoetzeBook,Szamel11,Ikeda12,Yoshino14,Klix12,Klix15} but 
in perfect agreement with all published experimental \cite{McKenna05,McKenna08,McKenna17} and 
computational \cite{Pablo05b,Riggleman13} studies.
%
%As shown elsewhere \cite{film18}, our data are compatible with a successful 
%time-temperature superposition scaling (TTS) \cite{FerryBook,GraessleyBook}. 
%The validity of the TTS is important as it supports the continuous nature of the solidlike 
%elasticity emergence at the glass transition \cite{lyuda19a,FerryBook,McKenna05,McKenna08,McKenna17}.
%
The presented $R(t)$-data is used below to obtain $\svgauss[R]$.

\begin{figure}[t]
\centerline{\resizebox{.9\columnwidth}{!}{\includegraphics*{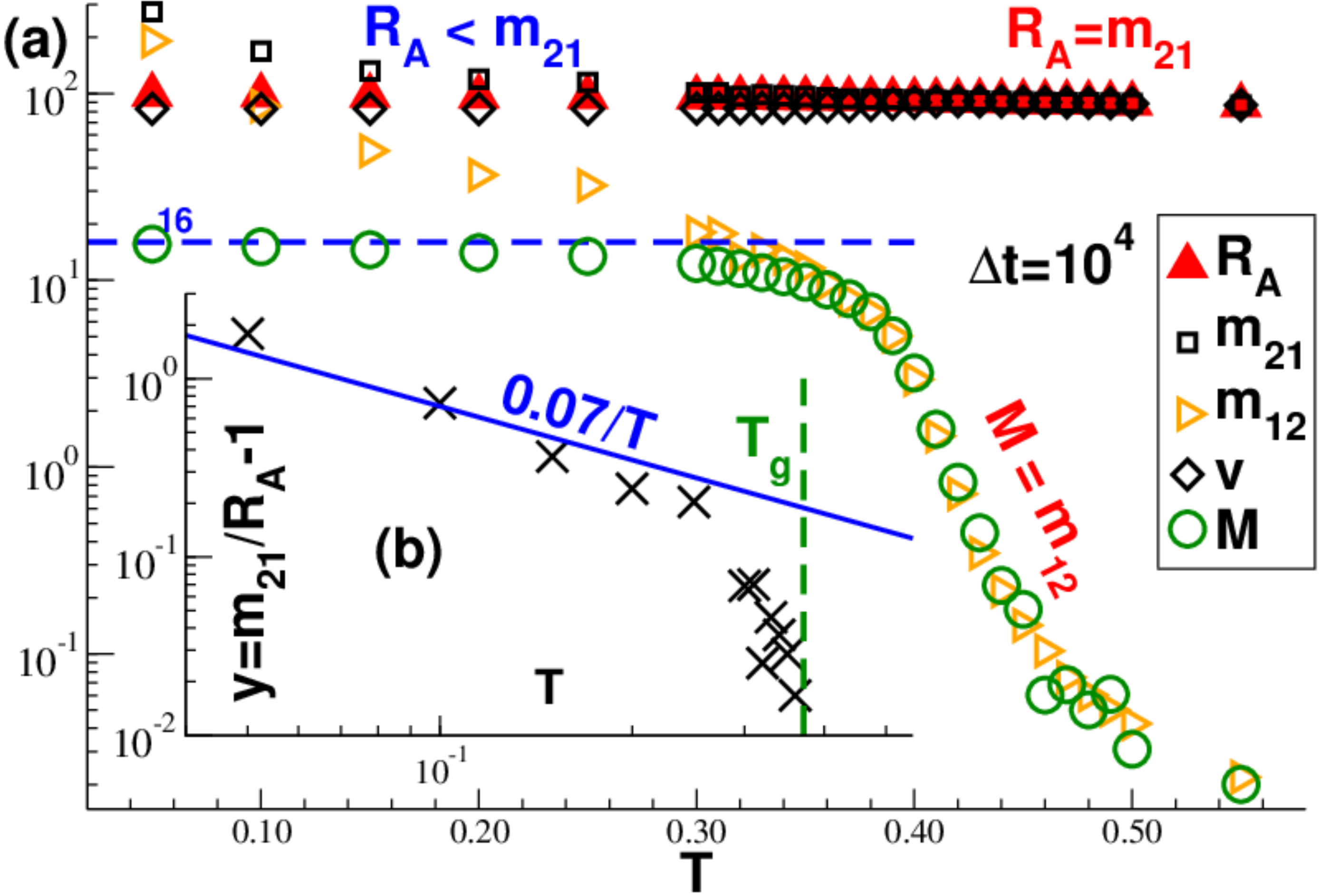}}}
\centerline{\resizebox{.9\columnwidth}{!}{\includegraphics*{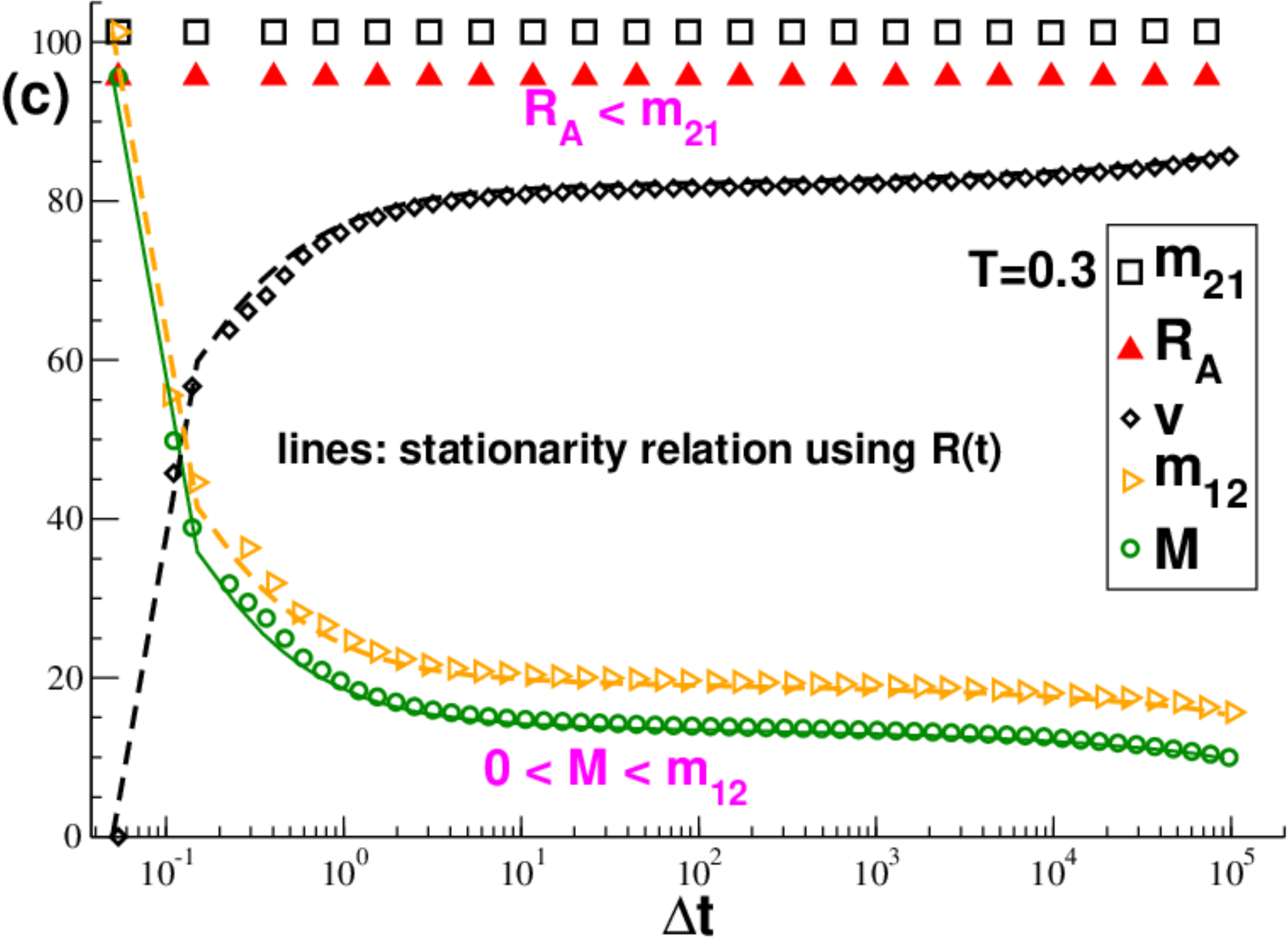}}}
%%%\vspace*{-1.5cm_
\caption{Contributions to $M=\RA-v$:
{\bf (a)} Temperature dependence for $\tsamp=10^4$.
{\bf (b)} Double-logarithmic representation of $m_{21}/\RA-1$ vs. $T$.
{\bf (c)} $\tsamp$-effects for $M$ and its contributions for $T=0.3$.
Only $\RA$ and $m_{21}$ are strictly $\tsamp$-independent.
The dashed lines have been obtained using Eq.~(\ref{eq_vc}), the solid line using Eq.~(\ref{eq_Rt2M})
taking advantage of the directly measured shear-stress relaxation function $R(t)$.
}
\label{fig_film_SFF}
\end{figure}

Figure~\ref{fig_film_SFF} presents various contributions to the generalized shear modulus 
$M=\RA-v=(\RA-m_{21})+m_{12}$. %, Eq.~(\ref{eq_Rt2M}).
At variance to Fig.~\ref{fig_pdLJ_dh} a half-logarithmic
representation is used in panel (a) to better show the decay of $M \approx m_{12}$
for temperatures above the glass transition where $\RA \approx m_{21}$.
Note that $\RA < m_{21}$ below $\Tglass$. As may be seen in panel (b)
frozen-in out-of-equilibrium stresses are observed upon cooling below $\Tglass$ as 
made manifest by the dramatic increase of the dimensionless parameter $y=m_{21}/\RA-1$.
The prefactor $\beta=1/T$ of $m_{21}$, Eq.~(\ref{eq_muFtwo}), implies due to the frozen stresses
the observed $y(T) \propto 1/T$ for $T \ll \Tglass$.
(Similar behavior has been reported for three-dimensional polymer bulks \cite{ivan18}.)
%
%%\paragraph*{Stationarity.}
%%\label{film_intro}
Since there is currently no algorithm comparable to the swap algorithm \cite{Berthier17}
allowing to equilibrate glass-forming polymer melts and films as for the pLJ model, 
our films are clearly not at thermal equilibrium below $\Tglass$,
at least not in the sense of an equilibrium liquid.
This does not mean that the stochastic process is not {\em effectively stationary}. 
This is addressed in panel (c) where we test for one temperature below $\Tglass$ 
(where $\RA < m_{21}$) that Eq.~(\ref{eq_vc}) and Eq.~(\ref{eq_Rt2M}) 
hold for $v$ (dashed lines) and $M$ (solid line). 
To test these relations $R(t)$ was integrated numerically.\footnote{The visible minor differences are due 
to numerical difficulties related to the finite time step and the inaccurate integration of the strongly
oscillatory $R(t)$ at short times.} % \cite{foot_film_stationary}.
Albeit time-translational invariance apparently holds,
this does not mean that no aging occurs (for $T < \Tglass$)
but just that this is irrelevant for the timescales considered here.

\begin{figure}[t]
\centerline{\resizebox{0.9\columnwidth}{!}{\includegraphics*{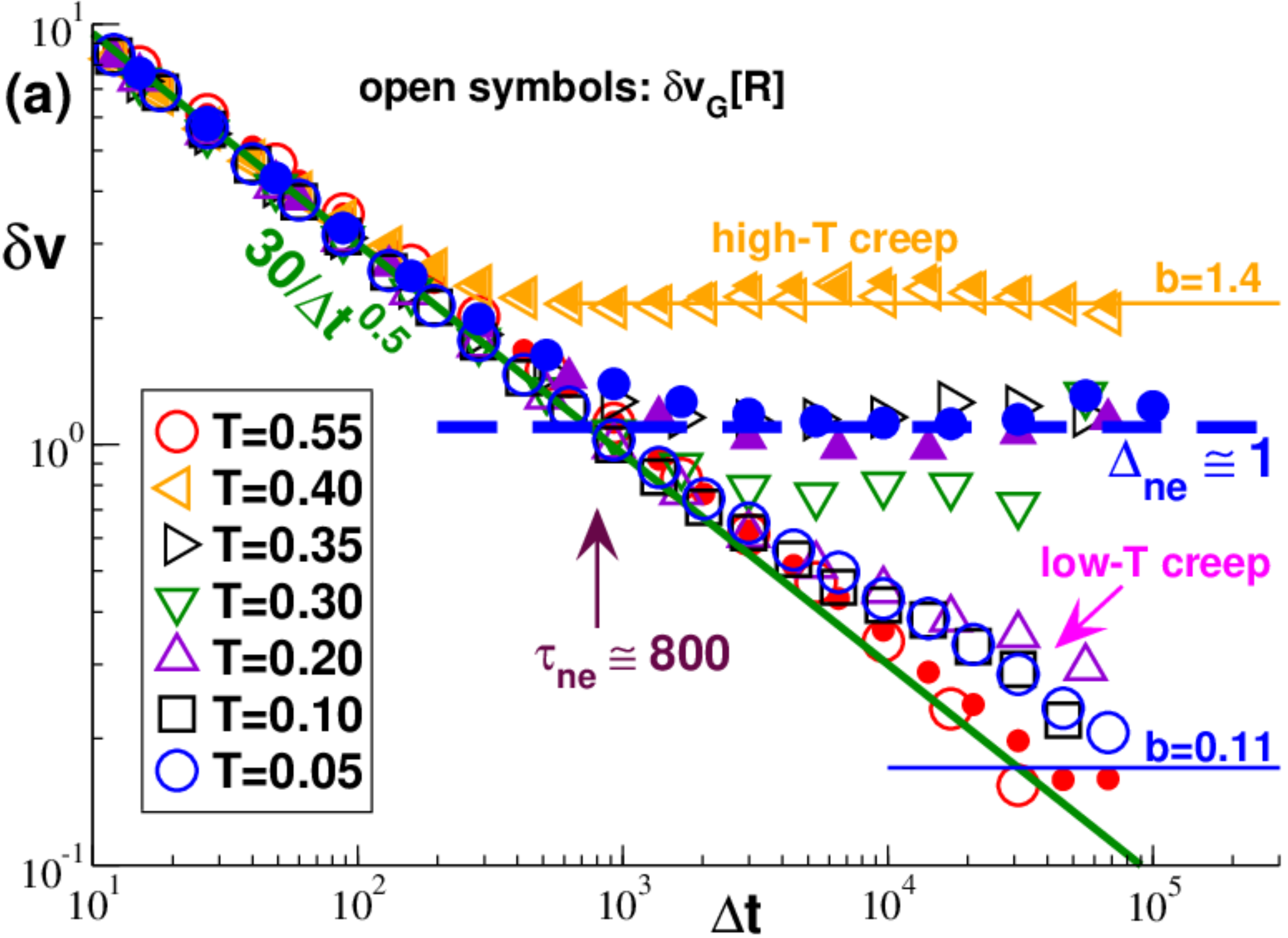}}}
\centerline{\resizebox{0.9\columnwidth}{!}{\includegraphics*{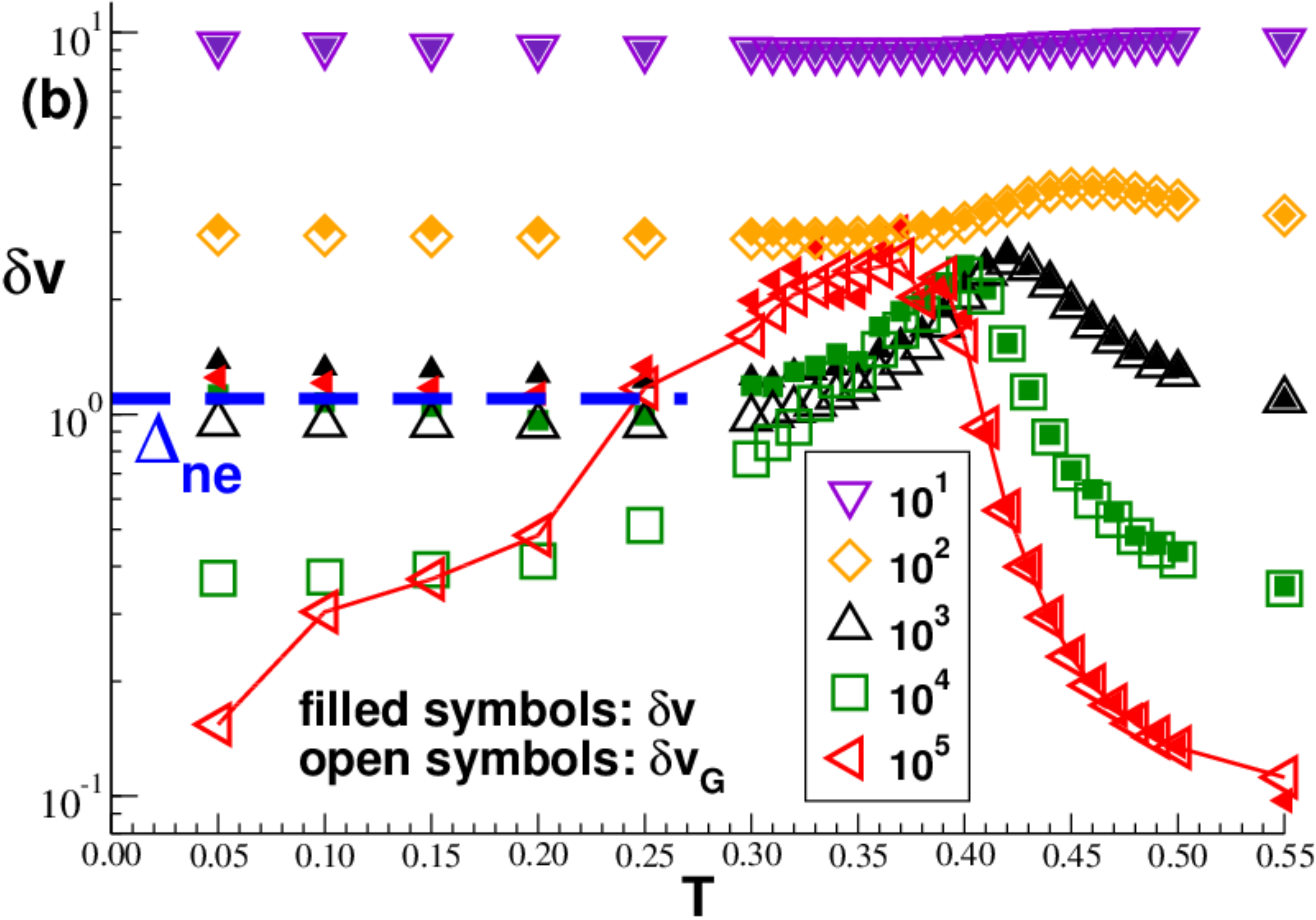}}}
\caption{Comparison of $\delta v$ (filled symbols) and $\svgauss[R]$ (open symbols) for free-standing films:
{\bf (a)}
$\tsamp$-dependence of $\svgauss[R]$ (open symbols) for all indicated $T$ and 
of $\delta v$ (filled symbols) for $T=0.55$, $0.4$, $0.2$ and $0.05$.
The $1/\sqrt{\tsamp}$-decay for small $\tsamp$ is shown by the bold solid line,
the plateau value $\Snonerg \approx 1$ of $\delta v$ for small $T$ and $\tsamp \gg \Tnonerg \approx 800$ 
by the bold dashed line and $\svgauss \approx 1.55|b|$
($b=1.4$ for $T=0.4$ and $b=0.11$ for $T=0.05$) 
expected for logarithmic creep by thin horizontal lines.
{\bf (b)}
Temperature dependence for different $\tsamp$ as indicated.
}
\label{fig_film_dv}
\end{figure}

%\paragraph*{Standard deviation $\delta v$.}
%\label{film_intro}

We address in panel (a) of Fig.~\ref{fig_film_dv} 
the $\tsamp$-dependence of $\delta v$ and $\svgauss$ (open symbols) 
for different temperatures as indicated. 
$\delta v \approx \svgauss$ holds again for $T$ above and around $\Tglass$. 
Note also that $\delta v \approx \svgauss \propto 1/\sqrt{\tsamp}$ 
(bold solid line) for all $\tsamp \le \tsampmax$ for the largest temperature $T=0.55$.
Interestingly, $T$-dependent shoulders appear for $T \approx 0.4$. 
This is a consequence of the creep-like decay of $R(t)$ in this regime 
which is approximately fitted by $R(t) \approx a - b \ln(t)$ as indicated in Fig.~\ref{fig_film_Rt}. 
According to Fig.~\ref{fig_ft_creep}, 
one expects a shoulder with $\svgauss \approx 1.55 |b|$.
That this holds is seen by the upper thin horizontal line using $b=1.4$ for $T=0.4$.  
In the low-$T$ limit we see again that $\delta v$ becomes constant, $\delta v \approx \Snonerg$
for $\tsamp \gg \Tnonerg \approx 800$, while $\svgauss$ continues to decrease with $\tsamp$.  
The small deviations of $\svgauss$ from the $1/\sqrt{\tsamp}$-asymptote 
are caused by the fact that $R(t)$ does {\em not} become rigorously constant, $R(t) \to R_p > 0$, 
even for our lowest temperatures.
As shown by the upper thin line in Fig.~\ref{fig_film_Rt}, $R(t)$ is fitted by a logarithmic creep.
The amplitude $b$ of this low-temperature creep are, however, too small to lead to a
clear-cut shoulder for $\svgauss$. 
As shown by the lower thin solid line with $b = 0.11$ for $T=0.05$ at least 
one decade longer production runs are required to make the low-$T$ creep manifest for $\svgauss$.
%
%\begin{figure}[h]
%\centerline{\resizebox{0.9\columnwidth}{!}{\includegraphics*{fig_film_dv_T}}}
%\caption{Temperature dependence of $\delta v$ (filled symbols) and $\svgauss$ (open symbols) for free-standing films.
%$\delta v \approx \svgauss$ holds accurately for large and intermediate $T$ for all $\tsamp$.}
%%%The maximum of $\delta v \approx \svgauss$ slightly below $\Tglass$ becomes larger with increasing $\tsamp$.
%%%At low temperatures $\svgauss$ decreases while $\delta v$ becomes constant upon cooling (dashed horizontal line).}
%\label{fig_film_dv_T}
%\end{figure}
%

A complementary representation is given in penal (b) of Fig.~\ref{fig_film_dv} 
focusing on the temperature dependence of $\delta v$ and $\svgauss$. 
A very similar behavior as in Fig.~\ref{fig_pdLJ_dv_T} for pLJ particles is seen.
Most importantly, $\delta v \approx \svgauss$ for large and intermediate $T$. 
The maximum of $\delta v \approx \svgauss$ slightly below $\Tglass$ becomes systematically larger with increasing $\tsamp$ and shifts to lower temperatures. 
Also it is confirmed that Eq.~(\ref{eq_key_2}) holds for sufficiently small $\tsamp$ for all $T$. 
As expected, Eq.~(\ref{eq_key_2}) breaks down for $\tsamp \gg \Tnonerg(T) \approx 800$ in the non-ergodic limit (lowest temperatures).
While $\svgauss$, measuring the fluctuations within each configuration, is seen to decrease upon cooling,
$\delta v$ becomes constant, $\delta v \to \Snonerg$ (dashed horizontal line), due to the finite dispersion 
of the quenched $v_c$ of the independent configurations.
%While $\delta v \approx \svgauss$ for small $\tsamp$ and high $T$, both differ for 
%$T \ll \Tglass$ and the more the larger $\tsamp$.
%We emphasize again that $\delta v$ differs from $\svgauss$ only for large $\tsamp$ and low $T$
%where $\delta v \to \Snonerg \approx 1$ (dashed horizontal line).

%\clearpage
%\newpage
\bibliographystyle{epj.bst}
%\bibliography{../../../bibl/BOOK,../../../bibl/JPW,../../../bibl/ANS,../../../bibl/Barrat,../../../bibl/Baschnag,../../../bibl/Berthier,../../../bibl/EM,../../../bibl/Einstein,../../../bibl/Friedrich,../../../bibl/Fuchs,../../../bibl/Karmakar,../../../bibl/Kob,../../../bibl/LAMMPS,../../../bibl/Leibler,../../../bibl/Ligoure,../../../bibl/Maret,../../../bibl/McKenna,../../../bibl/Pablo,../../../bibl/Porte,../../../bibl/Procaccia,../../../bibl/Provencher,../../../bibl/Riggleman,../../../bibl/Safran,../../../bibl/Szamel,../../../bibl/Terentjev}

\end{document}